\documentclass[
prd,
aps,
superscriptaddress,
showpacs,
preprintnumbers,
amsmath,
floatfix,
twocolumn,
nofootinbib
]{revtex4}

\usepackage[
]{graphicx}
\usepackage{epstopdf}
\usepackage[psamsfonts]{amssymb}
\usepackage{ulem}

\unitlength=1.2mm \preprint{RBRC-675,KEK-TH-1150}
\begin{document}

\newcommand{\tr}{\mbox{tr}\,}
\newcommand{\DDs}{$D$/$D_{ s}$\ }
\newcommand{\mres}{m_{\rm res}}
\newcommand{\simg}{\rlap{\raise -4pt \hbox{$\sim$}}
                   \raise 3pt \hbox{$>$}}
\newcommand{\siml}{\rlap{\raise -4pt \hbox{$\sim$}}
                   \raise 3pt \hbox{$<$}}

\title{Nucleon structure with two flavors of dynamical domain-wall
fermions}

\author{Huey-Wen Lin}\email{hwlin@jlab.org}
\affiliation{Jefferson Laboratory, 12000 Jefferson Avenue, Newport
News, VA 23606} \affiliation{RIKEN-BNL Research Center, Brookhaven
National Laboratory, Upton, NY 11973}

\author{Tom Blum} \email{tblum@phys.uconn.edu}
\affiliation{Physics Department, University of Connecticut, Storrs, CT 06269-3046}
\affiliation{RIKEN-BNL Research Center, Brookhaven National Laboratory, Upton, NY 11973}

\author{Shigemi Ohta} \email{shigemi.ohta@kek.jp}
\affiliation{Institute of Particle and Nuclear Studies, KEK, Tsukuba, 305-0801, Japan}
\affiliation{RIKEN-BNL Research Center, Brookhaven National Laboratory, Upton, NY 11973}
\affiliation{Physics Department, Sokendai Graduate U.\ Adv.\ Studies, Hayama, Kanagawa 240-0193, Japan}

\author{Shoichi Sasaki} \email{ssasaki@phys.s.u-tokyo.ac.jp}
\affiliation{Department of Physics, University of Tokyo, Hongo 7-3-1, Bunkyo-ku, Tokyo 113, Japan}

\author{Takeshi Yamazaki}\email{yamazaki@phys.uconn.edu}
\affiliation{Physics Department, University of Connecticut,
Storrs, CT 06269-3046} \affiliation{RIKEN-BNL Research Center,
Brookhaven National Laboratory, Upton, NY 11973}

\date{Feb. 4, 2008}
\begin{abstract}
We present a numerical lattice quantum chromodynamics calculation of
isovector form factors and the first few moments of the isovector structure functions of the nucleon. The calculation employs two degenerate dynamical flavors of domain-wall fermions, resulting in good control of chiral symmetry breaking. Non-perturbative renormalization of the relevant quark currents is performed where necessary. The DBW2 gauge action is used to further improve the chiral behavior while maintaining a reasonable physical lattice volume. The inverse lattice spacing, $a^{-1}$, is approximately 1.7~GeV. Degenerate up and down dynamical quark masses of approximately 1, 3/4 and 1/2 times the strange quark mass are used. The physical volume of the lattice is about $(1.9\mbox{ fm})^3$. The ratio of the isovector vector to axial charges, $g_A/g_V$, tends to a somewhat lower value than the experimental value as the quark mass is reduced toward the physical point. Momentum-transfer dependences of the isovector vector, axial, induced tensor and induced pseudoscalar form factors are calculated. The Goldberger-Treiman relation holds at low momentum transfer and yields an estimation of the pion-nucleon coupling, $g_{\pi NN} = 15.5(1.4)$, where the quoted error is only statistical.
We find that the flavor non-singlet quark momentum fraction $\langle x\rangle_{u-d}$ and quark helicity fraction $\langle x\rangle_{\Delta u - \Delta d}$ overshoot their experimental values after linear chiral extrapolation. We discuss possible systematic errors for this discrepancy. An estimate for transversity, $\langle 1 \rangle_{\delta u - \delta d} = 0.93(6)$ in $\overline{\rm MS}$ at 2~GeV is obtained and a twist-3 polarized moment, $d_1$, appears small, suggesting that the Wandzura-Wilczek relation holds approximately. We discuss in detail the systematic errors in the calculation, with particular attention paid to
finite volume, excited state contamination, and chiral extrapolations.
\end{abstract}
\maketitle

\section{Introduction}

In this paper we report numerical lattice quantum chromodynamics (QCD) calculations of the isovector form factors and moments of the isovector structure functions of the nucleon on gauge field configurations with two degenerate flavors of dynamical domain-wall fermions (DWF)~\cite{Kaplan:1992bt,Shamir:1993zy,Furman:1994ky} generated by the RIKEN-BNL-Columbia (RBC) Collaboration.

Four isovector form factors parameterize neutron $\beta$ decay: the vector and induced tensor form factors from the vector current,
\begin{equation}\label{eq:cont_vector}
\langle p| V^+_\mu(x) | n \rangle = \bar{u}_p \left[\gamma_\mu
g_{V}(q^2) -q_{\lambda} \sigma_{\lambda \mu} {g_{T}(q^2)} \right]
u_n e^{iq\cdot x},
\end{equation}
and the axial and induced pseudoscalar form factors from the axial current,
\begin{equation}\label{eq:cont_axial}
\langle p| A^+_\mu(x) | n \rangle = \bar{u}_p
            \left[\gamma_\mu  \gamma_5 g_{A}(q^2)
             +i q_\mu \gamma_5 {g_{P}(q^2)} \right]  u_n e^{iq\cdot x},
\end{equation}
which are given here in the Euclidean metric convention. Thus, $q^2$ as used in this paper stands for Euclidean four-momentum squared, and corresponds to the time-like momentum squared as $q_M^2=-q^2<0$ in Minkowski space. We also note a slight change in the sign convention of the pseudoscalar form factor from our earlier publications, such as Ref.~\cite{Sasaki:2003jh}. Here $q=p_n-p_p$ is the momentum transfer between the proton ($p$) and neutron ($n$). In the limit $|{\vec q}| \rightarrow 0$, the momentum transfer should be small because the mass difference of the neutron and proton is only about 1.3~MeV. This makes the limit $q^2 \rightarrow 0$, where the vector and axial form factors dominate, a good approximation. Their values in this limit are called the vector and axial charges of the nucleon: $g_{V} = g_{V}(q^2=0)$ and $g_{A} = g_{A}(q^2=0)$. Experimentally, $g_{V} = \cos \theta_C$ (with the Cabibbo mixing angle $\theta_C$), and $g_{A} = 1.2695(29) \times g_{V}$~\cite{PDBook}.

These form factors are calculable on the lattice~\cite{Liu:1994dr,Gockeler:2003ay,Alexandrou:2006ru,Hagler:2007xi,Alexandrou:2007xj,Gockeler:2007hj,Yamazaki:2007mk,Sasaki:2007gw}  but quite often are prohibitively complicated if one uses conventional staggered or Wilson fermions. The staggered fermions, with their complicated flavor/taste structure, make even definition of baryon operators difficult. The Wilson fermions make the necessary current renormalization complicated due to large explicit violation of chiral symmetry. The domain-wall fermions (DWF)~\cite{Kaplan:1992bt,Kaplan:1992sg,Shamir:1993zy,Furman:1994ky}, with their exponential suppression of chiral symmetry breaking, make such renormalizations easy. In particular, currents that are connected by chiral transformation such as vector and axial ones should share the same renormalization. Thus, the ratio $g_{A}/g_{V}$ is naturally renormalized in DWF lattice calculations~\cite{Sasaki:2003jh}.

Earlier (mostly quenched) lattice calculations~\cite{Fukugita:1995fh,Liu:1994ab,Gockeler:1996wg} of this ratio gave significant underestimations by up to 20\%. In a quenched calculation with DWF we found much of this deficit comes from the small lattice volumes employed in those earlier studies~\cite{Sasaki:2003jh}.
With the sufficiently large lattice volume of about $(2.4\mbox{ fm})^3$ used in our quenched calculation with DWF and the improved DBW2 gauge action~\cite{Takaishi:1996xj,deForcrand:1999bi}, we observed that the ratio does not depend strongly on the quark mass and obtained a value of $1.212 \pm 0.027({\rm stat}) \pm 0.024({\rm norm})$ in the chiral limit.
The value is almost consistent with experiment, and combined with the very weak dependence on the quark mass, does not require any chiral extrapolation but a linear one in quark mass. Thus an obvious question arises: do these observations hold when the quarks are treated dynamically?

There has been recent interest in the pseudoscalar form factor, $g_P(q^2)$, due to the MuCap Collaboration's new high-precision experiment studying ordinary muon capture (OMC) by protons, $\mu^{-} p \to \nu_{\mu} n$~\cite{Andreev:2007wg}. The OMC experiments determine the induced pseudoscalar coupling $g_P=m_{\mu}g_P(q^2)$ at $q^2=0.88m_{\mu}^2$, where $m_\mu$ is the muon mass. The new experiment yields $g_P=7.3\pm1.1$, which is consistent with the value predicted by chiral perturbation theory, $g_P^{\rm ChPT}=8.26\pm0.16$~\cite{Bernard:2001rs}, but much smaller than the last world average for OMC, $(g_P^{\rm OMC})_{\rm ave}=10.5\pm 1.8$ given in Ref.~\cite{Gorringe:2002xx}, and the value from a single experiment of radiative muon capture (RMC), $\mu^{-} p \to \nu_{\mu} n\gamma$, $g_P^{\rm RMC}=12.4 \pm 1.0$~\cite{Jonkmans:1996my}. Recently, a quenched DWF calculation~\cite{Sasaki:2007gw} reported a result, $g_P=8.15\pm0.54({\rm stat})\pm0.16({\rm norm})$, consistent with the MuCap experiment. It is interesting to see what value is obtained in a $n_f=2$ dynamical DWF calculation.

In this paper we also report on this form factor, $g_{P}(q^2)$. It is the induced part of Eq.~(\ref{eq:cont_axial}), directly related to the pion-nucleon coupling, $g_{\pi NN}$, which should satisfy
$\displaystyle
m_{N} g_{A} = F_\pi g_{{\pi NN}};$
at finite momentum transfer,
\begin{equation}\label{eq:cont_GT_qne0}
2 m_{N} g_{A}(q^2)-q^2 {g_{P}(q^2)} = \frac{2g_{{\pi NN}} F_\pi
m_\pi^2}{q^2+m_\pi^2}
\end{equation}
with a more traditional convention of $F_\pi=f_\pi/\sqrt{2}\sim 92$~MeV.
These should hold up to residual dependence on the momentum transfer $q^2$; the pion-nucleon coupling at high momentum transfer may significantly differ from its value near zero momentum transfer.
On the lattice there may also arise some corrections from the finite lattice volume.

Since we study the dependence of these form factors at relatively low momentum transfer, we can try to extract corresponding mean square radii,
$\langle r^2 \rangle_{{V, T, A}}$, defined by
 $\displaystyle g_{V, T, A} (q^2)/g_{V, T, A}(0)\sim 1 - \frac{1}{6}q^2
 \langle r^2\rangle_{{V, T, A}}  + \cdots$.

The structure functions are measured in deep inelastic scattering
of leptons from
nucleons~\cite{Breidenbach:1969kd,Friedman:1991nq,Kendall:1991np,Taylor:1991ew,Gluck:1995yr,Gehrmann:1995ag,Lai:1996mg,Adams:1997tq,Adeva:1997qz,Gluck:1998xa,Ackerstaff:1999ey,Martin:2001es},
the cross section of which is factorized in terms of leptonic
and hadronic tensors:
\begin{eqnarray}\label{eq:cross_section}
k_0^\prime\frac{d\sigma}{d^3k^\prime} = \frac{2 M}{s-M^2}
\frac{\alpha^2}{(q^2)^2} l^{\mu\nu}W_{\mu\nu}.
\end{eqnarray}
The leptonic tensor is known to be
\begin{equation}\label{eq:leptonic_tensor}
l_{\mu\nu}(k,k^\prime) = 2 \left( k_\mu k^\prime_\nu + k^\prime_\mu k_\nu
-\frac{1}{2} Q^2 g_{\mu\nu}\right).
\end{equation}
Hence, the cross section provides us with structure information
about the target nucleon through the hadronic tensor, $W_{\mu\nu}$,
which is decomposed into symmetric unpolarized and antisymmetric
polarized parts:
\begin{widetext}
\begin{equation}\label{eq:Wsym}
W^{\{\mu\nu\}}(x,Q^2) =
\left( -g^{\mu\nu} + \frac{q^\mu
q^\nu}{q^2}\right)  {F_1(x,Q^2)} +
\left(P^\mu-\frac{\nu}{q^2}q^\mu\right)\left(P^\nu-\frac{\nu}{q^2}q^\nu\right)
\frac{F_2(x,Q^2)}{\nu}
\end{equation}
\begin{equation}\label{eq:Want}
W^{[\mu\nu]}(x,Q^2) = i\epsilon^{\mu\nu\rho\sigma} q_\rho
\left(\frac{S_\sigma}{\nu}({g_1(x,Q^2)} \right. +
 \left. {g_2(x,Q^2)}) - \frac{q\cdot S
P_\sigma}{\nu^2}{g_2(x,Q^2)} \right).
\end{equation}
\end{widetext}
$\nu = q\cdot P$, $S^2 = -M^2$, and $x=Q^2/2\nu$, and \(Q^2=|q^2|\). The unpolarized structure functions are $F_1(x,Q^2)$ and $F_2(x,Q^2)$, and the polarized, $g_1(x,Q^2)$ and $g_2(x,Q^2)$. Their moments are described in terms of Wilson's operator product expansion:
\begin{widetext}
\begin{eqnarray}
\label{eq:moments}
2 \int_0^1 dx\,x^{n-1} {F_1(x,Q^2)} &=& \sum_{q=u,d}
c^{(q)}_{1,n}(\mu^2/Q^2,g(\mu))\: \langle x^n \rangle_{q}(\mu)
+{O(1/Q^2)},
\nonumber \\
\int_0^1 dx\,x^{n-2} {F_2(x,Q^2)} &=& \sum_{f=u,d}
c^{(q)}_{2,n}(\mu^2/Q^2,g(\mu))\: \langle x^n \rangle_{q}(\mu)
+{O(1/Q^2)},
\nonumber \\
2\int_0^1 dx\,x^n {g_1(x,Q^2)}
  &=& \sum_{q=u,d} e^{(q)}_{1,n}(\mu^2/Q^2,g(\mu))\: \langle x^n \rangle_{\Delta q}(\mu)
+{O(1/Q^2)},  \nonumber
 \\
2\int_0^1 dx\,x^n {g_2(x,Q^2)}
  &=& \frac{1}{2}\frac{n}{n+1} \sum_{q=u,d} \left[e^{q}_{2,n}(\mu^2/Q^2,g(\mu))\: d_n^{q}(\mu)
-  2 e^{q}_{1,n}(\mu^2/Q^2,g(\mu))\: \langle x^n \rangle_{\Delta
q}(\mu)\right] + {O(1/Q^2)},
\end{eqnarray}
\end{widetext}
where $c_1$, $c_2$, $e_1$, and $e_2$ are perturbatively known
Wilson coefficients and ${\langle x^n \rangle_{q}(\mu)}$,
${\langle x^n \rangle_{\Delta q}(\mu)}$ and $d_n(\mu)$ are
calculable on the lattice as forward nucleon matrix elements of
certain local operators.

Again, the conventional staggered or Wilson fermions would complicate such lattice calculations for the same reasons as discussed for the form factors. The DWF calculations are simpler because of easier renormalizations due to good chiral symmetry. In particular the first moments $\langle x \rangle_{u-d}$ (quark momentum fraction) and $\langle x \rangle_{\Delta u - \Delta d}$ (quark helicity fraction) share a common renormalization, and so their ratio is naturally renormalized in DWF calculations~\cite{Orginos:2005uy}.

During the last few years lattice computations have provided many interesting results for these structure function moments~\cite{Gockeler:1996wg,Gockeler:1999jb,Detmold:2001jb,Dolgov:2002zm,Gockeler:2002mk,Orginos:2002mn,Hagler:2003jd,Ohta:2003ux,Ohta:2004mg,Sasaki:2003jh,:2003is,Gockeler:2003jf,Khan:2004vw,Gockeler:2004wp,Orginos:2005uy,Edwards:2006qx,Gockeler:2006ui,Lin:2006xt,Orginos:2006pl,Schroers:2007qf,Hagler:2007xi,Lin:2007zzc,Lin:2007gv,Yamazaki:2007mk,Hagler:2007hu}, in both quenched and full QCD. These calculations provide first-principles values for the moments of structure functions at leading twist. One of the major unresolved issues in these previous calculations is significant overestimation of the moments $\langle x \rangle_{u-d}$ and $\langle x \rangle_{\Delta u-\Delta d}$ compared with results from fits to the experimental data~\cite{Gockeler:1996wg,Dolgov:2002zm,Gockeler:2004wp,Orginos:2005uy}.
In our quenched calculation~\cite{Orginos:2005uy} of these quantities we encountered similar overestimations. Interestingly, however, the ratio of these quantities,  $\langle x \rangle_{u-d}/ \langle x \rangle_{\Delta u-\Delta d}$ showed very weak dependence on the quark mass and agreed well with the experimental ratio in the chiral limit. Again an obvious question is whether this behavior survives when the quarks are treated dynamically.

We address these questions with a lattice QCD calculation using two degenerate dynamical flavors of domain-wall fermions and DBW2 rectangle-improved gauge action~\cite{Aoki:2004ht}. Three bare sea quark masses of 0.02, 0.03 and 0.04 in lattice units (corresponding to pion masses of about 0.5, 0.6 and 0.7~GeV) are used with about 200 gauge configurations each. The lattice cutoff is $a^{-1}\sim1.7 \mbox{ GeV}$, and the spatial volume is about $(1.9\mbox{ fm})^3$. The lattice cutoff is sufficiently high to allow us to take advantage of various benefits of DWF such as good chiral and flavor symmetries in performing fully-nonperturbative renormalization for our nucleon observables.
Unlike our past quenched calculations~\cite{Sasaki:2003jh,Orginos:2005uy} which did not show significant dependence on quark mass in most of the observables, we sometimes observe significant deviations at the lightest quark mass from the heavier quark mass results. The pion mass times lattice extent, $m_\pi L$, is slightly less than 5 at the lightest quark mass, which may be problematic for calculations involving a large hadron such as the nucleon. We must therefore perform quark-mass extrapolation on a case-by-case basis and with great caution.

The rest of the paper is organized as follows. In Section~\ref{Sec:Simulation}, we first briefly summarize the numerical ensembles used in the calculation. Then we discuss the choice of nucleon source and sink and the operators used to calculate the form factors and moments of the structure functions. Finally we briefly summarize the numerical nonperturbative renormalization methods. In Section~\ref{sec:Q2_0} we begin with discussion of the form factors at zero momentum transfer. We show the conservation of the isovector vector current is under good control and then give an evaluation of the nucleon axial charge. Then we discuss momentum dependence of various form factors. In particular the Dirac mean-squared charge radius is extracted. The Goldberger-Treiman relation is shown to hold and provides an estimate of the pion-nucleon coupling. Finally, we describe our calculations of some low moments of the structure functions. We summarize our conclusions and future plans in Section~\ref{sec:conclusion}.

\section{Numerical Method}
\label{Sec:Simulation}

We use DWF lattice ensembles generated by the RBC Collaboration~\cite{Aoki:2004ht}. These were generated with two degenerate flavors of dynamical quarks described by the domain-wall fermion~\cite{Kaplan:1992bt,Kaplan:1992sg,Shamir:1993zy,Furman:1994ky} action and gluons described by the doubly-blocked Wilson (DBW2) gauge action~\cite{Takaishi:1996xj,deForcrand:1999bi}. There are three such ensembles with sea quark masses of $m_{\rm sea}=0.02$,
0.03,  and 0.04 in lattice units, which respectively correspond to about 1/2, 3/4 and 1 times the strange quark mass, or degenerate pseudoscalar meson masses of about 500, 600 and 700~MeV. The inverse lattice spacing is about 1.7~GeV, set from the $\rho$-meson mass, yielding a physical volume of $(1.9\mbox{ fm})^3$.
\begin{table*}[htb]
\caption{
Summary of simulation parameters and the numbers of
configurations. The pion mass, the nucleon mass and
the pion decay constant ($F_{\pi}$) are also tabulated.
}\label{tab:SrcPar}
\begin{center}
\begin{tabular}{c|cc|c|c}
\hline\hline
$m_f$ & \multicolumn{2}{c|}{0.02} & 0.03 & 0.04 \\
\hline
$t_{\rm snk}-t_{\rm src}$ & 12 & 10 & 10 & 10\\
$t_{\rm src}$ &  $0,9,18$ & $0, 15$ &$0, 15$ & $0,15$ \\
$r_{\rm gauss}$ & 8 & 4.35 & 4.35 & 4.35 \\
\# of conf. & 185 each & 220 each & 220 each & 220 each \\
\hline
$m_{\pi}$~(GeV)\cite{Aoki:2004ht} & \multicolumn{2}{c|}{0.493(5)}
& 0.607(4) &  0.695(4)  \\
$m_{N}$~(GeV)\cite{Aoki:2004ht} &   \multicolumn{2}{c|}{1.28(2)}
&1.43(2) & 1.55(2)  \\
$F_{\pi}$~(GeV)\cite{Aoki:2004ht} & \multicolumn{2}{c|}{0.1141(7)}
&0.1232(7)&0.1329(7) \\
\hline\hline
\end{tabular}
\end{center}
\end{table*}

We study nucleon matrix elements using quark propagators obtained with Gaussian gauge-invariant sources with radius ($r_{\rm gauss}$) of 4.35 for all three quark masses and additionally with Gaussian smearing radius
of 8 for the lightest quark mass. The former parameter was previously used for the neutron dipole moments calculation on the $m_{\rm sea}=0.03$ and 0.04 ensembles~\cite{Berruto:2005hg}. The latter is a better size for overlap with the ground state of the nucleon. For details of our source definition see Ref.~\cite{Berruto:2005hg}. We use only unitary valence quark mass values, $m_{\rm valence}=m_{\rm sea}=m_f$, when computing quark propagators. We use two different sequential sources generated with source-sink locations, $[t_{\rm src}, t_{\rm sink}]=[0, 10]$ and [15, 25], where the source-sink separation $t_{\rm sep}$ is fixed as 10 lattice units, on a given gauge configuration for all three quark masses with the smaller $r_{\rm gauss}$. As for the larger $r_{\rm gauss}$, we use a different choice of source-sink separation, $t_{\rm sep}=12$. The longer separation causes the larger statistical noise. Therefore, we use three sources placed at $[t_{\rm src}, t_{\rm sink}]=
[0, 12]$, [9, 21] and [18, 30] to increase the statistics. Later, we will discuss the possibility of excited-state contaminations in our calculations by comparing results obtained from two sets of parameters,
$\{t_{\rm sep}, r_{\rm gauss}\}$, on the lightest quark ensemble.
The parameter values and the number of configurations used from each ensemble are summarized in Table~\ref{tab:SrcPar}. We also compile some basic physics results from Ref.~\cite{Aoki:2004ht} in the same table.

We define our nucleon two-point functions with  a nucleon interpolation field $\chi$ and smearing parameters $A$ and $B$ as follows:
\begin{widetext}
\begin{equation}
\langle \chi_A(t,{\vec p})\overline{\chi}_B(0,-{\vec p})\rangle
= \sum_{s}\langle 0|\chi_{A}| p,s\rangle \langle p,s|
\overline{\chi}_B|0\rangle  e^{-E(p)t} + \cdot\cdot\cdot
= \frac{E({\vec p})\gamma^t -i\vec\gamma\cdot \vec p+ m_N}{2E}
\sqrt{z_{A}(p)z_{B}(p)}\;e^{-E({\vec p})t}+ \cdot\cdot\cdot
\label{eq:two-pt}
\end{equation}
\end{widetext}
with normalized states defined as
\begin{eqnarray}
\langle 0 |\chi_{A}|p,s\rangle &=& \sqrt{z_{A}({\vec p})} u_s(\vec p),
\end{eqnarray}
and spinors satisfying
\begin{eqnarray}
\sum_{s} u_s(\vec p)\bar u_s(\vec p)&=& E(\vec{p})\gamma^t
-i\vec\gamma\cdot \vec p+ m_N, \label{eq:spinor}
\end{eqnarray}
where $E({\vec p})=\sqrt{m_N^2+{\vec p}^2}$. The ellipsis in Eq.~(\ref{eq:two-pt}) denotes excited-state contributions, which can be ignored in the case of $t\gg 1$.

We often use the projector $\displaystyle {\cal P}_+ = \frac{1+\gamma_t}{2}$ to the positive-parity states,
\begin{widetext}
\begin{equation}
\label{eq:two-pt_proj}
\Gamma^{(2)}_{AB} (t_{\rm snk},t_{\rm src};\vec{p})
=\frac{1}{4}{\rm Tr}({\cal P}_{+}\langle \chi_A(t_{\rm snk},{\vec p})\overline{\chi}_B
(t_{\rm src},-{\vec p})\rangle)
= \frac{E({\vec p})+m_N}{2E}\sqrt{z_{A}({\vec p})z_{B}({\vec p})} \;e^{-E({\vec p})(t_{\rm snk}-t_{\rm src})} +\cdot\cdot\cdot.
\end{equation}
\end{widetext}
These two-point functions provide appropriate normalization factors when we extract matrix elements from three-point correlation functions, as well as estimates for the nucleon mass.

Now we define the three-point functions:
\begin{widetext}
\begin{eqnarray}
\label{eq:three-pt}
G_{\mu, {AB}}(t_{\rm src},t,t_{\rm snk})
&=& \langle \chi_{A}(t_{\rm snk},\vec p_{\rm snk})\,J_\mu(t,\vec{q})\,
\overline{\chi}_{B}(t_{\rm src},\vec p_{\rm src})\rangle \nonumber \\
&=& \frac{E^\prime\gamma^t +i\vec\gamma\cdot \vec p^\prime+ m_N}{2E^{\prime}}
\cdot \Sigma_\mu \cdot \frac{E\gamma^t -i\vec\gamma\cdot
\vec p+ m_N}{2E}
\, \sqrt{z_{A}({\vec p}^{\prime})
z_{B}({\vec p})}\;e^{-E^\prime(t_{\rm snk}-t)}e^{-E (t-t_{\rm
src})}+\cdots,
\end{eqnarray}
\end{widetext}
where ${\vec q}={\vec p}^{\prime}-{\vec p}$, ${\vec p}={\vec p}_{\rm src}$, ${\vec p}^{\prime}={\vec p}_{\rm snk}$,
$E=E({\vec p}_{\rm src})$ and $E^\prime=E(\vec p_{\rm snk})$.
The operator $\Sigma_\mu$ is appropriately  selected for each observable of interest discussed in the following subsections.

We calculate the nucleon isovector form factors and some low moments of structure functions,
namely the isovector vector ($g_{V}$), induced tensor ($g_{T}$), isovector axial ($g_{A}$), and induced pseudoscalar ($g_{P}$), form factors at both zero and finite momentum transfer, and structure function moments corresponding to the momentum fraction $\langle x \rangle_{u-d}$, helicity fraction $\langle x \rangle_{\Delta u-\Delta d}$, a twist-3 moment $d_1$, and transversity $\langle 1\rangle_{\delta u -\delta d}$.
The respective choice of the lattice operators for these observables are the same as in earlier RBC reports~\cite{Sasaki:2001nf,Sasaki:2003jh,Orginos:2005uy,Sasaki:2007gw} on quenched calculations, and are briefly summarized in what follows.

\begin{figure}[htb]
\includegraphics[width=\columnwidth,clip]{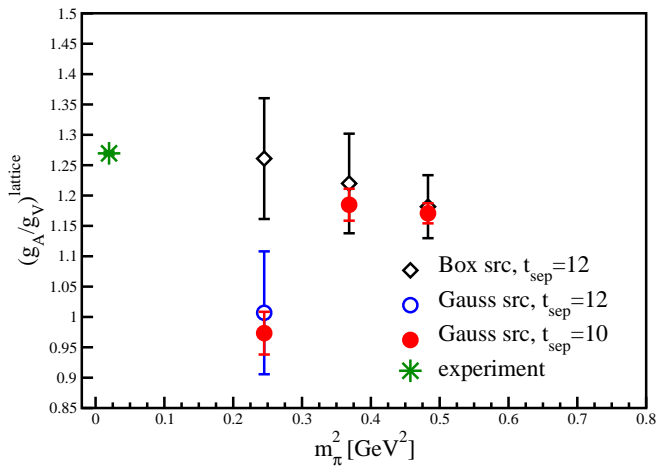}
\caption{Renormalized isovector axial charge $g_{A}/g_{V}$ calculated with box~~\cite{Ohta:2004mg} and Gaussian smeared sources with various source-sink separations ($t_{\rm sep}$). The value from the Gaussian source with $t_{\rm sep}=10$ and smearing radius 4.35 at the lightest quark mass of $m_f=0.02$ deviates significantly from the previous box-source calculation. In that case the box size is chosen for better overlap with the ground state of the nucleon and the longer source-sink separation ($t_{\rm sep}=12$) is adopted to avoid possible excited-state contaminations. This discrepancy is not resolved by using a Gaussian source of an radius 8 and the same source-sink separation. As discussed in the text, this is likely caused by the larger statistical fluctuation in the longer source-sink separation.
}\label{fig:gAgV_src2}
\end{figure}

Before we move on to discuss the operator choices for our observables of interest, we would like to briefly explain our different choice of source-sink separation, $t_{\rm sep}$, and Gaussian smearing radius, $r_{\rm gauss}$, for the lightest quark mass ensemble (see~\cite{Lin:2006xt,Ohta:2004mg,Orginos:2002fr} for the earlier calculations). We started our calculation by setting $t_{\rm sep}=10$ time units and the Gaussian source smearing radius to 4.35, parameters previously used for the neutron dipole moments calculation on the $m_{f}=0.03$ and 0.04 ensembles\cite{Berruto:2005hg}. Although the new results of the Gaussian source in the heaver quark-mass region agree with the heavier quark-mass points of the box source, for the same calculation with $m_f=0.02$, we find a significant decrease in the axial charge compared to our previous box-source calculation, which used a separation of 12 time units as shown in Fig.~\ref{fig:gAgV_src2}. This discrepancy prompted us to examine whether there is a systematic error due to excited-state contamination in the new results.

\begin{figure*}[ht]
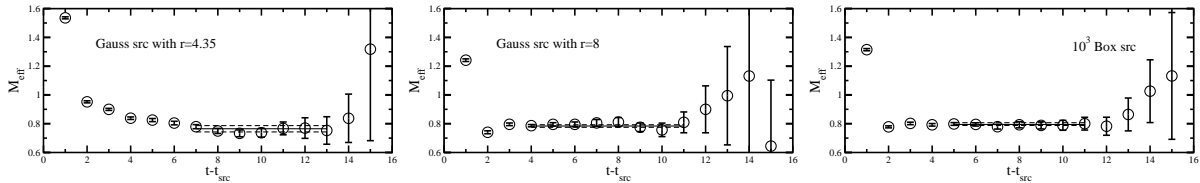

\begin{center}
\begin{tabular}{ccc}
\includegraphics[width=0.6\columnwidth,clip]{figs/meff_002_G10.eps} &
\includegraphics[width=0.6\columnwidth,clip]{figs/meff_002_G12.eps} &
\includegraphics[width=0.6\columnwidth,clip]{figs/meff_002_B12.eps} \\
\end{tabular}
\end{center}
\caption{Nucleon effective mass plots from two-point functions. Gaussian source with smearing radius, $r_{\rm gauss}=4.35$ (left), 8 (center), and box source (right) from the calculation in Ref.~\cite{Ohta:2004mg}.}
\label{fig:meff_src}
\end{figure*}

The effective masses plotted in Fig.~\ref{fig:meff_src} show that
the Gaussian source with smearing radius 4.35 has a ground state plateau that begins at $t-t_{\rm src}=7$ or 8 (top figure), while the box source appears to plateau sooner (bottom figure). Thus one may suspect that excited-state contamination accounts for the effect observed in Fig.~\ref{fig:gAgV_src2}. Since the major difference between the two calculations is the source type and separation, we extend the source-sink separation to $t_{\rm sep}=12$ for the Gaussian source
(the same as in the previous box-source calculation) and also increase its radius to 8 in order to reduce excited-state contamination.
Indeed, the Gaussian source with radius 8 provides similar quality of  plateau in the effective mass plot, as shown in the middle panel of Fig.~\ref{fig:meff_src}. However, as shown in Fig.~\ref{fig:gAgV_src2},  the resulting $g_A/g_V$ with the second parameter set $\{t_{\rm sep},r_{\rm gauss}\}=\{12, 8\}$ is consistent with that of the first one $\{10, 4.35\}$ rather than the previous result of the box source. So, the discrepancy between the results is probably not caused by excited-state contamination.

\begin{figure*}[ht]
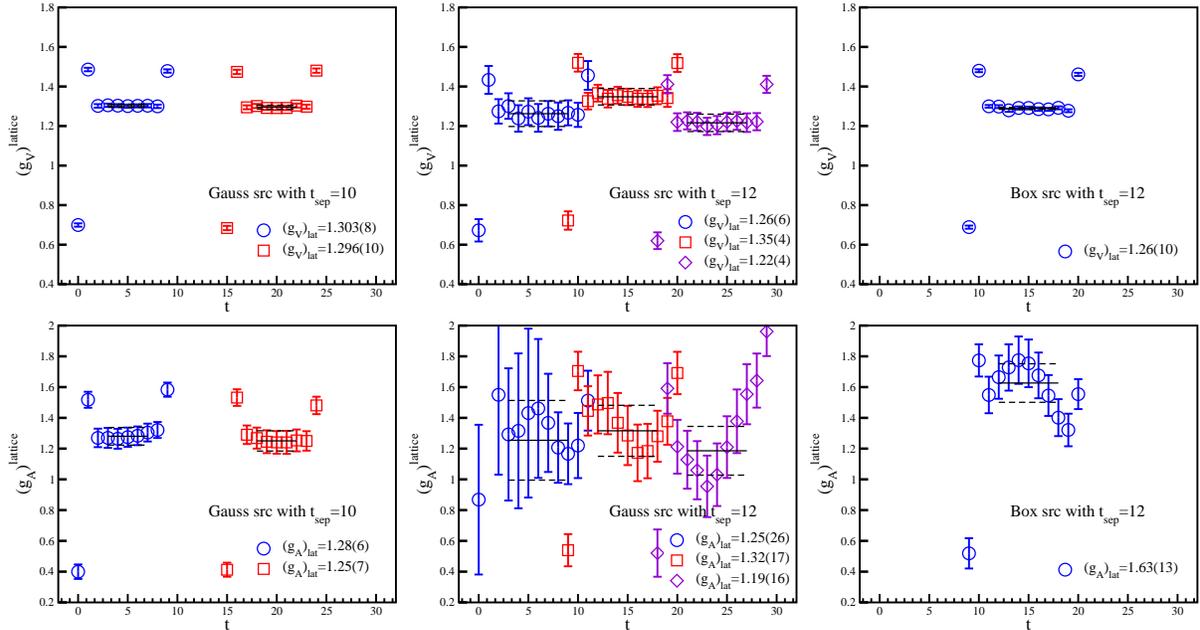

\begin{center}
\begin{tabular}{ccc}
\includegraphics[width=0.6\columnwidth,clip]{figs/gv_tdep_G10.eps} &
\includegraphics[width=0.6\columnwidth,clip]{figs/gv_tdep_G12.eps} &
\includegraphics[width=0.6\columnwidth,clip]{figs/gv_tdep_B12.eps} \\
\includegraphics[width=0.6\columnwidth,clip]{figs/ga_tdep_G10.eps} &
\includegraphics[width=0.6\columnwidth,clip]{figs/ga_tdep_G12.eps} &
\includegraphics[width=0.6\columnwidth,clip]{figs/ga_tdep_B12.eps} \\
\end{tabular}
\end{center}
\caption{Comparison of the bare isovector vector charge $g_{V}$ (top) and axial charge $g_{A}$ (bottom) from source-sink separation $t_{\rm sep}=10$ Gaussian source (left panels),
$t_{\rm sep}=12$ Gaussian source (middle panels) and
$t_{\rm sep}=12$ box source (right panels).}
\label{fig:gA_src}
\end{figure*}

As we mentioned before, to reduce statistical fluctuations, we use multiple sources in the present studies. This is another difference from the previous calculation, where only a single source was utilized. To make this point clear, we show the dependence of bare $g_V$ and $g_A$ on the location of the current insertion $t$ for each choice of the source location in Fig.~\ref{fig:gA_src}. In the top panels, all calculations of the bare $g_V$ show clear plateaus between source and sink locations. Although the slight dependence on the source location may be observed among the three sources in the $\{t_{\rm sep}, r_{\rm gauss}\}=\{12, 8\}$ case, all the values of these plateaus agree with one another within the statistical errors. We see no evidence of excited-state contamination in the bare $g_V$.

On the other hand, for the bare $g_A$ (bottom panels), both the results
obtained from box and Gaussian sources with $t_{\rm sep}=12$ exhibit  larger fluctuations and less clear plateaus, while two consistent plateaus clearly appear in the cases of $t_{\rm sep}=10$. This indicates that the larger separation causes the larger fluctuations, since the absolute values of three-point functions are exponentially suppressed as a function of $t_{\rm sep}$. Indeed, the $t_{\rm sep}=12$ Gaussian source results obtained from a single source-sink location $[t_{\rm src}, t_{\rm sink}]=[9, 21]$ agree with those of the box source due to the large statistical errors, while the average from all three sources for $\{t_{\rm sep},r_{\rm gauss}\}=\{12,8\}$ provides good agreement with the results of $\{t_{\rm sep},r_{\rm gauss}\} =\{10, 4.35\}$.

In this context, the lower value of the bare $g_A$ obtained from the present studies is more statistically significant, since the box source results have been obtained only for a single source. Although a systematic error from excited-state contaminations in the result of $\{t_{\rm sep},r_{\rm gauss}\}=\{10, 4.35\}$ might be hidden, owing to
the large fluctuations in that of $\{t_{\rm sep},r_{\rm gauss}\}=\{12,8\}$, we may choose to live with the former for the final result. This is mainly because the finite momentum calculation suffers much from such large fluctuations. We obtain results for the form factors only from the case of $t_{\rm sep}=10$.


\subsection{Form factors} \label{sec:formfactor}

For numerical convenience, we redefine the form factors as dimensionless quantities. Thus, the isovector vector current in Eq.~(\ref{eq:cont_vector}) is rewritten as
\begin{equation}\label{eq:lat_vector}
\langle p| V^+_\mu(0) | n \rangle = \bar{u}_p \left[\gamma_\mu
G_{V}(q^2)- q_{\lambda} \sigma_{\lambda\mu} \frac{G_{T}(q^2)}{2m_{N}}
\right] u_n
\end{equation}
in terms of the dimensionless vector $G_{V}$ and induced-tensor $G_{T}$  form factors. Here the $\Sigma_\mu$ in Eq.~(\ref{eq:three-pt}) is $\gamma_\mu G_{V}(q^2)-q_{\lambda} \sigma_{\lambda\mu} \frac{G_{T}(q^2)}{2m_N}$.

Likewise the isovector axial current in Eq.~(\ref{eq:cont_vector}) is rewritten as
\begin{equation}\label{eq:lat_axial}
\langle p| A^+_\mu(0) | n \rangle = \bar{u}_p \left[\gamma_\mu
\gamma_5 G_{A}(q^2) +i q_\mu \gamma_5 \frac{G_{P}(q^2)}{2m_{N}}\right]
u_n
\end{equation}
with the dimensionless axial $G_{A}$ and induced-pseudoscalar $G_{P}$ form factors. Note that the latter is normalized with twice the nucleon mass $2 m_{N}$ unlike in some muon-capture literature where it is normalized with the muon mass $m_\mu$. This is for numerical convenience in this paper, where the nucleon mass calculated on the lattice is heavier than its physical value. Here the $\Sigma_\mu$ in Eq.~(\ref{eq:three-pt}) is  $\gamma_\mu \gamma_5 G_{A}(q^2) +i q_\mu \gamma_5 \frac{G_{P}(q^2)}{2m_N}$. (In the above, $g_{V,A}(q^2) \equiv G_{V,A}(q^2)$.)

The right-hand sides of the above two equations have the most general form consistent with Lorentz covariance. The momentum transfer $q=p_n-p_p$ becomes very small in the forward limit, because of the small mass difference between the neutron and proton.

We use two projection operators to help us extract the momentum dependence of the form factor:
\begin{equation}\label{eq:three-pt_proj}
 {\Gamma^{(3),{\cal P}}_{\mu, {AB}} (t_{\rm snk},t;p_{\rm src},p_{\rm snk})} =
\frac{1}{4}{\rm Tr}({\cal P} G_{\mu, {AB}}(t_{\rm snk},t;p_{\rm
src},p_{\rm snk})),
\end{equation}
where we choose ${\cal P}={\cal P}_{A_z}={\cal P}_+\gamma_5\gamma_z$  for both axial and vector currents and ${\cal P}={\cal P}_+$ for vector current. We combine three- and two-point functions to remove time dependence and redundant source or sink normalization ($z$) factors~\cite{Hagler:2003jd}:
\begin{widetext}
\begin{equation}\label{eq:three-pt_R}
R_{j_\mu} = \frac{
      \Gamma^{(3),{\cal P}}_{\mu,GG}(t_{\rm src},t,t_{\rm snk},\vec p_{\rm src},\vec p_{\rm snk})}
      {\Gamma^{(2)}_{GG}(t_{\rm src},t_{\rm snk},\vec p_{\rm snk})}
      \times
      \left(
      \frac{\Gamma^{(2)}_{LG}(t,t_{\rm snk},\vec p_{\rm src})
      \Gamma^{(2)}_{GG}(t_{\rm src},t,\vec p_{\rm snk})
      \Gamma^{(2)}_{LG}(t_{\rm src},t_{\rm snk},\vec p_{\rm snk})}
      {\Gamma^{(2)}_{LG}(t,t_{\rm snk},\vec p_{\rm snk})
      \Gamma^{(2)}_{GG}(t_{\rm src},t,\vec p_{\rm src})
      \Gamma^{(2)}_{LG}(t_{\rm src},t_{\rm snk},\vec p_{\rm src})}
      \right)^{\frac{1}{2}}
\end{equation}
\end{widetext}
where $L$ denotes a local (point) source or sink and $G$ stands for a Gaussian-smeared one.
In this work, we fix the sink momentum to zero.
Therefore, in the axial-current case, Eq.~(\ref{eq:three-pt_R}) gives
\begin{eqnarray}\label{eq:solve_axial}
{\cal R}_{\rm A_i}&=&
\frac{1}{\sqrt{2E(\vec{p})(E(\vec{p})+m_N)}}\nonumber\\
&\times&\left[
\delta_{iz}(E(\vec{p})+m_N)G_{A}  - \frac{p_z p_i }{2m_N}G_{P}
\right]
\end{eqnarray}
with $i\in\{x,y,z\}$ for corresponding insertion of operator ${A_i}$. In the vector current case, we use the projection operator $P_+$ on $V_4$ and $P_{A_z}$ on $V_1$ and $V_2$ to extract $G_{V}$ and induced $G_{T}$; the necessary equations are respectively
\begin{eqnarray}\label{eq:solve_vector}
&& \sqrt{\frac{E(\vec{p})+m_N}{ 2E(\vec{p})}}
\left(
G_{V} -\frac{E(\vec{p})-m_N}{2m_N}G_{T}\right), \\
&&\frac{-i  p_y }{ \sqrt{2E(\vec{p})
(E(\vec{p})+m_N)} }(G_{V} + G_{T}),\\
&&\frac{+i  p_x }{ \sqrt{2E(\vec{p})
(E(\vec{p})+m_N)} }(G_{V} + G_{T}).
\end{eqnarray}
One can in principle solve for $G_{A}(q^2)$, $G_{P}(q^2)$, $G_{V}(q^2)$, and $G_{T}(q^2)$ with overconstrained data for each $q^2$.

\subsection{Moments of structure functions}

Of the moments of the structure functions summarized in Eq.~(\ref{eq:moments}), we calculate those which do not require finite momentum transfer: the quark momentum fraction $\langle x \rangle_q$, helicity fraction $\langle x \rangle_{\Delta q}$, transversity $\langle 1 \rangle_{\delta q}$ and the twist-3 moment $d_1$.  Of these we calculate the isovector contribution for the former three: $q=u-d$. This simplifies their renormalization as the flavor-singlet contribution drops out. Thus we fully nonperturbatively renormalize these three so the results are compared with the experiments under an assumption of good isospin symmetry. On the other hand we do not renormalize the twist-3 moment: here our interest is to look at its individual up and down components to see if the Wandzura-Wilczek relation~\cite{Wandzura:1977qf} holds. Our choice of operators ($\Sigma_\mu$ in Eq.~(\ref{eq:three-pt})) for the three-point functions follow an earlier RBC paper~\cite{Orginos:2005uy} and are summarized in Table~\ref{tab:OptDef}.

\begin{table}[h]
\caption{
Operators used in our structure-function moments calculations, including the notation for the operator, the explicit operator form, the hypercubic group representation, the correlator ratios and the
projection operators used in the numerical nonperturbative renormalization in Eq.~(\ref{eq:Zri}).}
\label{tab:OptDef}
\begin{center}
\begin{tabular}{lcc}
\hline\hline
\multicolumn{2}{c}{quark momentum fraction $\langle x \rangle_q$} \\
\hline
$\Sigma_\mu$&$\displaystyle {\cal O}^q_{44} = \overline{q} \left[\gamma_4
\stackrel{\leftrightarrow}{D_4} - \frac{1}{3}\sum_k \gamma_k
\stackrel{\leftrightarrow}{D_k} \right] q$ \\
hypercubic group rep.&${\bf 3}^+_1$\\
correlator ratio&$\displaystyle R_{\langle x \rangle_{q}} = \frac{C^{\Gamma,{\cal O}^q_{44}}_{\rm
3pt}}{C_{\rm 2pt}} = m_N \langle x \rangle_q$\\
NPR projection&$
{{\cal P}^q_{44}}^{-1} =
\gamma_4 p_4 - \frac{1}{3}\sum_{i=1,3} \gamma_i p_i$\\
\hline\hline
\multicolumn{2}{c}{quark helicity fraction $\langle x \rangle_{\Delta q}$}\\
\hline
$\Sigma_\mu$&$\displaystyle {\cal O}^{5q}_{\{34\}} = i \overline{q} \gamma_5 \left[\gamma_3
\stackrel{\leftrightarrow}{D_4} + \gamma_4
\stackrel{\leftrightarrow}{D_3}  \right] q$\\
hypercubic group rep.&${\bf 6}^-_3$ \\
correlator ratio&$\displaystyle R_{\langle x \rangle_{\Delta q}} = \frac{C^{\Gamma,{\cal
O}^{5q}_{\{34\}}}_{\rm 3pt}}{C_{\rm 2pt}} = m_N \langle x
\rangle_{\Delta q}$ \\
NPR projection&$
{{\cal P}^{5q}_{34}}^{-1} = i \gamma_5
\left( \gamma_3 p_4 + \gamma_4 p_3 \right)$\\
\hline\hline
\multicolumn{2}{c}{transversity $\langle 1 \rangle_{\delta q}$}\\
\hline
$\Sigma_\mu$&$\displaystyle {\cal O}^{\sigma q}_{34} = \overline{q} \gamma_5 \sigma_{34} q$\\
hypercubic group rep.&${\bf 6}^+_1$\\
correlator ratio&$\displaystyle R_{\langle 1 \rangle_{\delta q}} = \frac{C^{\Gamma,{\cal
O}^{\sigma q}_{\{34\}}}_{\rm 3pt}}{C_{\rm 2pt}} =
\langle 1 \rangle_{\delta q}$\\
NPR projection& $
{{\cal P}^{\sigma q}_{34}}^{-1} = \gamma_5 \sigma_{34}$\\
\hline\hline
\multicolumn{2}{c}{twist-3 matrix element $d_1$}\\
\hline
$\Sigma_\mu$& $\displaystyle {\cal O}^{5q}_{[34]} = i \overline{q} \gamma_5 \left[\gamma_3
\stackrel{\leftrightarrow}{D_4} - \gamma_4
\stackrel{\leftrightarrow}{D_3}  \right] q$
\\
hypercubic group rep.& ${\bf 6}^+_1$\\
correlator ratio& $\displaystyle R_{d_1} = \frac{C^{\Gamma,{\cal O}^{5q}_{[34]}}_{\rm 3pt}}{C_{\rm 2pt}} = d_1$\\
NPR projection& ${{\cal P}^{5q}_{[34]}}^{-1} = i \gamma_5
\left( \gamma_3 p_4 - \gamma_4 p_3 \right)$\\
\hline\hline
\end{tabular}
\end{center}
\end{table}

\subsection{Nonperturbative renormalization} \label{subsec:NPR}

In order to compare our calculation with the experimental values, we need to establish the proper connection to the continuum through renormalization. Fortunately, the well-preserved chiral and flavor symmetries of the domain-wall fermions (DWF) make this task much easier than in the cases of more conventional fermions~\cite{Blum:2001xb,Blum:2001sr,Dawson:2002nr,Aoki:2002vt}.

For the form factors, the chiral symmetry of DWF assures the isovector vector and axial local currents which are used in the present calculation share a common renormalization: $Z_{V}=Z_{A}$, up to higher order discretization errors, $O(a^2)$.
Further, since the vector current is conserved, the vector renormalization is calculated from the vector charge, $g_{V}$, as $Z_{V}=1/g_{V}$.
These will be demonstrated in the numerical results section.
Note the induced tensor and pseudoscalar form factors share this common renormalization as well and do not require any additional calculation for renormalization.

For structure function moments, we follow Ref.~\cite{Martinelli:1994ty} and implement the regularization-independent momentum-subtraction (RI/MOM) scheme to nonperturbatively renormalize the operators. This was first applied to structure functions by Gimenez {\it et~al.}~\cite{Gimenez:1998ue}. The procedure may be described as follows.
First, the Fourier transform of the Green function with our operator of interest is constructed:
\begin{eqnarray}
G_{O_\Gamma}(p;a) = \sum_{x,y} e^{-i p \cdot (x-y)} \langle
\psi(x) {O_\Gamma}(0) \overline{\psi}(y) \rangle,
\label{eq:greenFuc}
\end{eqnarray}
where $O_\Gamma$ is  one of ${\cal O}^q_{44}$, ${\cal
O}^{\sigma q}_{34}$, ${\cal O}^{5q}_{\{34\}}$ or ${\cal
O}^{5q}_{[34]}$ for the first-moment operator renormalization. In the lattice calculation, we need to produce Fourier-transformed point-source $S(pa;0)$ and point-split--source $D_{\mu}S(pa;0)$ propagators,
\begin{eqnarray}
S(pa;0)&=&\sum_x e^{-ip\cdot x}S(x;0)\\
D_{\mu}S(pa;0)&=&\sum_x \frac{1}{2}e^{-ip\cdot
x}\left[S(x;-\hat{\mu})U_\mu(-\hat{\mu})\right. \nonumber \\ &-&
\left. S(x;\hat{\mu})U_\mu^\dagger(0)\right],
\end{eqnarray}
to construct the Green function in Eq.~(\ref{eq:greenFuc}). In this
work, we use momenta $(p_x,p_y,p_z,p_t)$ ranging from $(0,0,0,0)$
to $(2,2,3,5)$ in units of $2\pi/L$.

The next step is to truncate the external legs
from the Green function:
\begin{eqnarray}
  \Lambda_{O_\Gamma}(p;a) = S(p;a)^{-1} G_{O_\Gamma}(p;p^\prime;a)
  S(p^\prime;a)^{-1},
\end{eqnarray}
where $Z^{\rm RI}$ is obtained after the projection
\begin{eqnarray}\label{eq:Zri}
 Z_{O_\Gamma}(\mu; a)^{-1} Z_q(\mu; a)  =
  \frac{1}{12}{{\rm Tr}\left(\Lambda_{O_{\Gamma}}(p;a)
  P_\Gamma\right)}|_{p^2=\mu^2}.
\end{eqnarray}
The projection operators for various structure function operators
are listed in Table~\ref{tab:OptDef}. Note that $\mu$ must fall
inside the renormalization window $\Lambda_{\rm QCD} \ll \mu \ll
1/a$.

The following steps allow us to convert the renormalization constants into the continuum $\overline{\rm MS}$ scheme at 2~GeV. See Appendix~\ref{subsec:renorm} for specific details on the strong-coupling scheme-matching coefficients, RGI running equation and anomalous dimensions for various operators.

\begin{enumerate}

\item Obtain $Z^{\rm RI}(\mu)$:\\
The ratio of $Z_{O_\Gamma}(\mu; a)/Z_q(\mu; a)$ to $Z_A/Z_q(\mu; a)$ is computed and yields  $Z_{O_\Gamma}(\mu; a)/Z_A$. Each of the factors in the ratio is first exptrapolated to the chiral limit, $m_f=-m_{\rm res}$, at fixed
momentum.
Using $Z_V=1/g_V$, we can determine
$Z_{O_\Gamma}(\mu; a)$ in Eq.~(\ref{eq:Zri}), the
renormalization constant in the RI scheme, which we denote as
$Z^{\rm RI}$. (See the lightly-filled circles in Fig.~\ref{fig:npr}.)
\begin{figure}[htb]
\includegraphics[width=\columnwidth,clip]{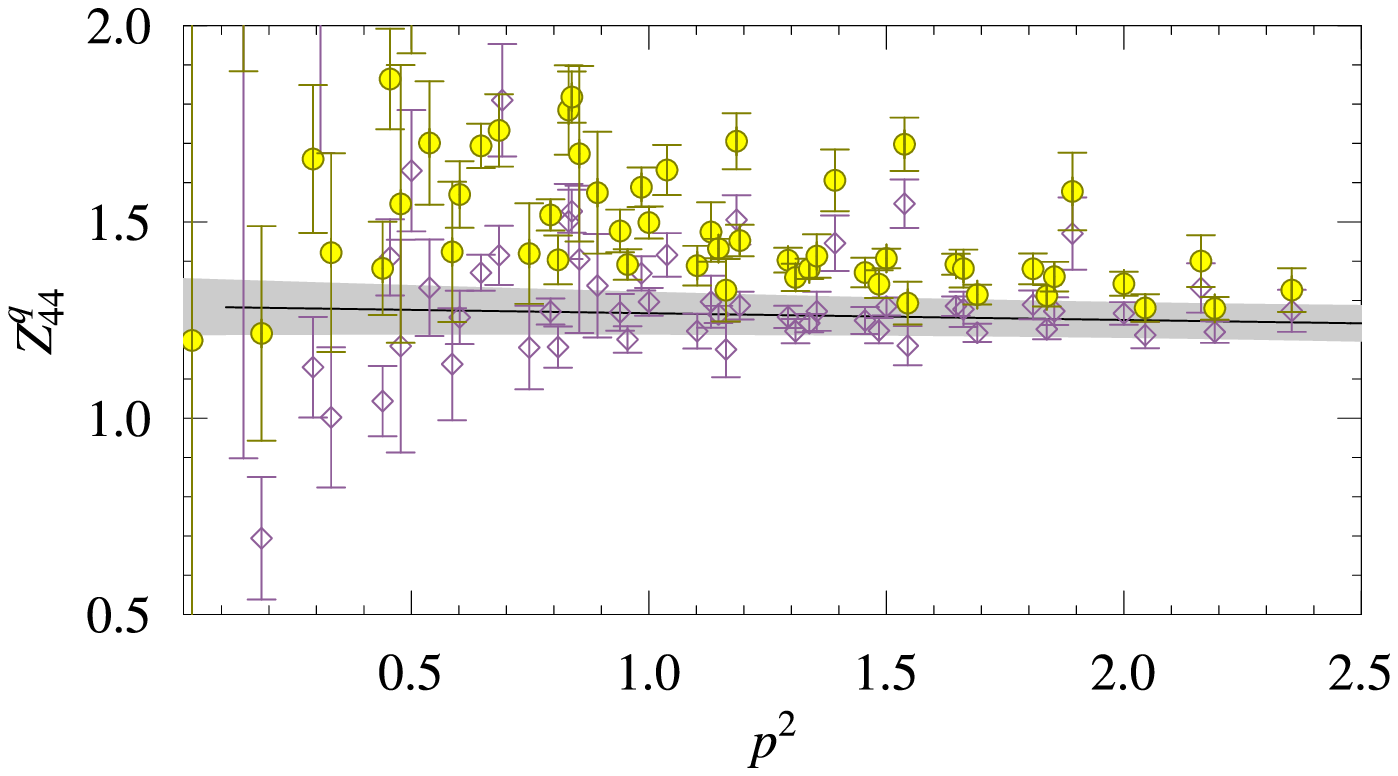}
\includegraphics[width=\columnwidth,clip]{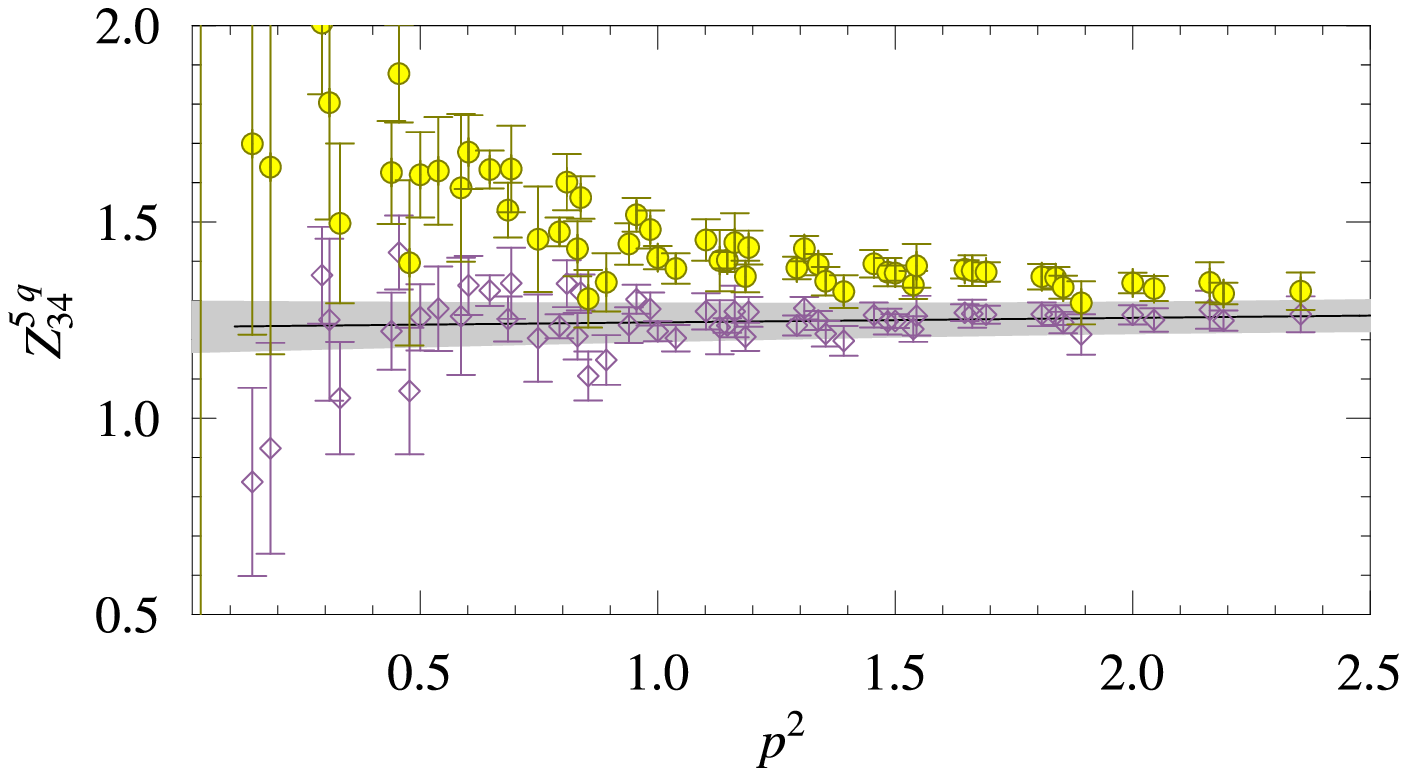}
\includegraphics[width=\columnwidth,clip]{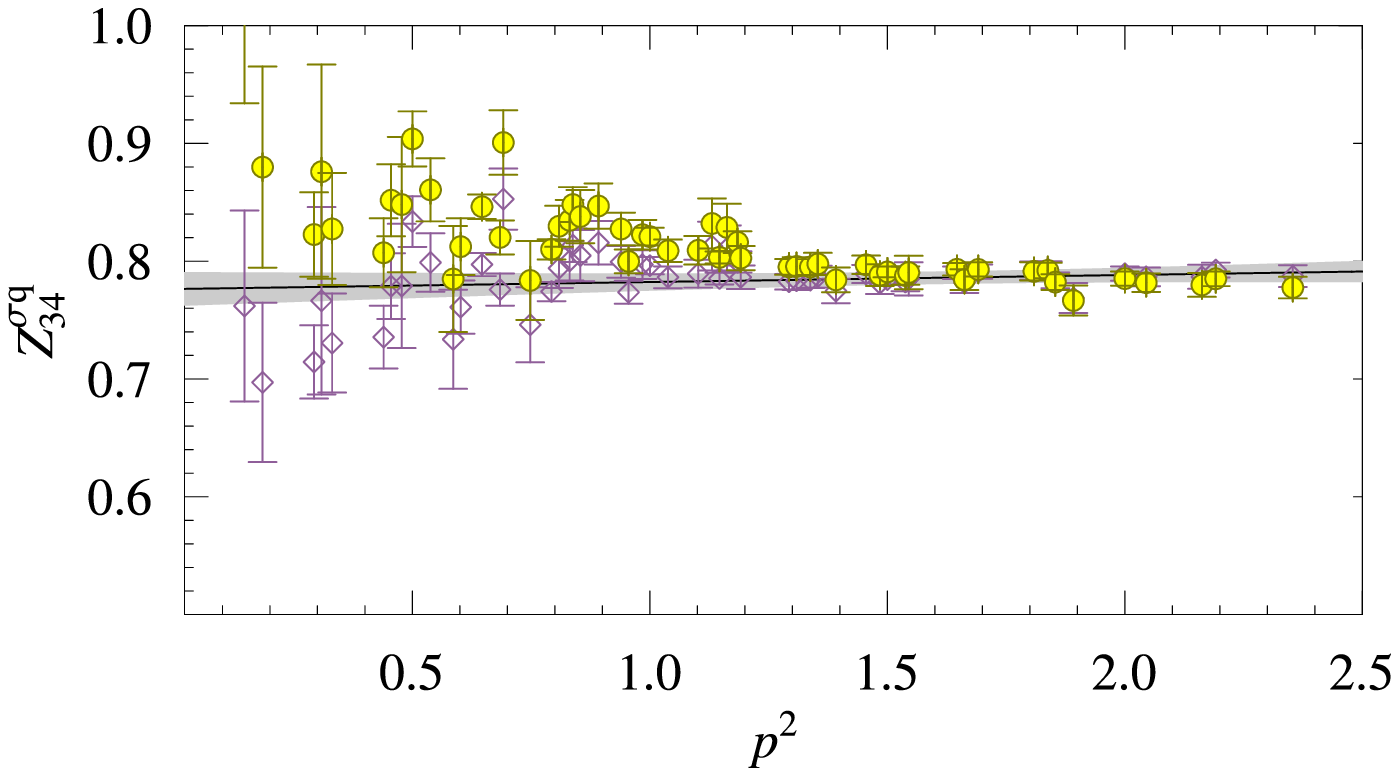}
\caption{Renormalization constants in the chiral limit. The lightly-filled circles are the renormalization constants in RI-MOM scheme, and the diamonds are $\overline{\rm MS}$-scheme at 2~GeV. The fits (solid lines) are used to remove residual $(ap)^2$ artifacts. } \label{fig:npr}
\end{figure}

\item Convert to $\overline{\rm MS}$ scheme:\\
We are interested in continuum quantities, mostly calculated in the
$\overline{\rm MS}$ scheme. The conversion factors between RI and
$\overline{\rm MS}$ schemes for the operators discussed here
have been calculated in
Refs.~\cite{Gockeler:1998ye,Floratos:1977au}. To get $Z^{\rm
\overline{MS}}(\mu)$, we use the three-loop $\alpha_s(\mu)$
running coupling constants defined in Ref.~\cite{Blum:2001sr} with
$\Lambda(n_f=2)=300$~MeV\cite{Izubuchi:2003rp}.

\item Remove $(ap)^2$ lattice artifacts:\\
To remove ${\cal O}((ap)^2)$ errors that might mimic continuum scale dependence, we first divide out the continuum running factor. The resulting renormalization factors should be scale-independent, so any remaining $(ap)^2$ dependence must be a lattice artifact. Hence, $Z^{\overline{\rm MS}, {\rm RGI}}$ can be obtained from a fit to the form $f=A(pa)^2+B$. (See Fig.~\ref{fig:npr}.)

\item Running to 2~GeV:\\
Finally, we use the RGI formula to obtain $Z^{\rm
\overline{MS}}(2\mbox{ GeV})$.
\end{enumerate}
Table~\ref{tab:Zfactors} summarizes the renormalization factors in $\overline{\rm MS}$ scheme at 2~GeV for each operator.
\begin{table}[hbt]
\caption{
Summary of renormalization factors in $\overline{\rm MS}$ scheme at 2~GeV in the chiral limit. The number of configurations is about 100.
}
\label{tab:Zfactors}
\begin{center}
\begin{tabular}{ccll}
\hline\hline
\multicolumn{1}{c}{$m_f$}&
\multicolumn{1}{c}{${\cal O}^q_{44}$}&
\multicolumn{1}{c}{${\cal O}^{5q}_{34}$}&
\multicolumn{1}{c}{${\cal O}^{\sigma q}_{34}$}\\
\hline
\multicolumn{1}{c}{$-m_{\rm res}$}&
1.28(7)&1.23(7)&0.776(14)\\
\hline\hline
\end{tabular}
\end{center}
\end{table}
These renormalization constants will be applied in Section~\ref{subsec:moments}.

\section{Numerical Results}\label{sec:Q2_0}

\subsection{Vector and axial charge}\label{subsec:gA}

The isovector vector and axial charges, $g_{V}$ and $g_{A}$, are defined as the zero-momentum-transfer limits of the corresponding form factors, $G_{V}(q^2)$ and $G_{A}(q^2)$.  Because of chiral symmetry, a Takahashi-Ward identity ensures that the two currents, which are related by chiral transformation, share a common renormalization: $Z_{A} = Z_{V}$ up to a lattice discretization error of $O(a^2, m_f^2\,a^2)$.
Since the vector current is conserved, its renormalization is easily obtained as the inverse of the vector charge $g_{V} = G_{V} (q^2=0)$.
Thus, by calculating the ratio of the three-point functions for $g_{A}/g_{V}$, we get the renormalized axial charge, $(g_A)^{\rm ren}$\cite{Sasaki:2003jh}.

\begin{figure}[htb]
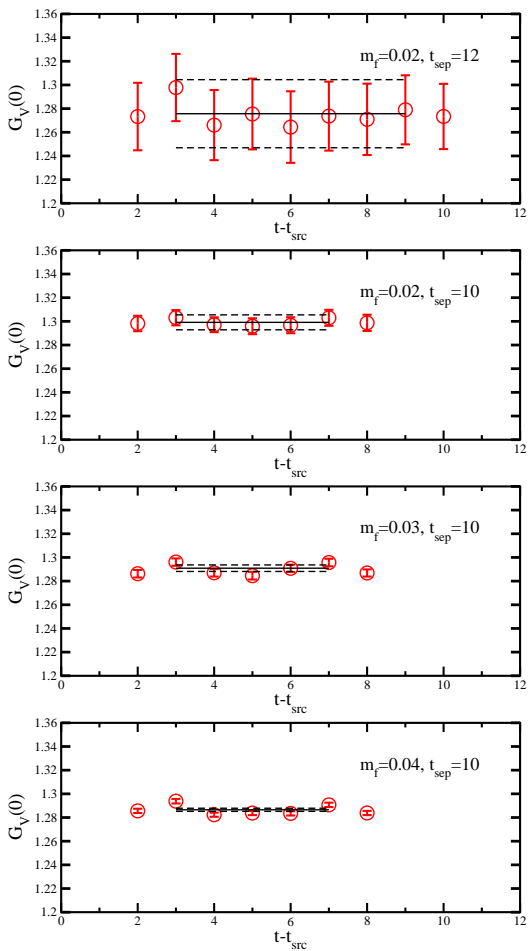

\includegraphics[width=0.8\columnwidth,clip]{figs/gv_002_sep12.eps}
\includegraphics[width=0.8\columnwidth,clip]{figs/gv_002_sep10.eps}
\includegraphics[width=0.8\columnwidth,clip]{figs/gv_003_sep10.eps}
\includegraphics[width=0.8\columnwidth,clip]{figs/gv_004_sep10.eps}
\caption{Bare vector charge as a function of the current insertion
time. Results are averaged over all sources. The top figure depicts $t_{\rm sep}=12$ calculation at the lightest quark mass, $m_f=0.02$.
The other three figures are obtained from $t_{\rm sep}=10$ calculations at all three quark masses (in order of increasing mass from top to bottom). The fit ranges are shown by horizontal lines.}
\label{fig:rawgv}
\end{figure}

\begin{figure}[htb]
\includegraphics[width=0.9\columnwidth,clip]{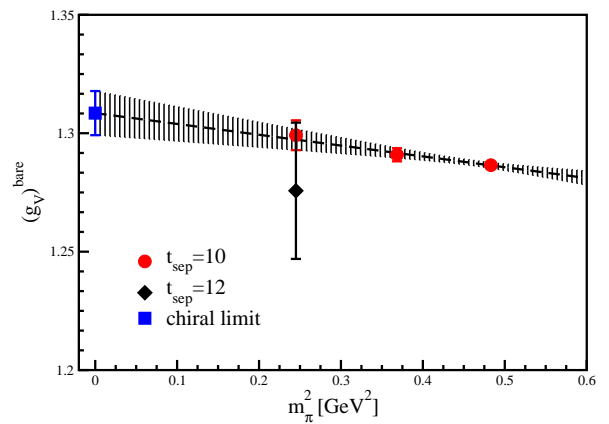}
\caption{The isovector vector-charge $(g_V)^{\rm bare}$
as a function of the pion mass squared. }
\label{fig:gvvsmq}
\end{figure}

Let us first discuss the vector charge. We note, however, that since the vector charge is conserved, it can not provide a transition matrix element from one state to another, in particular from any excited state to the ground state. Thus the precocious plateau does not necessarily mean that excited state contamination is absent in the signal after one time unit. As mentioned earlier, to avoid excited-state contributions at the lightest quark mass, we have lengthened the source-sink separation to 12 time units and used a larger smearing radius for the Gaussian-smeared source and sink. From the fitting ranges of 3--9 for $t_{\rm sep}=12$ and 3--7 for $t_{\rm sep}=10$ (see also Fig.~\ref{fig:rawgv}),
we obtain the plot in Fig.~\ref{fig:gvvsmq}.
This figure shows that larger fluctuations appear in the case of the longer time separation. By linearly extrapolating three points of $t_{\rm sep}=10$ to zero quark mass, we obtain an estimate of $g_{V}=1.308(9)$ and $Z_{V}=0.764(5)$. The resulting value agrees reasonably well with the value of $Z_A=0.75734(55)$ found in Ref.~\cite{Aoki:2004ht}, which was determined from the ratio of two-point meson correlation functions. Thus we conclude that excited state contamination in our value of $g_V$ is negligible.

\begin{figure}[htb]
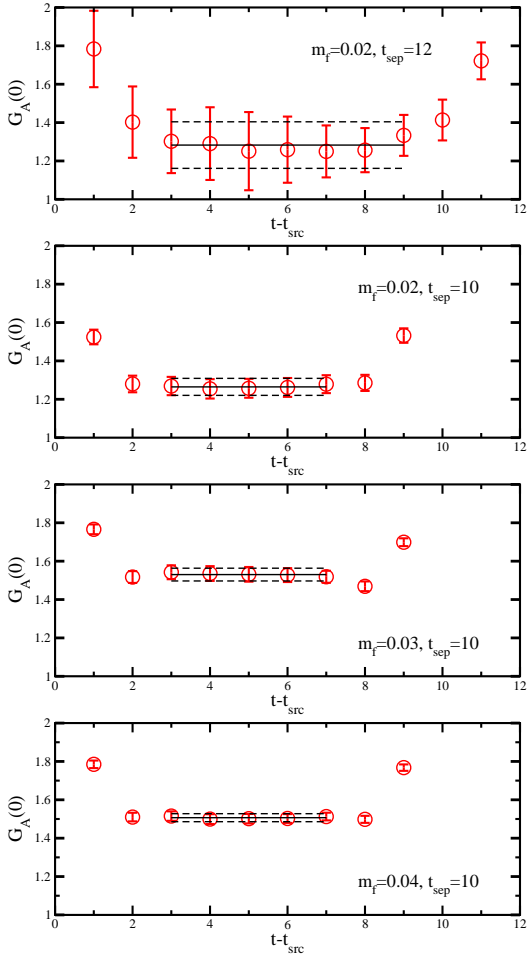

\includegraphics[width=0.8\columnwidth,clip]{figs/ga_002_sep12.eps}
\includegraphics[width=0.8\columnwidth,clip]{figs/ga_002_sep10.eps}
\includegraphics[width=0.8\columnwidth,clip]{figs/ga_003_sep10.eps}
\includegraphics[width=0.8\columnwidth,clip]{figs/ga_004_sep10.eps}
\caption{The bare axial charge as in Fig.~\ref{fig:rawgv}}
\label{fig:rawga}
\end{figure}

\begin{figure}[htb]
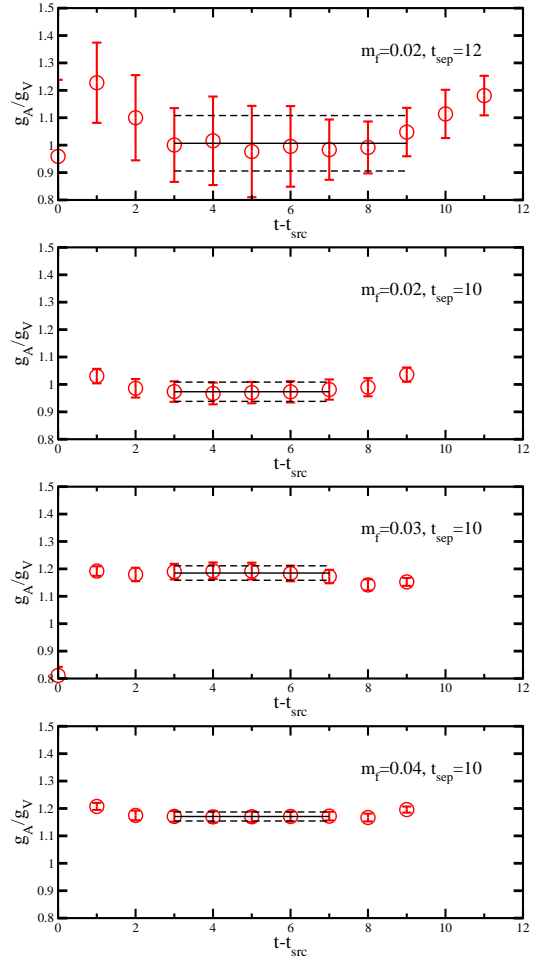

\includegraphics[width=0.8\columnwidth,clip]{figs/gagv_002_sep12.eps}
\includegraphics[width=0.8\columnwidth,clip]{figs/gagv_002_sep10.eps}
\includegraphics[width=0.8\columnwidth,clip]{figs/gagv_003_sep10.eps}
\includegraphics[width=0.8\columnwidth,clip]{figs/gagv_004_sep10.eps}
\caption{The axial-to-vector charge ratio as in Fig.~\ref{fig:rawgv}
} \label{fig:rawgagv}
\end{figure}

\begin{figure}[htb]
\includegraphics[width=0.9\columnwidth,clip]{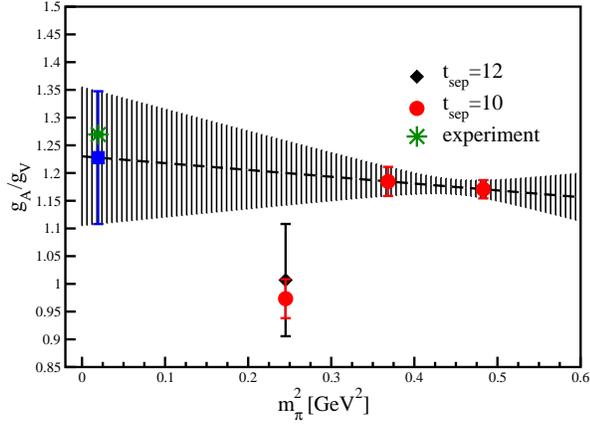}
\caption{The axial-to-vector charge ratio as a function of the pion mass squared. The lightest mass points with either $t_{\rm sep}=10$ or $t_{\rm sep}=12$ deviate significantly from the heavier mass points.
Omitting the lightest mass point, a constrained linear fit is performed.
}\label{fig:gAvsmpi2}
\end{figure}

In Fig.~\ref{fig:rawga}
we present the raw results for the bare axial charge.
After taking an average of all sources, the a plateau emerges, even in the case of $t_{\rm sep}=12$ (top figure), albeit with larger statistical errors. As can be seen from this figure, the plateaus settle in after about three time slices for both $t_{\rm sep}$ cases at the lightest quark mass. This indicates that excited-state contamination is negligible compared to the statistical error. Corresponding renormalized values are obtained from the ratio of correlation functions, and are presented in Fig.~\ref{fig:rawgagv}.
As expected, we also find that plateaus settle in after about three time slices for all cases. By using the fitting ranges shown in this figure, we obtain the values in Table~\ref{tab:gagv}
\begin{table}[b]
\caption{Bare isovector vector and axial charges, $g_{V}^{\rm bare}$ and $g_{A}^{\rm bare}$ and their ratio $(g_{A}/g_{V})^{\rm bare}$
which is by-definition renormalized, extracted from data summarized in Figs.~\ref{fig:rawgv}, \ref{fig:rawga} and \ref{fig:rawgagv}.}
\label{tab:gagv}
\begin{center}
\begin{tabular}{cclll}
\hline\hline
$m_f$  & $\{t_{\rm sep}, r_{\rm gauss}\}$ &
\multicolumn{1}{c}{$g_{V}^{\rm bare}$}&
\multicolumn{1}{c}{$g_{A}^{\rm bare}$}&
\multicolumn{1}{c}{$(g_{A}/g_{V})^{\rm ren}$}  \\
                  \hline
0.02  &\{12, 8\} &1.28(3)    &1.28(12) & 1.01(10) \\
0.02  &\{10, 4.35\} &1.299(6) &1.27(4)    & 0.97(4)\\
0.03  &\{10, 4.35\} &1.291(3) &1.53(3)    & 1.19(3)\\
0.04  &\{10, 4.35\} &1.2865(13) &1.51(2)    & 1.171(16) \\
\hline\hline
\end{tabular}
\end{center}
\end{table}
and Fig.~\ref{fig:gAvsmpi2}.

We observe that our axial charge value at the lightest quark mass deviates significantly from the heavier quark mass points. This deviation causes the linear extrapolation to the physical point to deviate from experiment by more than five standard deviations. (The obtained value is given as $g_A/g_V=0.89(6)$ with a poor value of $\chi^2/{\rm dof}=11.3$) Omitting the lightest mass point, a constrained linear fit gives $g_A/g_V=1.23(12)$ at the physical point, in good agreement with experiment. As mentioned in Section~\ref{Sec:Simulation}, our physical volume is about $(1.9\mbox{ fm})^3$. In light of our previous quenched calculation~\cite{Sasaki:2003jh} where we saw that a small volume resulted in a small axial charge, one may worry that a similar finite-volume effect occurs here.

Since our quenched calculation~\cite{Sasaki:2003jh} was published, a way to interpret the finite-volume effect has been proposed by Beane and Savage~\cite{Beane:2004rf} within the small scale expansion (SSE)
scheme~\cite{Hemmert:1997ye}, which is one possible extension of HBChPT
with explicit $\Delta$ degrees of freedom. The SSE scheme can provide milder quark-mass dependence for the axial charge, which seems to be consistent with lattice results, while either leading order (LO) or next-to-leading order (NLO) HBChPT shows strong quark-mass dependence in the vicinity of the chiral limit~\cite{Procura:2006gq}. There is a caveat that we have to arbitrarily fix at least one of the four parameters in SSE. However, if one considers only the finite-volume correction $\delta g_A$ (which is $g_A(L) - g_A(L=\infty)$) within this model, $\delta g_A$ depends only on two phenomenological parameters, the $N$-$\Delta$ coupling ($c_A$) and $\Delta$-$\Delta$ coupling ($g_1$)~\footnote{
The correspondence between the SSE notation and the parameters used in Ref.~\cite{Beane:2004rf} can be found in Ref.~\cite{Khan:2006de}. Here, we prefer to use the original SSE notation. }, which can be barely fixed. Furthermore, if the $N$-$\Delta$ coupling is set to zero, the Beane-Savage formula for the finite-volume correction to $g_A$ reduces to that of HBChPT at leading order. In the original paper~\cite{Beane:2004rf}, $\delta g_A$ is predicted to be positive. This remains true in the case of $c_A=0$, where SSE reduces to LO HBChPT. In this sense, this formula fails to account for the negative $\delta g_A$ observed in the quenched DWF calculation.

In Ref.~\cite{Khan:2006de}, a finite-volume study of $g_A$ has been done with $n_f=2$ dynamical clover simulations. They also observe negative $\delta g_A$ in both quenched and dynamical simulations. Although their lattice data points are outside of the range of applicability of SSE (or LO HBChPT), they can fit their data using the SSE formula for $g_A$ with the finite-volume corrections, which should be negative. They insisted that the parameters adopted in the
Beane-Savage paper do not fully reflect decoupling constraints, which are guaranteed to reduce SSE to LO HBChPT when $c_A=0$. Indeed, their adopted parameter set yields $\delta g_A<0$.

\begin{figure}[t]
\includegraphics[width=0.9\columnwidth,clip]{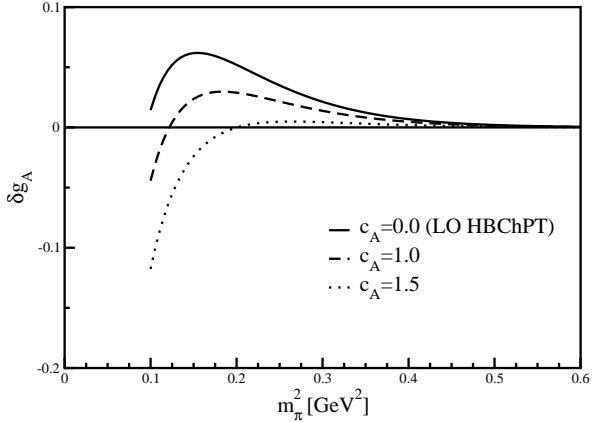}
\caption{Finite volume correction of the axial charge $g_A$ is predicted within ${\cal O}(\varepsilon^3)$ SSE. The sign of $\delta g_A$ is very sensitive to the choice of $c_A$.
} \label{fig:fvc_gA_SSE}
\end{figure}

Apart from the concept of SSE, this suggests that the sign of $\delta g_A$ predicted from SSE is very sensitive to adopted parameters. We find that the choice of $c_A$ strongly affects the sign of $\delta g_A$, as shown  in Fig.~\ref{fig:fvc_gA_SSE}.
We fix the chiral limit value of the axial charge $g_A$, the pion decay constant $F_{\pi}$ and the $N$-$\Delta$ mass splitting $\Delta_0$ to their physical values and set $g_1=3.0$ as in Ref~\cite{Beane:2004rf}.  We also set $L=1.9$~fm. We vary the remaining parameter of the $N$-$\Delta$ coupling $c_A$ from 0 to 1.5. Here, we recall that $c_A=1.5$ is adopted in Ref.~\cite{Khan:2006de}, while $c_A=1$ in Ref.~\cite{Beane:2004rf}.
The predicted correction at our simulation point, less than 5\%, is negligible. Furthermore, the sign of $\delta g_A$ seems to be positive, against our expectation. Therefore, the finite-volume correction from SSE can not account for our data either qualitatively or quantitatively. Again, it is likely that the heavy quark masses in our simulation do not allow use of such formulae. We will come back to this finite-volume question when we discuss the momentum dependence of the form factors and quark momentum and helicity fractions.

\subsection{Momentum dependence of form factors}
As described in the previous subsection, data sets with $\{t_{\rm sep}, r_{\rm gauss}\}=\{10, 4.35\}$ do not seem to suffer from excited-state contamination. Therefore, we focus on those data sets for analysis of the form factors at finite momentum transfer in this subsection.

\subsubsection{Vector current}

Under the present assumption of exact isospin symmetry that ignores strange quark contribution, a straight-forward exercise in SU(2) Lie algebra leads to the following well known relations between the weak vector-current matrix elements in neutron beta decay and the isovector part of the nucleon electromagnetic current matrix elements:
\begin{equation}
\langle p|\bar{u}\gamma_{\alpha}d|n\rangle=
\langle p|\bar{u}\gamma_{\alpha}u
-\bar{d}\gamma_{\alpha}d|p\rangle
=\langle p|j_{\alpha}^{\rm em}|p\rangle-\langle n|j_{\alpha}^{\rm em}|n\rangle
\end{equation}
where $j_{\alpha}^{\rm em}=\frac{2}{3}\bar{u}\gamma_{\alpha}u-\frac{1}{3}\bar{d}
\gamma_{\alpha}d$.
This relates the weak vector and induced tensor form factors with the isovector part of the nucleon electromagnetic form factors:
\begin{eqnarray}
F_1^{p}(q^2)-F_1^{n}(q^2)&=&G_V(q^2)\\
F_2^{\it p}(q^2)-F_2^{\it n}(q^2)&=&G_T(q^2).
\end{eqnarray}
These are respectively are called the Dirac and Pauli form factors.
They are related to the more conventional Sachs electric $G_E$ and magnetic $G_M$ form factors:
\begin{eqnarray}
G_{E}^{N}(q^2)&=&F_1^N(q^2)-\frac{q^2}{4m^2_N}F_2^N(q^2)\\
G_{M}^{N}(q^2)&=&F_1^N(q^2)+F_2^N(q^2)
\end{eqnarray}
where $N$ represents $p$ (proton) or $n$ (neutron). The isovector ones $G_{E,M}^{v}$
are defined by $G_{E,M}^{v}=G_{E,M}^{p}-G_{E,M}^{n}$.

As such we can compare our lattice results computed at finite momentum
transfer with the experiments. The easiest quantity for such a comparison is the charge radius, which is conventionally defined as $\langle r^2\rangle=-6{d F_1}/{d q^2}$. The proton mean-squared charge radius is known experimentally~\cite{PDBook} as $0.7656(119)\mbox{ fm}^2$ and the neutron, -0.1161(22), resulting in an
estimate of $0.636(12)\mbox{ fm}^2$ for the mean-squared ``Dirac'' radius of the nucleon. (For details, see Appendix~A of Ref~\cite{Sasaki:2007gw}).

\begin{figure}[htb]
\includegraphics[width=\columnwidth,clip]{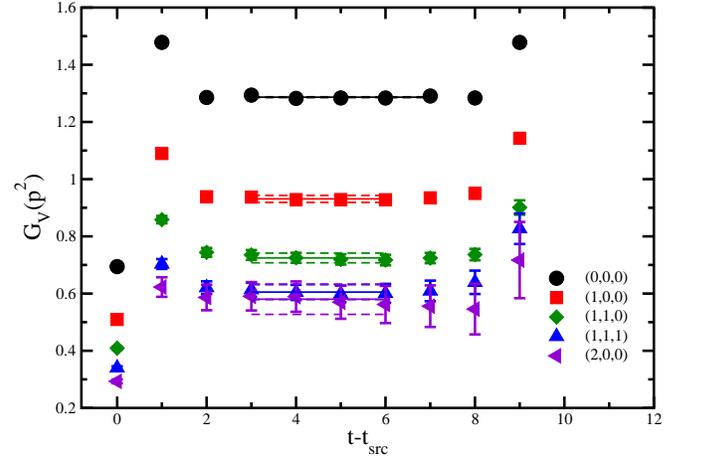}
\caption{Bare Dirac isovector form factor as a function of the current insertion time at the heaviest pion mass ($m_f=0.04$).  The horizontal lines represent the average value (solid lines) and their one standard deviations (dashed lines) in the fitted range.} \label{fig:rawDiracFF}
\end{figure}

\begin{figure}[h]
\includegraphics[width=0.9\columnwidth,clip]{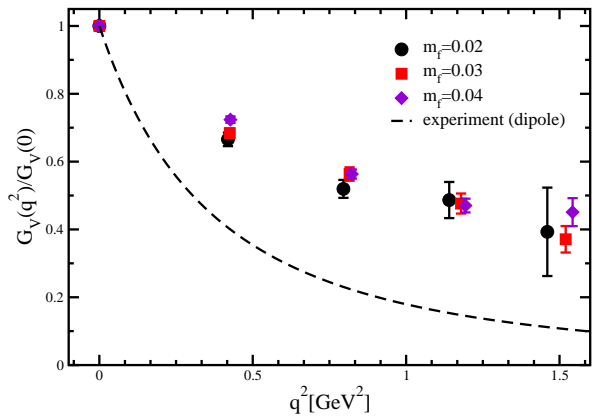}
\caption{Momentum-transfer dependence of vector form factors.
A dashed curve corresponds to the dipole form with the empirical value
of the dipole mass $M_V=0.857(8)$~GeV, which is evaluated
from the electric charge and magnetization radii of the proton and
neutron.}
\label{fig:allgQ2-3}
\end{figure}

The raw results from the lattice are presented in Fig.~\ref{fig:rawDiracFF}.
From these we extract the values of the Dirac isovector form factor, normalized by the value at $q^2=0$, as listed in Table~\ref{tab:q^2_FF}.
\begin{table}[t]
\caption{Momentum transfer dependence of the isovector form factors. $n^2$ refers to the momenta carried by the vertex in units of $(2\pi/L)^2$.}
\label{tab:q^2_FF}
\begin{center}
\begin{tabular}{crlllll}
\hline\hline
&
\multicolumn{1}{c}{$m_f$}&
\multicolumn{1}{c}{$n^2=0$}&
\multicolumn{1}{c}{$n^2=1$}&
\multicolumn{1}{c}{$n^2=2$}&
\multicolumn{1}{c}{$n^2=3$}&
\multicolumn{1}{c}{$n^2=4$}\\
\hline
&0.02& 1.299(6)&  0.86(3)&  0.68(3)& 0.63(7)& 0.51(17) \\
$G_{V}$
&0.03& 1.291(3)&  0.882(18)&  0.73(3)& 0.62(4)& 0.48(5)\\
&0.04& 1.2865(13)&  0.930(12)&  0.725(17)& 0.61(3)& 0.58(5)\\
\hline
&0.02&N/A & 2.13(15)  & 1.53(11)& 1.28(18) &  1.1(4)\\
$G_{T}$
&0.03&N/A & 2.67(13)  & 2.05(10)& 1.57(12) &  1.11(13)\\
&0.04&N/A & 2.81(9)     & 2.04(7)   & 1.69(8)   &  1.18(13)\\
\hline
&0.02& 1.27(4)&  1.09(4)&  0.92(5)&  0.87(9)&  0.58(19)\\
$G_{A}$
&0.03& 1.53(3)&  1.16(2)&  0.97(4)&  0.87(5)&  0.69(8)\\
&0.04& 1.51(2)&  1.21(2)&  0.99(3)&  0.88(4)&  0.82(8) \\
\hline
&0.02& N/A &9.3(8)& 4.9(5)& 3.3(5)&  1.8(7) \\
$G_{P}$
&0.03& N/A &9.5(6)& 6.1(4)& 4.2(4)&  2.6(4)\\
&0.04& N/A &9.7(5)& 6.1(4)& 4.4(3)&  3.1(4)\\
\hline\hline
\end{tabular}
\end{center}
\end{table}
In Fig.~\ref{fig:allgQ2-3},
we plot all three $m_{\rm f}$ results for $G_V(q^2)$ as a function of Euclidean four-momentum squared together with the dipole form of
$(1+q^2/M^2)^{-2}$ with the empirical value of the isovector Dirac dipole mass $M=0.857(8)$~GeV, which is evaluated from the electric charge and magnetization radii of the proton and neutron. Different symbols represent the values obtained from different quark masses. As can be seen, there is no large quark-mass dependence. Our calculation points are located far from the empirical curve.

The slope of the form factor at $q^2=0$ determines the mean-squared radius, which can be related to the corresponding dipole mass $M$ as $\langle r^2\rangle = 12/M^2$. To extract the mean-squared radius from our data, we simply adopt the dipole form for fitting three lower $q^2$ points including the $q^2=0$ value. The fitted values of the Dirac dipole mass $M_V$ and corresponding mean-squared Dirac radii
$\langle r_V^2\rangle$ are listed in Table~\ref{tab:rms_Dirac}.
\begin{table}[h]
\caption{Dirac mean-squared charge radius obtained from fitting to the conventional dipole form.
}
\label{tab:rms_Dirac}
\begin{center}
\begin{tabular}{lll}
\hline\hline
$m_f$& $M_V$ (GeV)&$\langle r_V^2\rangle ({\rm fm}^2)$ \\
\hline
0.02&1.40(5)&0.239(19)\\
0.03&1.49(4)&0.209(12)\\
0.04&1.57(3)&0.190(8)\\
\hline
Expt. & 0.857(8)& 0.636(12) \\
\hline\hline
\end{tabular}
\end{center}
\end{table}

\begin{figure}[htb]
\includegraphics[width=0.9\columnwidth,clip]{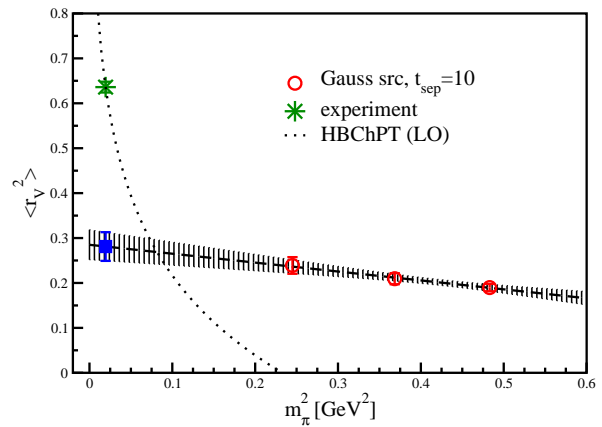}
\caption{The mean-squared Dirac radii $\langle r_V^2\rangle$ from $G_{V}(q^2)$. A simple linear extrapolation with respect to the pion mass squared underestimates the experimental point (asterisk). A dotted curve is the predicted quark-mass dependence of  mean-squared Dirac radius from LO HBChPT in the vicinity of the physical point.
}\label{fig:Diracr2}
\end{figure}

We show the quark-mass dependence of measured mean square Dirac
radius in Fig.~\ref{fig:Diracr2}.
Although it is difficult to extrapolate the lattice estimates at the three heavy pion masses to the physical pion mass, the mild $m_f$ dependence in our observed range of $m_{\pi}^2$ allows us to perform a simple linear extrapolation. Then we obtain the mean-squared radius of the vector form factor at the physical point: $\langle r_V^2\rangle=0.281(32)\mbox{ fm}^2$, which is much smaller than the experimental value of $0.636(12)\mbox{ fm}^2$.

However, it is well known that this particular quantity, as well as the pion charge radius, has a logarithmic divergence in heavy baryon chiral perturbation theory (HBChPT) as one approaches the chiral limit~\cite{Beg:1973sc}. The dotted curve plotted in Fig.~\ref{fig:Diracr2}, is the expected quark-mass dependence of $\langle r_V^2\rangle$ in the vicinity of the physical point within leading order (LO) HBChPT. (See Appendix~\ref{subsec:MSR-HBChPT} for more details.) As can be seen, the very steep $m_{\pi}^2$ dependence, which is associated with the logarithmic divergence, near the physical point is predicted by HBChPT at leading order. This may account for the smaller values of our measured mean-squared radius, which are calculated in the heavy quark mass region ($m_{\pi}>0.49$~GeV).

Thus, we may expect that there is non-linear behavior in terms of $m_\pi^2$ in the vicinity of the physical point. This expectation may look like somewhat contradictory to what we argue for $g_A$ with HBChPT and SSE. However, we recall that the logarithmic divergence can not be cured by higher-order corrections, whereas the strong quark-mass dependence of $g_A$, which simultaneously implies slower convergence of the chiral expansion, should be modified by higher-order corrections. Indeed, both the one-loop effective field theory approach with explicit $\Delta$ degrees of freedom (SSE)~\cite{Procura:2006gq} and two loop HBChPT calculation~\cite{Bernard:2006te} present flat quark-mass dependence at least down to $m_\pi \sim 0.3$~GeV. To observe this, we have to at least push the pion mass down to 0.3~GeV, which is approximately the location of the intersection between the LO HBChPT curve and the extrapolated line obtained from our three data points in Fig.~\ref{fig:Diracr2}. We leave this issue to future calculations.

\begin{figure}[h]
\includegraphics[width=\columnwidth,clip]{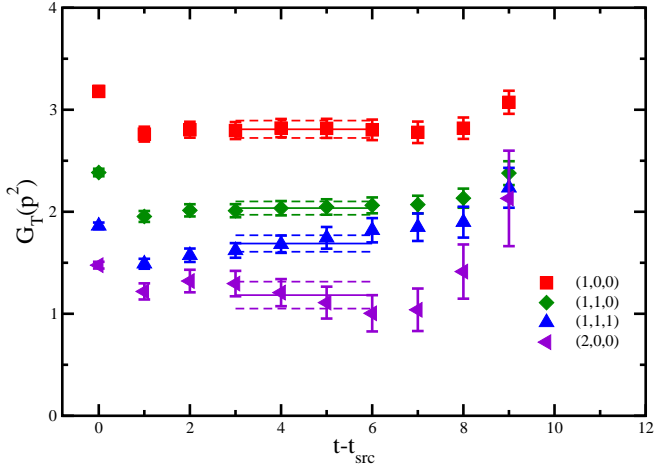}
\caption{The bare Pauli isovector form factor as in Fig.~\ref{fig:rawDiracFF}}
\label{fig:rawPauliFF}
\end{figure}

\begin{figure}[htb]
\includegraphics[width=0.9\columnwidth,clip]{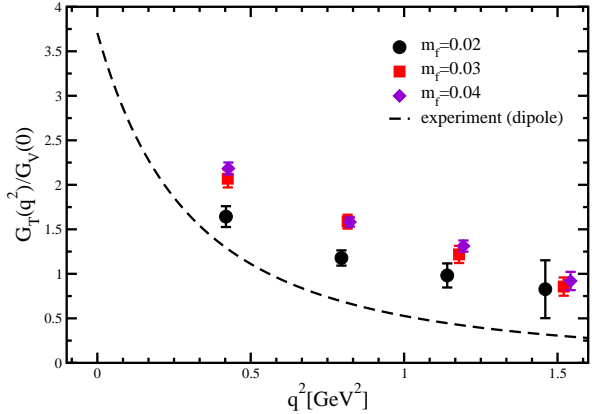}
\caption{
Momentum-transfer dependence of the induced-tensor form factors.
The dashed curve corresponds to the dipole form with the experimental value of the dipole mass $M_T=0.778(23)$~GeV and nucleon magnetic moment $F_2(0)=3.70589$.}\label{fig:gTdipole}
\end{figure}

Next, we extract the induced-tensor (Pauli) form factor, $G_T$, from the vector current; the raw data is listed in Table~\ref{tab:q^2_FF}, and an example of $m_f=0.04$ at various transfer momenta is shown in Fig.~\ref{fig:rawPauliFF}.
This corresponds to the combination $F_2^p-F_2^n$ of the nucleon electromagnetic form factors. The renormalized Pauli form factor can be obtained from $G_T^{\rm ren}(q^2)=G_T(q^2)/G_V(0)$ with vector current renormalization  $Z_V=1/G_V(0)$. Fig.~\ref{fig:gTdipole}
shows a comparison of our data and the dipole form $G_T^{\rm ren}(0)/(1+q^2/M_T^2)^2$ with the experimental values of $G_T^{\rm ren}(0)=F_2(0)=3.70589$ and $M_T=0.778(23)$~GeV, as described in Appendix~A of Ref~\cite{Sasaki:2007gw}. The heaviest two pion mass points are almost degenerate, as was the case with the Dirac form factor. However, the data from lightest pion mass here is closer to the experimental values. Again, this could either be interpreted as a trend toward the experimental values with decreasing pion mass or merely finite-volume effects.

Furthermore, we fit our $G_T^{\rm ren}(q^2)$ to the dipole form to extract the Pauli mean squared radius, which is related to the corresponding dipole mass $M_T$ by $\langle r_T^2 \rangle=12/M_T^2$. In contrast to the dipole fit
on $G_V(q^2)$, it is a two-parameter fit since we do not have data on the value of $G_T^{\rm ren}(q^2)$ at $q^2=0$ without the $q^2$ extrapolation. Here, the value of the $G_T^{\rm ren}(0)$ is associated with the difference between the proton and neutron magnetic moments $\mu_p -\mu_n = 1+ F_2(0)$. The fitted results are summarized in Table~\ref{tab:F2_r}. Here we perform the dipole fit with the two lower $q^2$ points.
\begin{table}[h]
\caption{Pauli mean-squared charge radius obtained from fitting to the conventional dipole form and the extrapolated value of the ratio $G_M^v(q^2)/G_E^v(q^2)$ at $q^2=0$ with the linear $q^2$ fitting form.
}\label{tab:F2_r}
\begin{center}
\begin{tabular}{lllll}
\hline\hline
$m_f$& $F_2(0)$ & $M_T ({\rm GeV})$ & $\langle r_T^2\rangle ({\rm fm}^2)$ & $G^v_M(0)/G^v_E(0)$\\
\hline
0.02& 2.57(37)& 1.29(16) & 0.28(7) & 3.69(41)\\
0.03& 2.89(22)& 1.53(12) & 0.20(3) & 4.23(26)\\
0.04& 3.32(17)& 1.36(6)   & 0.25(2) & 4.24(17)\\
\hline
\hline
Expt. & 3.70589 &0.778(23) &  0.773(32)  & 4.70589\\
\hline\hline
\end{tabular}
\end{center}
\end{table}

\begin{figure}[htb]
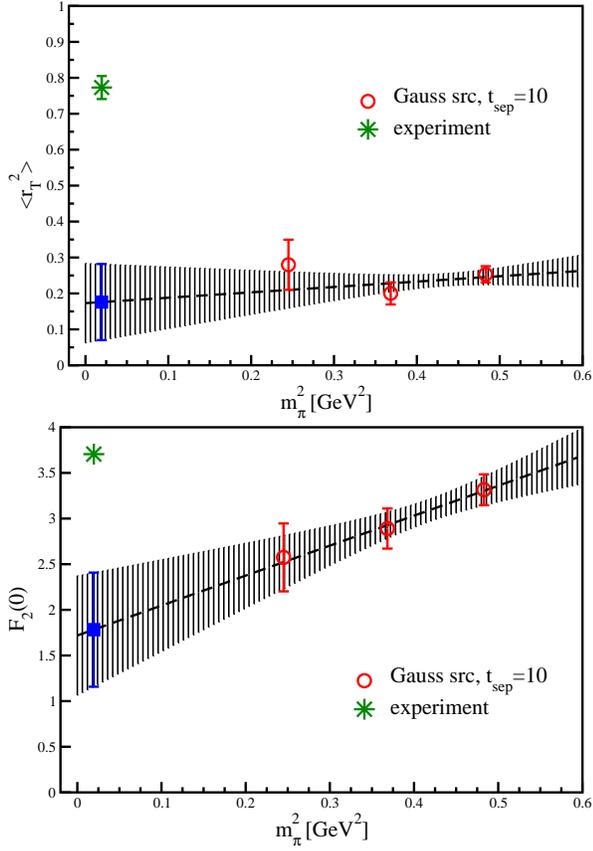

\includegraphics[width=0.9\columnwidth,clip]{figs/msr_t_chi.eps}
\includegraphics[width=0.9\columnwidth,clip]{figs/F2_chi.eps}
\caption{
The mean-squared Pauli radius $\langle r_T^2\rangle$ (top) and the value of $G_T^{\rm ren}(0)=F_2(0)$, which is associated with the difference between the proton and neutron magnetic moments, $\mu_p -\mu_n = 1 + F_2(0)$, (bottom) and  as a function of pion mass squared. A simple linear extrapolation with respect to the pion mass squared underestimates the experimental point (asterisk).
}\label{fig:F2_r}
\end{figure}

In Fig.~\ref{fig:F2_r}, we plot values of $\langle r_T^2\rangle$ and
$F_2(0)$ as a functions of pion mass squared
along with a naive linear extrapolation to the physical pion mass. In the upper panel of Fig.~\ref{fig:F2_r}, we linearly extrapolate through all three points, finding $\langle r_T^2 \rangle=0.17(11)\mbox{ fm}^2$. This is much smaller than the experimental measurement, $0.773(32)\mbox{ fm}^2$, and similar to what we observed for the Dirac mean-squared radius. Lighter pions in future measurements would be desirable to see whether there is an increase in radius as we approach the physical pion mass.

In the lower panel of Fig.~\ref{fig:F2_r}, although the value measured at the heaviest point is close to the experimental value, the extrapolated value of $F_2(0)$ at the physical point also tends to somewhat underestimate the experimental value. However, we recall that the value of $F_2(0)$ is highly dependent on our adopted fitting form. For example, a monopole fit yields a larger $F_2(0)$ which lies closer to the experimental value. Therefore, our estimation of $F_2(0)$ should carry a large systematic uncertainty due to the $q^2$ extrapolation.

We also have an alternative way to evaluate $F_2(0)=\mu_p-\mu_n-1$. The ratio of the isovector magnetic form factor $G^v_M(q^2)$ and the isovector electric form factor $G^v_E(q^2)$ provides the difference between the proton and neutron magnetic moments in the forward limit, $\mu_p-\mu_n=G^v_M(0)/G_E^v(0)$, which is related to $1+F_2(0)$. Experimentally, it is known that this ratio shows no $q^2$ dependence at low $q^2$ since both form factors are well fitted by the dipole form with the comparable dipole masses. Therefore, this ratio may have milder $q^2$ dependence than $G_T^{\rm ren}(q^2)$~\cite{Sasaki:2007gw}. Some RBC results on related ratios for the heavier quark masses were reported in Ref.~\cite{Berruto:2005hg}.

\begin{figure}[htb]
\includegraphics[width=0.9\columnwidth,clip]{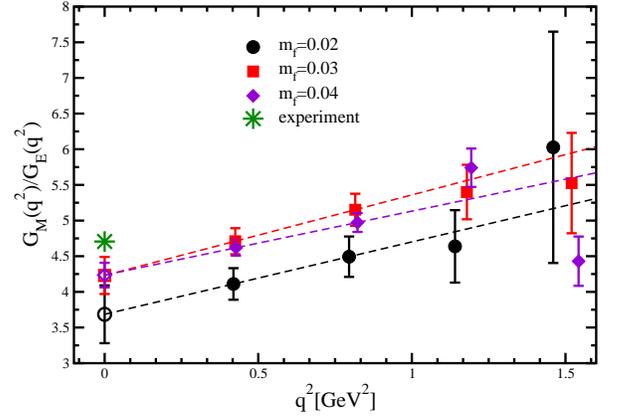}
\caption{The $q^2$ dependence of the ratio  $G^{v}_M(q^2)/G^{v}_E(q^2)$.
A simple linear extrapolation with respect to $q^2$ is utilized for
an alternative evaluation of $\mu_p-\mu_n$ thanks to its mild $q^2$ dependence.}\label{fig:Gm_ov_Ge}
\end{figure}

\begin{figure}[htb]
\includegraphics[width=0.9\columnwidth,clip]{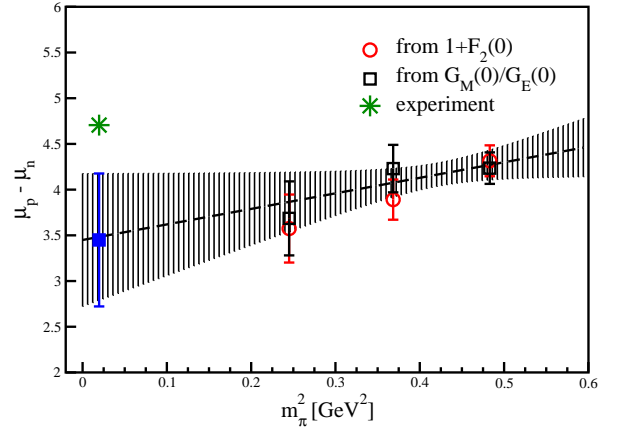}
\caption{Comparison of two determinations of $\mu_p-\mu_n$.
There is no appreciable $m_{\pi}^2$ dependence for the results
from $G^{v}_M(0)/G^{v}_E(0)$. A simple linear extrapolation
is applied to them. The resulting extrapolated value at the physical point
shows the better agreement with the experimental value (asterisk).}
\label{fig:MagMoment}
\end{figure}

In Fig.~\ref{fig:Gm_ov_Ge}, we show the $q^2$ dependence of the ratio $G^v_M(q^2)/G_E^v(q^2)$, which clearly exhibits mild dependence. We therefore use a simple linear fitting form with respect to $q^2$ for an alternative evaluation of the value $\mu_p-\mu_n$. In Fig.~\ref{fig:MagMoment}, we plot values of $\mu_p-\mu_n$, from two determinations, $1+F_2(0)$ and $G_M(0)/G_E(0)$, as functions of pion mass squared. Both determinations are roughly consistent with each other within the statistical error. Weak $m_{\pi}^2$ dependence is observed; the heavier two points in the case of $G_M(0)/G_E(0)$ are the same within error, while slight downward $m_{\pi}^2$ dependence appears in the results of $1+F_2(0)$. The simple linear extrapolation of $G_M(0)/G_E(0)$ yields $\mu_p-\mu_n=3.4(7)$, which gives the better agreement with the experimental value. We may quote this value for our final value of $\mu_p-\mu_n$, since the latter approach appears to have smaller systematic error in the $q^2$-extrapolation.

\subsubsection{Axial-vector current}

\begin{figure}[htb]
\includegraphics[width=\columnwidth,clip]{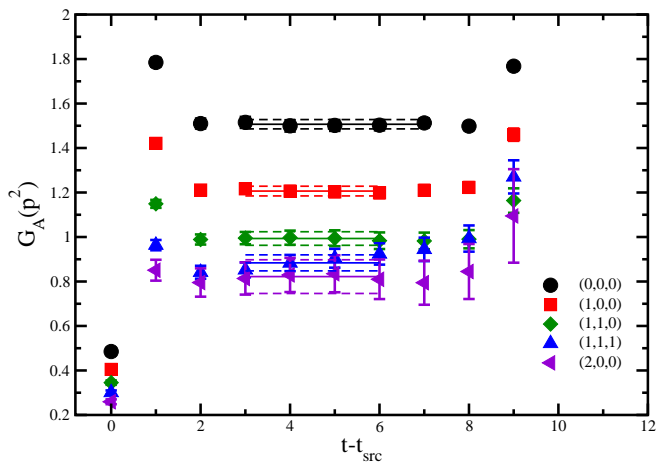}
\caption{The bare axial form factor as in Fig.~\ref{fig:rawDiracFF}}
\label{fig:rawgavsq}
\end{figure}

\begin{figure}[h]
\includegraphics[width=0.9\columnwidth,clip]{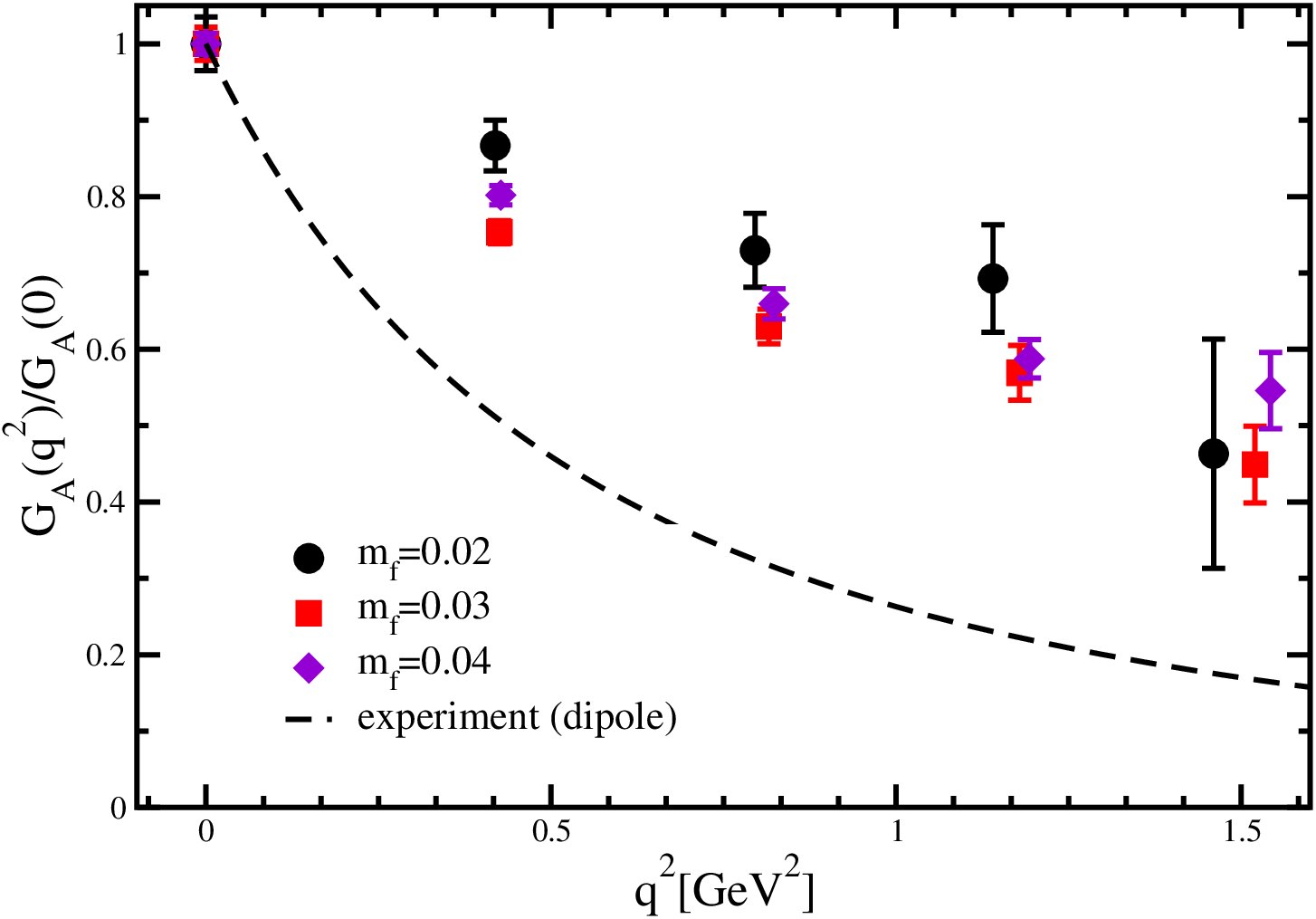}
\includegraphics[width=0.9\columnwidth,clip]{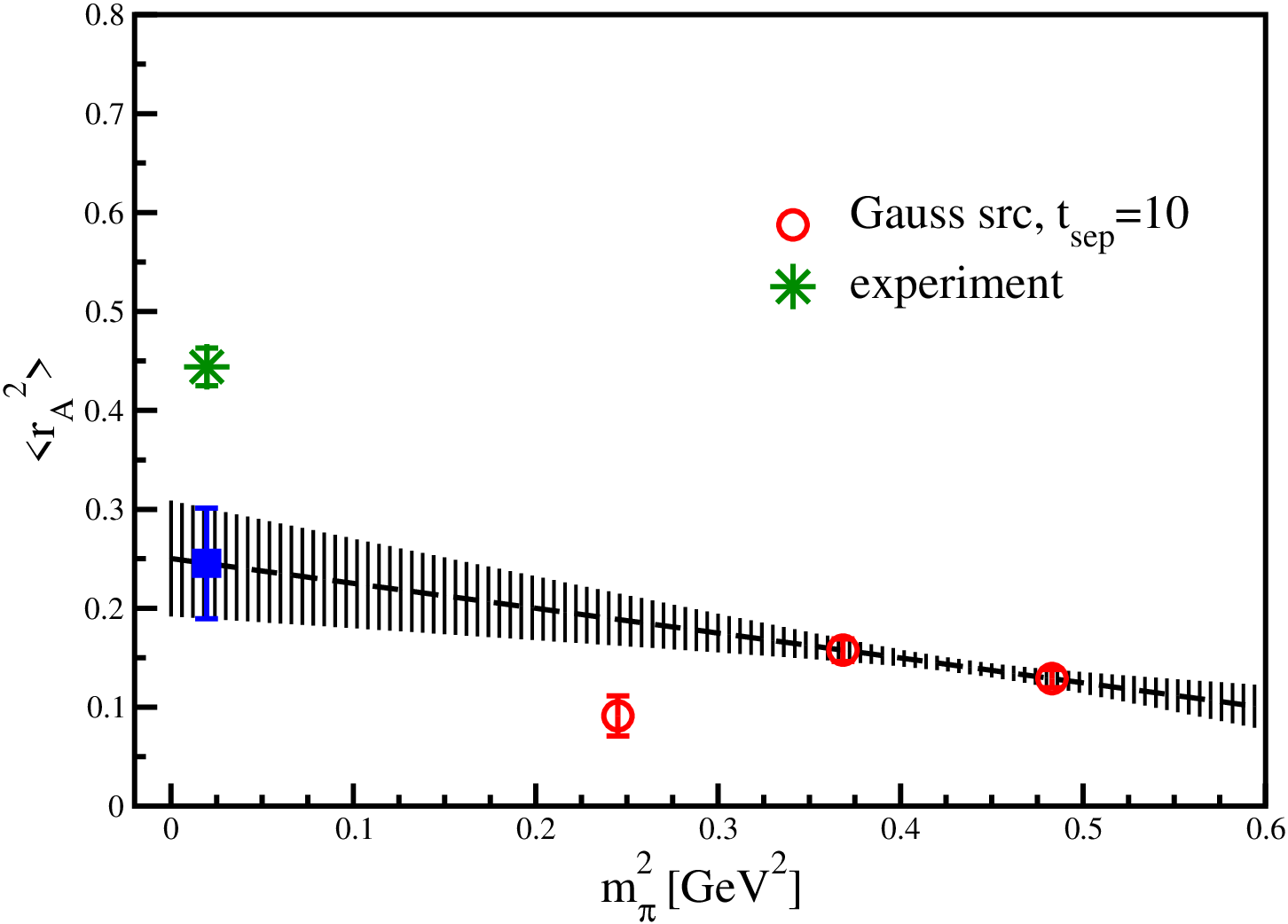}
\caption{
(Top) Momentum-transfer dependence of vector form factors.
The dashed curve corresponds to the dipole form with the experimental value of the dipole mass $M_{A}=1.026(21)$~GeV, evaluated
from pion electroproduction experiments. \\
(Bottom) Mean-square axial radii $\langle r^2_A \rangle$
from $G_A(q^2)$. The shaded regions are naive linear extrapolations (with errors) of our data points to the physical pion mass. An unexpected reduction of the measured $\langle r^2_A \rangle$ at the lightest quark mass is observed.}
\label{fig:gAdipole}
\end{figure}

In Fig.~\ref{fig:rawgavsq}
we present the raw data for the axial form factor at finite momentum transfer. These allow us to extract the momentum-transfer dependence of the form factor as listed in Table~\ref{tab:q^2_FF} and plotted in the top of Fig.~\ref{fig:gAdipole}.
The axial form factor is phenomenologically fitted with the dipole form, at least at low $q^2$; so are the Dirac and Pauli form factors~\cite{Bernard:2001rs}. The dashed curve in the top of Fig.~\ref{fig:gAdipole} shows the dipole form with an experimental value of the axial dipole mass $M_A=1.026(21)$~GeV~\cite{Bernard:2001rs}. Our lattice data lie above the experimental curve. Notice that the likely finite volume effect observed earlier for the axial charge leads to the non-monotonic behavior of the form factor with quark mass, and leads to the downward curvature of the mean-square axial radius.

To extract the mean-squared radius of the axial vector form factor, we perform the dipole fit to the three lowest $q^2$ points and convert the axial dipole mass $M_A$ into the mean-squared axial radius $\langle r_A^2\rangle$. The obtained values are listed in Table~\ref{tab:MA} and the bottom of Fig.~\ref{fig:gAdipole}.
\begin{table}[t]
\caption{Dipole fits for momentum-transfer dependence of the axial form factor.}
\label{tab:MA}
\begin{center}
\begin{tabular}{lll}
\hline\hline
$m_f$& $M_A$ ({\rm GeV})&$\langle r_A^2\rangle$ $({\rm fm}^2)$ \\
\hline
0.02&2.26(25)&0.091(20)\\
0.03&1.72(7)  &0.158(11)\\
0.04&1.90(7)  &0.129(9)\\
\hline
Expt. & 1.026(21)& 0.444(19) \\
\hline\hline
\end{tabular}
\end{center}
\end{table}
We also plot the mean-squared axial radius as a function of the pion mass squared. As can be seen, the value of $\langle r_A^2\rangle$ at the lightest pion mass shows a large reduction beyond statistical fluctuations. This large reduction in the axial radius suggests finite-volume effects could be significant at the lightest quark mass.

There is renewed interest in the pseudoscalar form factor, $G_P(q^2)$, due to the recent MuCap Collaboration~\cite{Andreev:2007wg} high-precision experiment studying ordinary muon capture by protons, $\mu^-p \rightarrow \nu_\mu n$, and because of improved electroweak radiative correction calculations\cite{Czarnecki:2007th} that allow precise extraction of the form factor from these experiments.
Using the new MuCap results, a value of $g_P=\frac{m_{\mu}}{2m_{N}}
G_P(0.88 m_\mu^2) = 7.3\pm1.1$ is found\cite{Andreev:2007wg,Czarnecki:2007th} which, using PCAC and chiral perturbation theory, is now in good agreement with QCD, $g_P=8.26\pm0.12$~\cite{Bernard:2001rs}. Earlier, a TRIUMF group obtained a value of $12.4 \pm 1.0$~\cite{Jonkmans:1996my}. Clark {\it et~al.}~\cite{Clark:2005as} got $g_P = 10.6 \pm 1.1$ after reanalyzing the TRIUMF data. Including the new MuCap result, the ``world average" is $8.7 \pm 1.0$~\cite{Czarnecki:2007th}.

\begin{figure}[t]
\includegraphics[width=\columnwidth,clip]{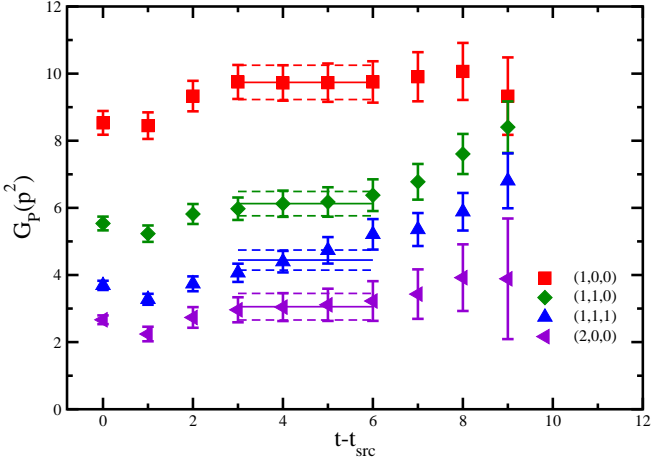}
\caption{The bare pseudoscalar form factor as in Fig.~\ref{fig:rawDiracFF}}
\label{fig:rawgpvsq}
\end{figure}

\begin{figure}[htb]
\includegraphics[width=0.9\columnwidth,clip]{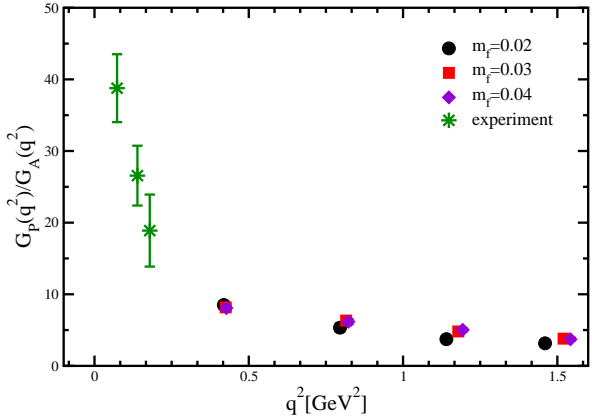}
\caption{Momentum-transfer dependence of the induced pseudoscalar form
factor compared with experiments.
}
\label{fig:gPgA}
\end{figure}

The induced pseudoscalar form factor $G_P$ is obtained from Eq.~(\ref{eq:solve_axial}), and is shown in Fig.~\ref{fig:rawgpvsq}.
This allows us to extract its momentum dependence which is listed in Table~\ref{tab:q^2_FF} and plotted in Fig.~\ref{fig:gPgA}.
As is noted in the previous section, we use a dimensionless definition for this, as well as the other three form factors, and use twice the nucleon mass $2m_{N}$ estimated for each ensemble of configurations.
This is different from some experimental analyses where the muon mass is used for normalization. Momentum-transfer dependence of the pseudoscalar form factor $G_{P}(q^2)$ has so far only been studied by one pion electroproduction experiment at Saclay in 1993~\cite{Choi:1993vt} with range below $0.2\mbox{ GeV}^2$.

Fig.~\ref{fig:gPgA} shows the $q^2$ dependence of the ratio of the induced pseudoscalar to axial form factors, $G_{P}(q^2)/G_A(q^2)$. Though our lattice momenta transfer are much higher than experimentally explored values, they trend upward toward the experiments.

Another topic regarding the induced pseudoscalar form factor is its relation with the axial form factor through the PCAC relation and pion-pole dominance (PPD) model~\cite{Braun:2006hz}:
\begin{equation}
G^{\rm PPD}_P (q^2) = \frac{4m_N^2 G_A (q^2)}{q^2 + m_\pi^2}.
\label{Eq:PPD_form}
\end{equation}
To see how the pion-pole behavior is preserved in measured $G_P(q^2)$, we consider the following ratio:
\begin{equation}
\alpha_{_{\rm PPD}}=\frac{G_P(q^2)}{G^{\rm PPD}_P(q^2)},
\end{equation}
which is inspired by the above PCAC prediction.
If the measured $G_P(q^2)$ has exactly the same form described in Eq.~(\ref{Eq:PPD_form}), this ratio yields unity in the entire $q^2$ region.

\begin{figure}[h]
\includegraphics[width=0.9\columnwidth,clip]{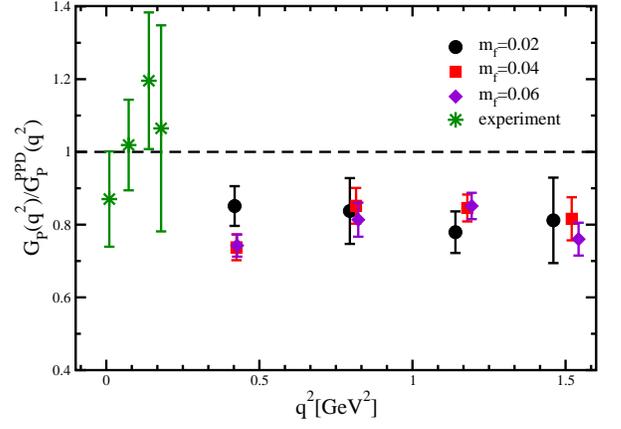}
\caption{The ratio of $G_P(q^2)$ and $G_P^{\rm PPD}(q^2)$ as a function of four-momentum squared $q^2$. Deviation of the ratio from unity indicated by dashed line is the deviation from the
PPD model. However, the deviation is rather modest. At least, the $q^2$ dependence of the PPD model approximately accounts for that of measured $G_P(q^2)$.}
\label{fig:gPgAcomp}
\end{figure}

\begin{figure}[h]
\includegraphics[width=0.9\columnwidth,clip]{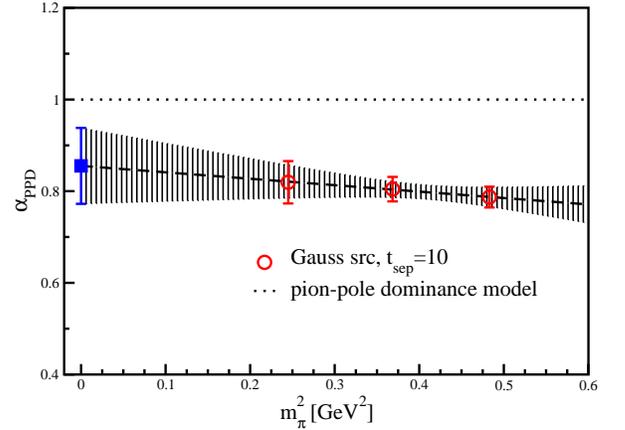}
\caption{The quenching factor $\alpha_{_{\rm PPD}}$ is plotted as a
function of the pion mass squared.}
\label{fig:alphaPPD}
\end{figure}

In Fig.~\ref{fig:gPgAcomp},
we plot the above ratio as a function of four-momentum squared $q^2$. There is no appreciable $q^2$ dependence. Thus, the $q^2$ dependence of the PPD model approximately accounts for that of measured $G_P(q^2)$. Four different $q^2$ points of the ratio $\alpha_{_{\rm PPD}}$ reveal a $q^2$ independent plateau within the statistical errors. We simply take the weighted average of $\alpha_{_{\rm PPD}}$ within four measured $q^2$ points, then plot them against the pion mass squared. As shown
in Fig.~\ref{fig:alphaPPD},
the average values of $\alpha_{_{\rm PPD}}$ gradually approach unity as the pion mass decreases. A simple linear extrapolation of the quenching factor $\alpha_{_{\rm PPD}}$ to the chiral limit yields 0.85(8), which is only about 2 standard deviations away from the theoretical expectation. One can obtain $g_P$ with input from $\alpha_{_{\rm PPD}}$:
\begin{equation}
g_P = \alpha_{_{\rm PPD}} \frac{g_A}{(1+q_0^2/M_A^2)^2}  \frac{2 m_N m_\mu}{(m_\pi^2+q_0^2)},
\end{equation}
where $q_0^2 = 0.88 m_\mu^2$ with $m_\mu= 0.10568$~GeV and $m_\pi = 0.13957$~GeV. $M_A$ and $g_A$ are 1.38(16)~GeV and 1.23(12) respectively  from naive linear fit using two heavier pion mass points. We found
\begin{equation}
g_P = 7.68 \pm 1.03
\end{equation}
with statistical error evaluated by jackknife analysis. This is consistent with the ``world average'' value of $8.7 \pm 1.0$~\cite{Czarnecki:2007th}.

The pseudoscalar form factor $G_{P}$ is related to the pion-nucleon coupling, $g_{\pi NN}$, through the Goldberger-Treiman relation ~\cite{Goldberger:1958tr}:
\begin{equation}
2 m_{N} G_{A}(q^2) -q^2 \frac{G_{P}(q^2)}{2m_N} = \frac{2g_{{\pi NN}} F_\pi m_\pi^2}{q^2+m_\pi^2}.\label{eq:TG_Qnq0}
\end{equation}
At zero momentum transfer, the relation reduces to the following form:
\begin{equation}
m_{N} g_{A} = F_\pi g_{\pi NN}.
\end{equation}
Testing the relation experimentally in this zero-momentum form is a little  tricky because the value of the pion-nucleon coupling, $g_{\pi NN}$, is not known at zero momentum transfer, but at the pion pole.
Worse, the pion-pole value of the coupling varies from analysis to analysis. A conventional partial-wave analysis of the pion-nucleon elastic scattering data to 2~GeV provides $13.75 \pm 0.15$~\cite{Arndt:1994bu} while a more recent reanalysis in Ref.~\cite{Bugg:2004cm} gives $g_{\pi NN} = 13.169 \pm 0.057$. Substituting the latter, we find there is a slight discrepancy from the Goldberger-Treiman relation at zero momentum transfer, $\Delta_{\rm GT}$,~\cite{Coon:1981jp} defined as
\begin{equation}
\Delta_{\rm GT} = 1-{m_N g_{A}}/ {F_\pi g_{\pi NN}}
\end{equation}
which is $(2.259 \pm 0.591)\%$ if we use the mean nucleon mass $m_N= 938.9$~MeV.

\begin{figure}[h]
\includegraphics[width=0.9\columnwidth,clip]{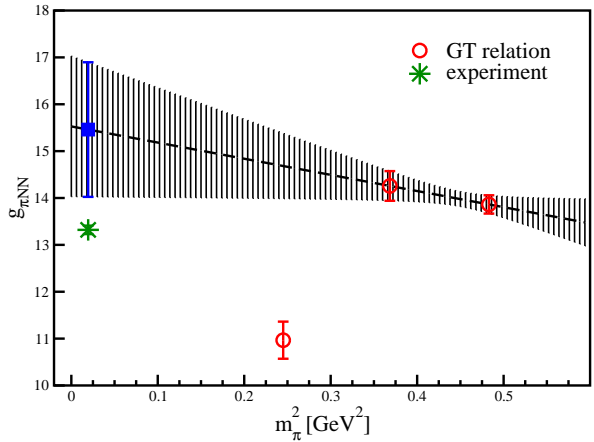}
\caption{Lattice estimate for the pion-nucleon coupling, $g_{\pi NN}^{\rm lat}$, from the Goldberger-Treiman relation as defined in Eq.~(\ref{eq:gPiNN}).
Omitting the lightest mass point, a constrained linear fit is performed.
}\label{fig:gPiNN}
\end{figure}

Testing the relation at finite momenta transfer is more difficult, as the pion-nucleon coupling is even more poorly known from experiment.
Thus it is useful to extract the pion-nucleon coupling, $g_{{\pi NN}}$, from our lattice data:
\begin{equation}
g_{\pi NN, {\rm lat}} = \frac{m_{N,{\rm lat}} g_{A, {\rm lat}}}
{F_{\pi, {\rm lat}} }.\label{eq:gPiNN}
\end{equation}
See Fig.~\ref{fig:gPiNN}
for $g_{\pi NN, {\rm lat}}$ results. A linear fit is performed without the lightest mass point to give an estimate of $g_{\pi NN}=15.5 \pm 1.4$ at the physical point. This should be compared with the experimental estimates at the pion pole such as $13.75\pm0.15$ or $13.169\pm0.057$ in the above.

\begin{figure}[t]
\includegraphics[width=0.9\columnwidth,clip]{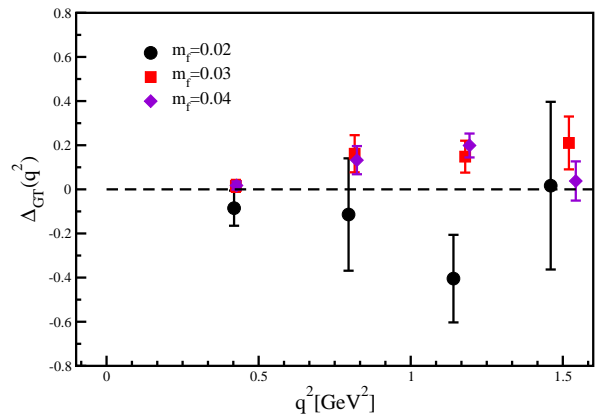}
\caption{Momentum-transfer dependence of the Goldberger-Treiman discrepancy as defined in Eq.~(\ref{eq:TG_Qnq0_2}).}
\label{fig:GTd}
\end{figure}

Now let us examine the Goldberger-Treiman relation by looking at the discrepancy $\Delta_{\rm GT}(q^2)$ defined by the lattice quantities as follows:
\begin{equation}
1-
\frac{q^2+m_{\pi,{\rm lat}}^2}{2 m_{N, {\rm lat}}}
\frac{4 m_{N,{\rm lat}}^2 G_{A,{\rm lat}}(q^2)-q^2 G_{P,{\rm lat}}(q^2)}
{2g_{{\pi NN,{\rm lat}}} F_{\pi, {\rm lat}} m_{\pi,{\rm lat}}^2};\label{eq:TG_Qnq0_2}
\end{equation}
see Fig.~\ref{fig:GTd}
for the value of $\Delta_{\rm GT}(q^2)$.
Note again that we are using the pion-nucleon coupling value determined at zero-momentum transfer, and cannot account for its variation at finite momentum transfer. From our data for the two heavier quark mass values, we observe the discrepancy is strongly dependent on the momentum transfer beyond  about $q^2 \approx 0.5~\mbox{ GeV}^2$. This suggests the pion-nucleon coupling at such high momentum transfer values is very much different from the low momentum-transfer region. Note our pion mass squared is roughly $m_\pi^2$ = 0.25, 0.36 and $0.49\mbox{ GeV}^2$, or below the lowest $q^2$ in the plot, and the physical pion mass is even lower.
Thus our data also suggest the Goldberger-Treiman relation holds at low momentum transfer, $q^2 \le 0.5\mbox{ GeV}^2$. In that region the lightest quark mass result is in broad agreement with the heavier masses, albeit with large statistical error. It is desirable to investigate this further at lower momentum transfer.

\subsection{Momentum and helicity fractions}
\label{subsec:moments}

\begin{figure}[b]
\includegraphics[width=0.9\columnwidth,clip]{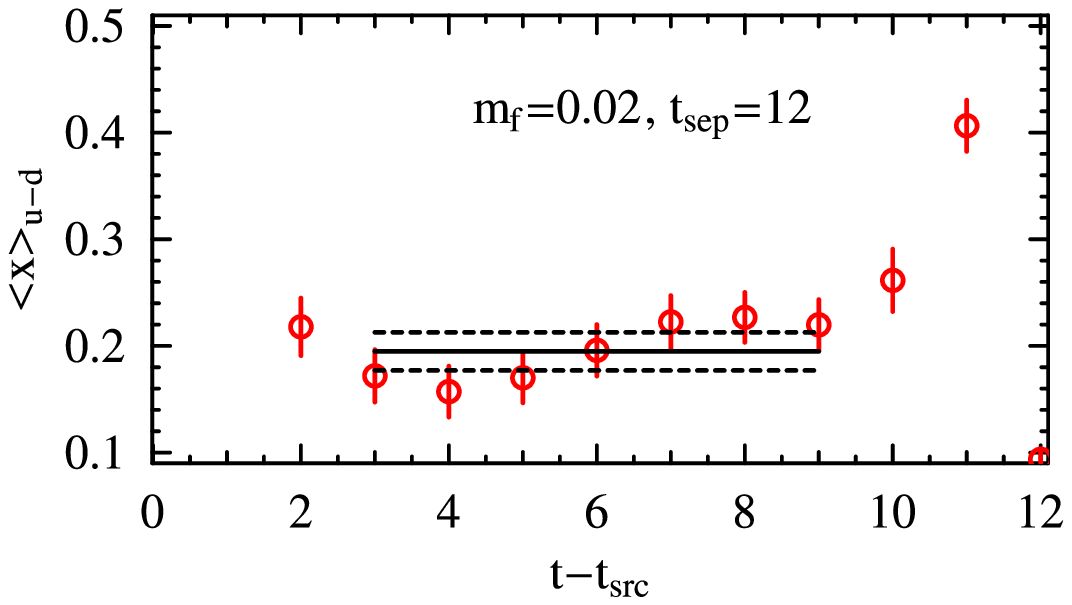}
\includegraphics[width=0.9\columnwidth,clip]{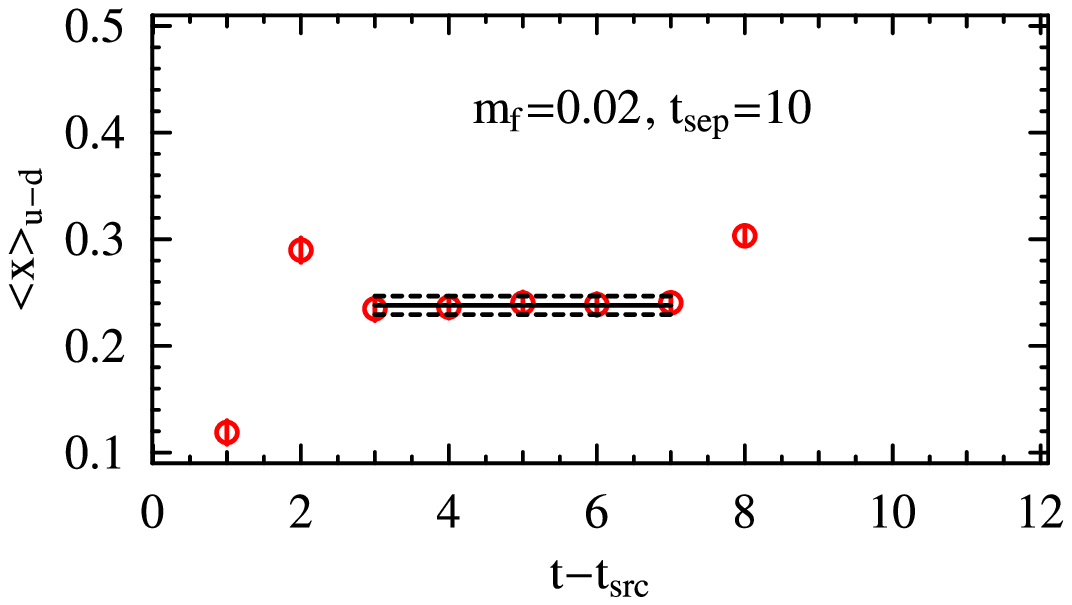}
\includegraphics[width=0.9\columnwidth,clip]{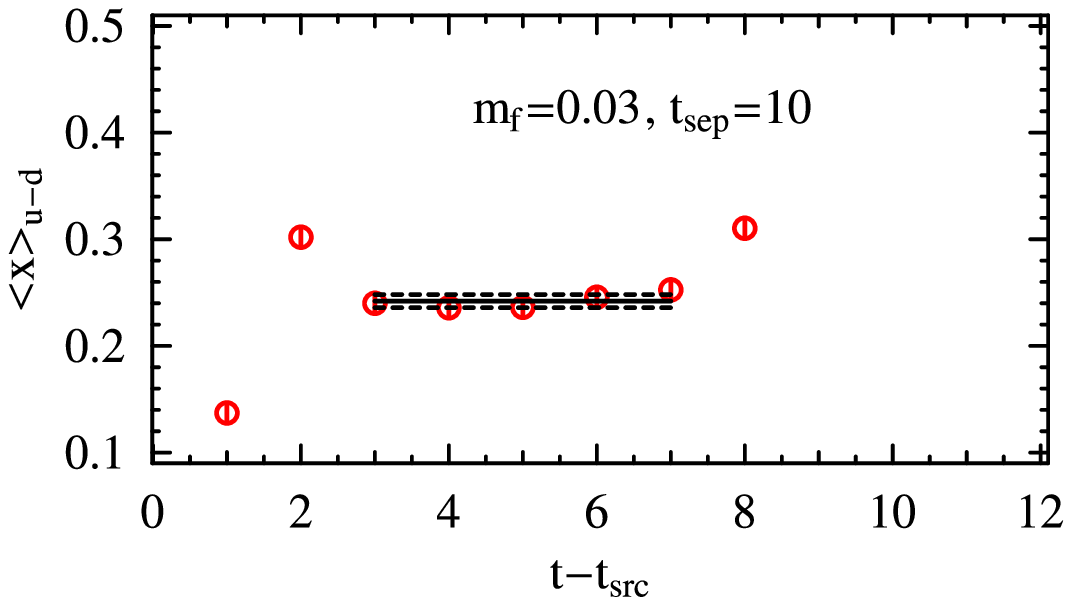}
\includegraphics[width=0.9\columnwidth,clip]{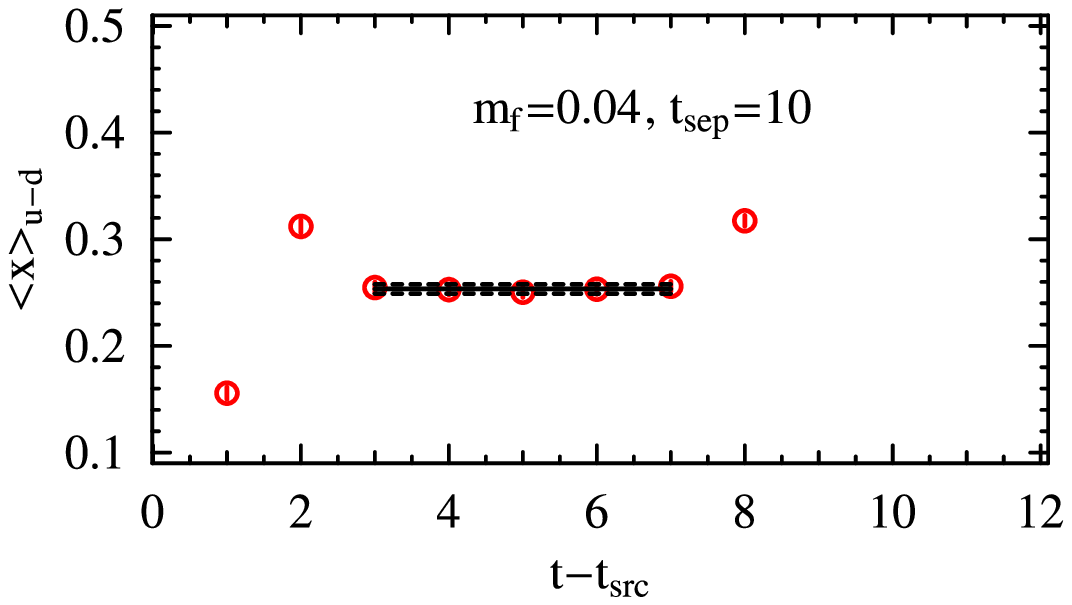}
\caption{Bare momentum fraction as a function of the operator insertion time. Results are averaged over all sources.
The top figure depicts the $t_{\rm sep} = 12$ calculation at the lightest quark mass, $m_f = 0.02$. The other three figures are obtained
from $t_{\rm sep} = 10$ calculations at all three quark masses (in order
of increasing mass from top to bottom). The fit ranges are
shown by horizontal lines.}
\label{fig:rawxq}
\end{figure}
Let us now turn our attention from the elastic form factors to the deep inelastic structure functions. As is well known and has been summarized in the introduction, what are calculable on the lattice in regard to the structure functions are their low-order moments. In this work we limit ourselves to the lowest order non-trivial moments that are calculable with zero momentum transfer. As in previous RBC reports, the operator renormalizations at the chiral limit are obtained nonperturbatively using the RI/MOM scheme as described in Sec.~\ref{subsec:NPR}.

We first discuss the quark momentum fraction, $\langle x \rangle_{u-d}$, or the first moment of the unpolarized structure function.
The bare lattice three-point correlators are shown in Fig.~\ref{fig:rawxq} and listed in Table~\ref{tab:Xud}.
The results for $t_{\rm sep}=12$ and $t_{\rm sep}=10$ at the lightest pion mass differ by about two standard deviations. This may suggest excited state contamination in this quantity, even though its plateau looks very nice and flat over the range $3\le t \le 7$ for $t_{\rm sep}=10$. From the top panel in Fig.~\ref{fig:rawxq}, the plateau for $t_{\rm sep}=12$ is not so clear or flat, owing to the much larger
statistical fluctuations in the data for this larger source-sink separation. However, since it is low on the left and high on the right, the average value is insensitive to the choice of (a symmetric) fit range. In the earlier calculation, using a box source on a single time slice and a source-sink separation of 12 \cite{Ohta:2004mg,Orginos:2002fr},
the momentum fraction was consistent, within large statistical errors, with the $t_{\rm sep}=10$ result. In this study, we can not conclude whether this is a true systematic effect or just statistics, especially given the quality of the $t_{\rm sep}=12$ plateau.

\begin{figure}[t]
\includegraphics[width=0.9\columnwidth,clip]{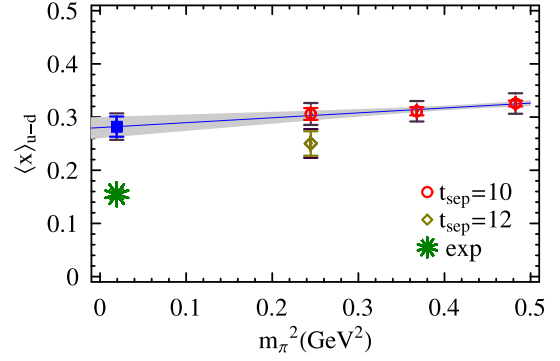}
\caption{Renormalized first moment of the unpolarized structure functions or the quark momentum fraction, $\langle x \rangle_{u-d}$. Note that the circles indicate the data at each pion mass with two errorbars: the inner are statistical errorbars and the outer are errorbars propagated with renormalization factors. The solid square is the extrapolated point; the star is the experimental value; and the band shows the chiral extrapolation with jackknife-calculated uncertainty.}
\label{fig:zxq}
\end{figure}

We show the results for all the measurements but only use the result from $t_{\rm sep}=10$ at the lightest pion mass for extrapolation. Similar results would be obtained using the $t_{\rm sep}=12$
result at the lightest mass since the larger error on this point does not tightly constrain the fit.
Using the fit ranges indicated in the figures and the renormalizations presented in subsection~\ref{subsec:NPR}, we arrive at the values of $\langle x\rangle_{u-d}$ shown in Fig.~\ref{fig:zxq}.
Note that there is only small pion-mass dependence; therefore, we use a simple linear fit. The quality of this fit in terms of $\chi^2$ per degree of freedom is reasonable at 0.28.  The chiral extrapolated value is $\langle x\rangle_{u-d} = 0.282(19)$, overshooting the experimental value, 0.154(3)~\cite{Lai:1996mg,Gluck:1998xa,Martin:2001es}, by  five to six standard deviations.
Here, the trend of the $t_{\rm sep}=12$ data point is suggestive, but again, not conclusive
because of the relatively large statistical error.

\begin{figure}[t]
\includegraphics[width=0.9\columnwidth,clip]{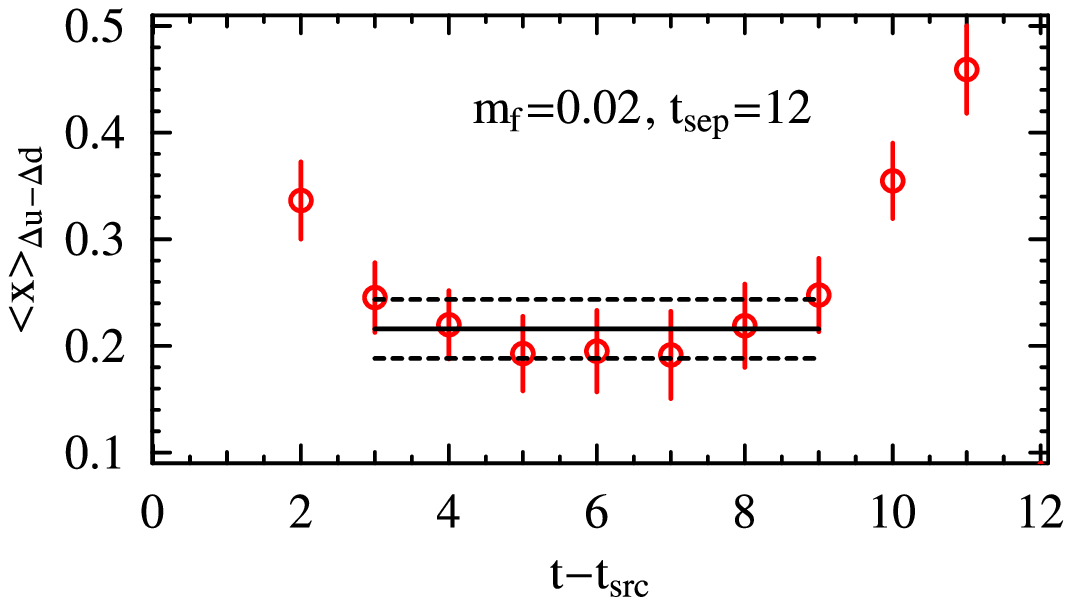}
\includegraphics[width=0.9\columnwidth,clip]{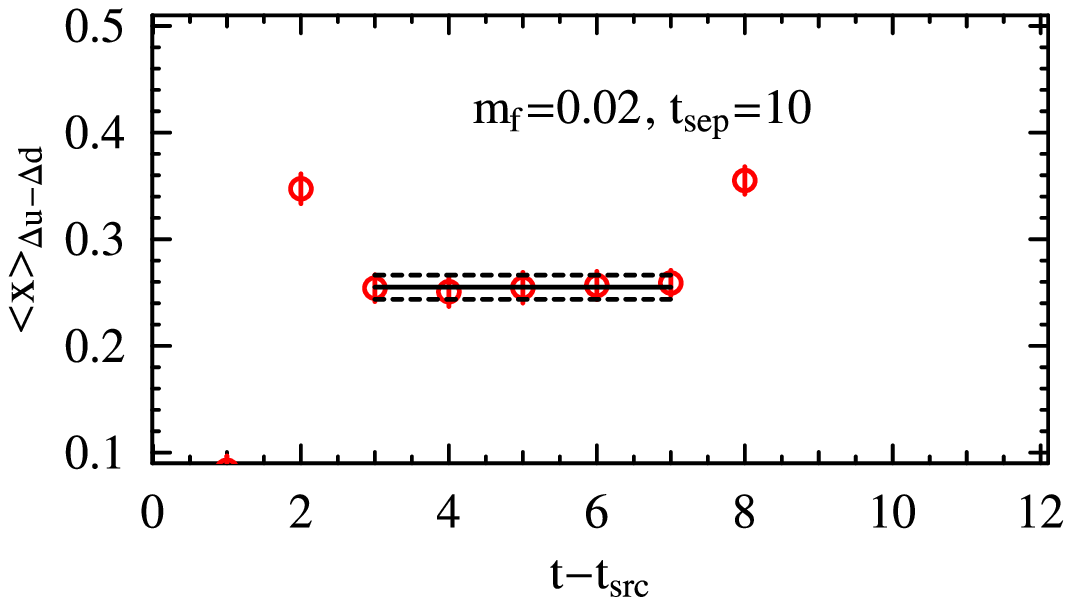}
\includegraphics[width=0.9\columnwidth,clip]{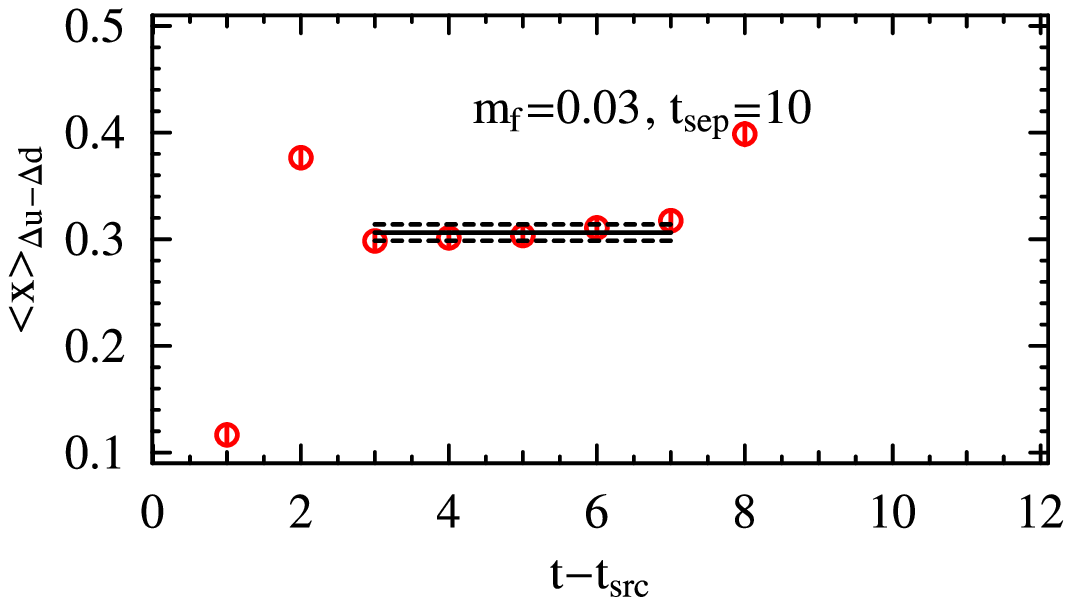}
\includegraphics[width=0.9\columnwidth,clip]{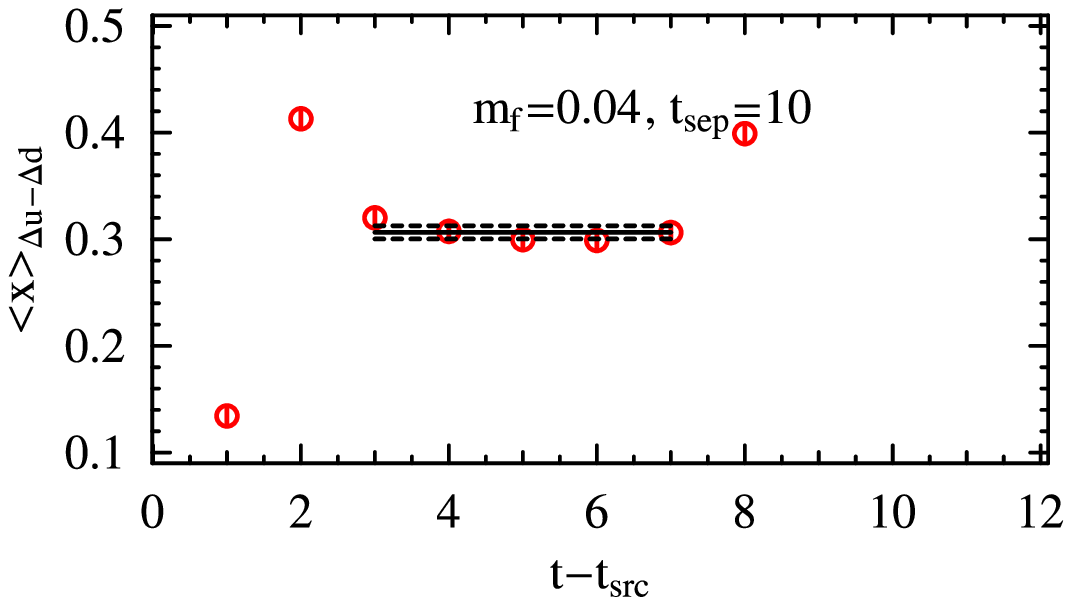}
\caption{The bare helicity fraction; symbols as in Fig.~\ref{fig:rawxq}.}
\label{fig:rawDxq}
\end{figure}

The quark helicity fraction, $\langle x \rangle_{\Delta u - \Delta d}$, is the first moment of the polarized structure function.  The bare three-point correlator values are shown in Fig.~\ref{fig:rawDxq} and the values are listed in Table~\ref{tab:Xud}. Similar remarks and conclusions
about the plateaus for $t_{\rm sep}=10$ and 12 as for the momentum fraction hold here as well.
\begin{figure}[htb]
\includegraphics[width=0.9\columnwidth,clip]{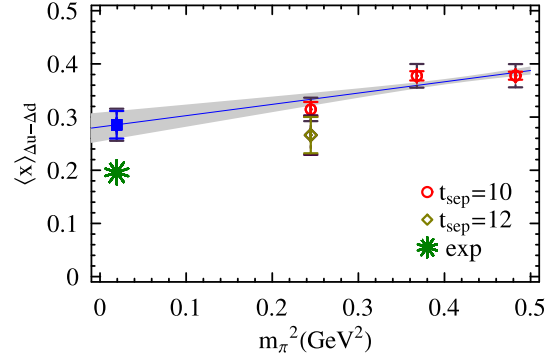}
\caption{Renormalized first moment of the polarized structure functions or quark helicity fraction, $\langle x \rangle_{\Delta u - \Delta d}$, as a function of $m_\pi^2$. (Similar graphics description as in Fig.~\ref{fig:zxq}.)}
\label{fig:zxdq}
\end{figure}
The renormalization factors at the chiral limit for the helicity fraction can be found in Sec.~\ref{subsec:NPR}. Fig.~\ref{fig:zxdq} shows the renormalized quantities at each pion mass point. A linear fit to these values versus $m_\pi^2$ gives a chiral extrapolation of 0.286(25), overshooting the experimental value of 0.196(4)~\cite{Gluck:1995yr,Gehrmann:1995ag,Dolgov:2002zm} by almost four standard deviations. The quality of the fit is poor, as $\chi^2$ per degree of freedom is 6.78. In contrast to the momentum fraction, the result at the light mass deviates from the heavier pion mass points, even at $t_{\rm sep}=10$, a situation similar to the deviation observed in the axial charge.  This trend is welcome since the experimental value apparently must be approached from above and  suggests trying fits different than linear, to which we turn below, after discussing the ratio of the momentum fraction to the helicity fraction.

\begin{figure}[t]
\includegraphics[width=0.9\columnwidth,clip]{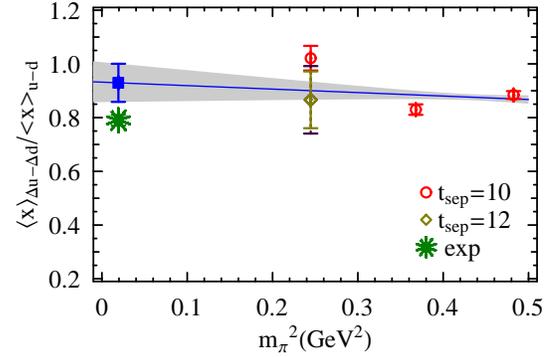}
\caption{Renormalized ratio $\langle x\rangle_{\Delta u - \Delta d}/\langle x \rangle_{u-d}$ of the quark momentum and helicity fraction. A linear fit is shown with an error estimate at the physical pion mass. Experiment gives 0.78(2).}
\label{fig:xdxratio}
\end{figure}

In an earlier quenched calculation~\cite{Orginos:2005uy} we reported that the ratio of the momentum and helicity fractions, $\langle x \rangle_{u-d}/\langle x\rangle_{\Delta u - \Delta d}$, agrees well with the experimental value, 0.78(2). The quark mass dependence of this ratio is presented in Fig.~\ref{fig:xdxratio}. Unlike the quenched case where the ratio is only weakly dependent on the quark mass and is in rough agreement with experiment, there is a noticeable deviation for the lightest pion mass. A naive linear extrapolation gives very poor fit quality, $\chi^2/{\rm dof}=15.7$, and a value at the physical pion mass of $0.93(7)$, which is about two standard deviations above the experimental one.  This may be due to systematic effects like
finite volume, or excited state contamination; the $t_{\rm sep}=12$ point is consistent with experiment and shows the weak mass dependence observed in the quenched case.

\begin{table}[t]
\caption{
Bare quark momentum and helicity fractions and their naturally renormalized ratio at each sea quark mass.}
\label{tab:Xud}
\begin{center}
\begin{tabular}{cclll}
\hline\hline
$m_f$  & $t_{\rm sep}$ &
\multicolumn{1}{c}{$\langle x \rangle_{u-d}$} &
\multicolumn{1}{c}{$\langle x \rangle_{\Delta u-\Delta d}$}&
\multicolumn{1}{c}{$\langle x \rangle_{u-d}/\langle x \rangle_{\Delta u-\Delta d}$}\\
\hline
0.02  &12&  0.195(17)& 0.22(3)& 0.99(14)\\
0.02  &10&  0.236(9)& 0.255(11)& 0.93(4)\\
0.03  &10&  0.242(6) & 0.306(6)& 0.790(17)\\
0.04  &10&  0.253(4) & 0.306(6)& 0.827(12)\\
\hline\hline
\end{tabular}
\end{center}
\end{table}

Since our pion masses are rather heavy (700, 600 and 500~MeV), it is not surprising that the chiral extrapolation to the physical pion mass (140~MeV) sometimes misses the experimental values. Also our heavy pion masses probably invalidate the application of a leading-order chiral form. The chiral extrapolation should ultimately be studied with more realistic lattice QCD ensembles such as the 2+1-flavor ones being generated jointly by RBC and UKQCD Collaborations\cite{Antonio:2006px,Allton:2007hx}, a work which is in progress. Nevertheless, it is a worthwhile exercise to try to extend the extrapolations beyond linear forms. We now discuss how some such attempts fair with our data.

\begin{figure}[t]
\includegraphics[width=0.8\columnwidth,clip]{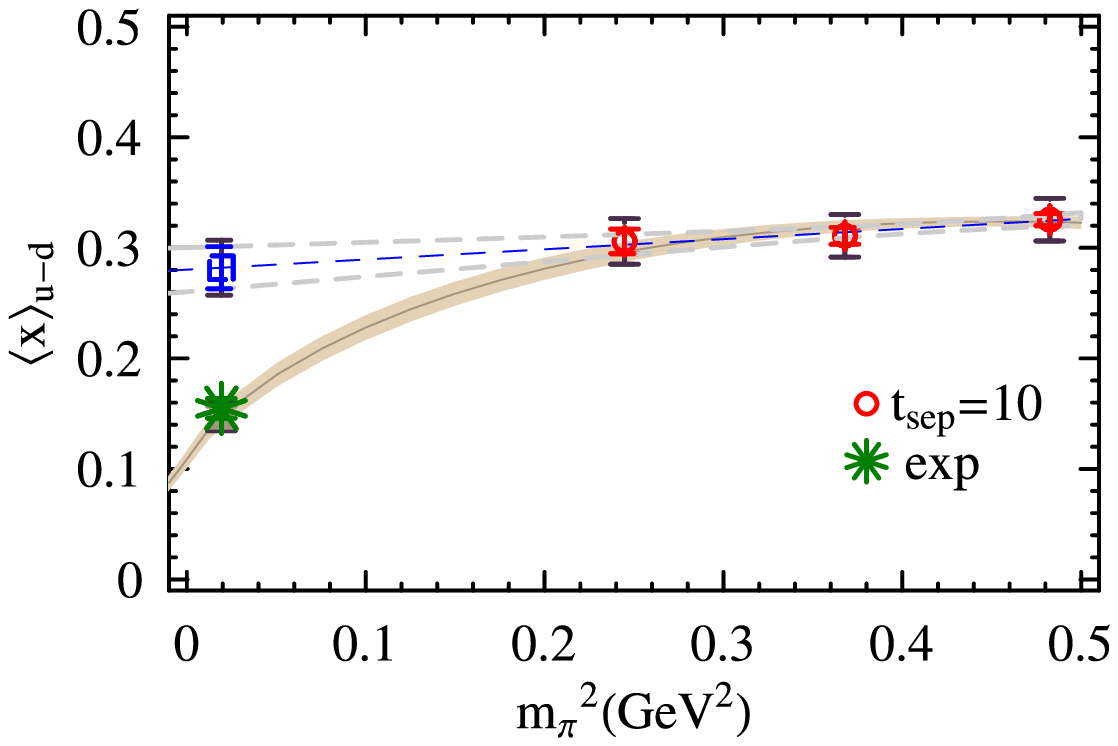}
\includegraphics[width=0.8\columnwidth,clip]{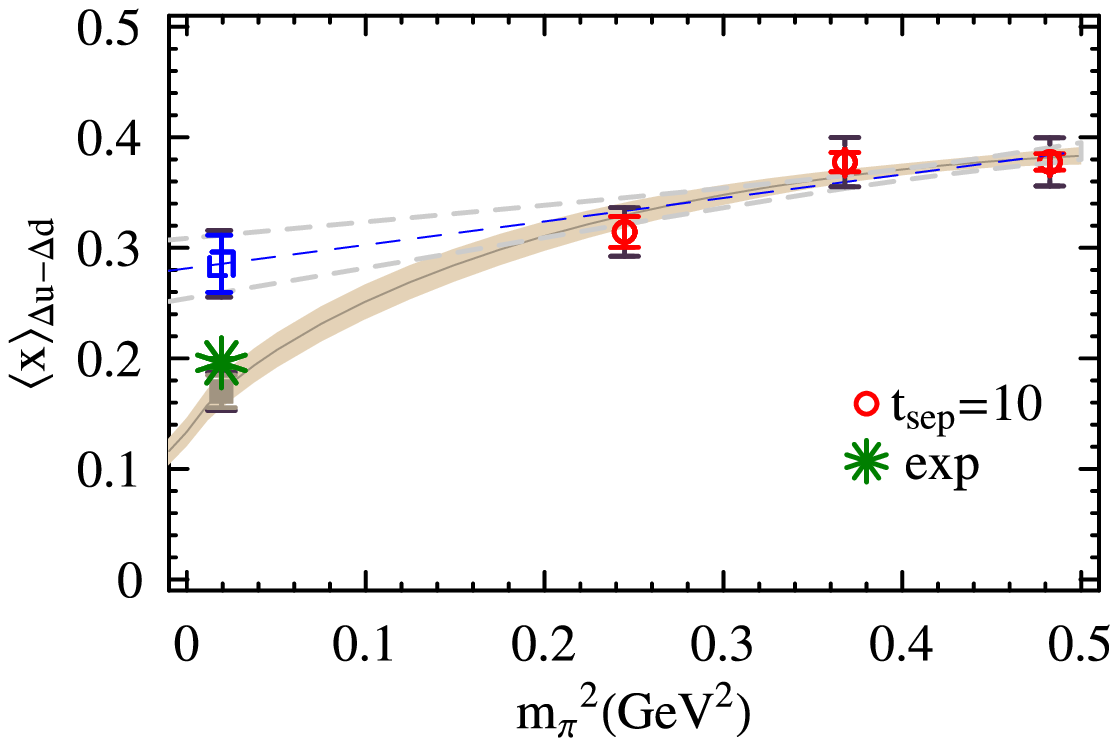}
\caption{Renormalized quark momentum fraction, $\langle x \rangle_{u-d}$, and helicity fraction, $\langle x \rangle_{\Delta u - \Delta d}$.
The phenomenological fit (brown band) is according to Refs.~\cite{Chen:2001gr,Chen:2001eg,Detmold:2002nf}. The dashed lines are the linear extrapolation for comparison.  (Graphics conventions as in Fig.~\ref{fig:zxq}.)}
\label{fig:zxqxdqchen}
\end{figure}

We begin with a  parametrization by Chen, {\it et al.}~\cite{Chen:2001gr,Chen:2001eg,Detmold:2002nf} (the relevant equations are summarized in appendix~\ref{subsec:Chen}.) Their prescription is to fix three parameters, the axial charge $g_{A,{\rm exp}}$, pion decay constant $F_{\pi,{\rm exp}}$ and the scale $\mu$ (set to the pion mass), to their experimental values and fit the coefficients $C$ and $e(\mu^2)$. The results are summarized in Fig.~\ref{fig:zxqxdqchen}. The fit yields momentum fraction 0.147(9) after extrapolation, which is consistent with the experimental value. However, the quality of the fit is poor, $\chi^2/{\rm dof}=2.19$; this is caused by a strong downward curvature in the fit form due to log terms. The corresponding helicity fraction extrapolation results in a value of 0.170(14), again, consistent with the experimental value. The fit for the helicity fraction, while still poor, is improved over the linear one, with a $\chi^2$ per degree of freedom of 3.57.

\begin{figure}[t]
\includegraphics[width=0.8\columnwidth,clip]{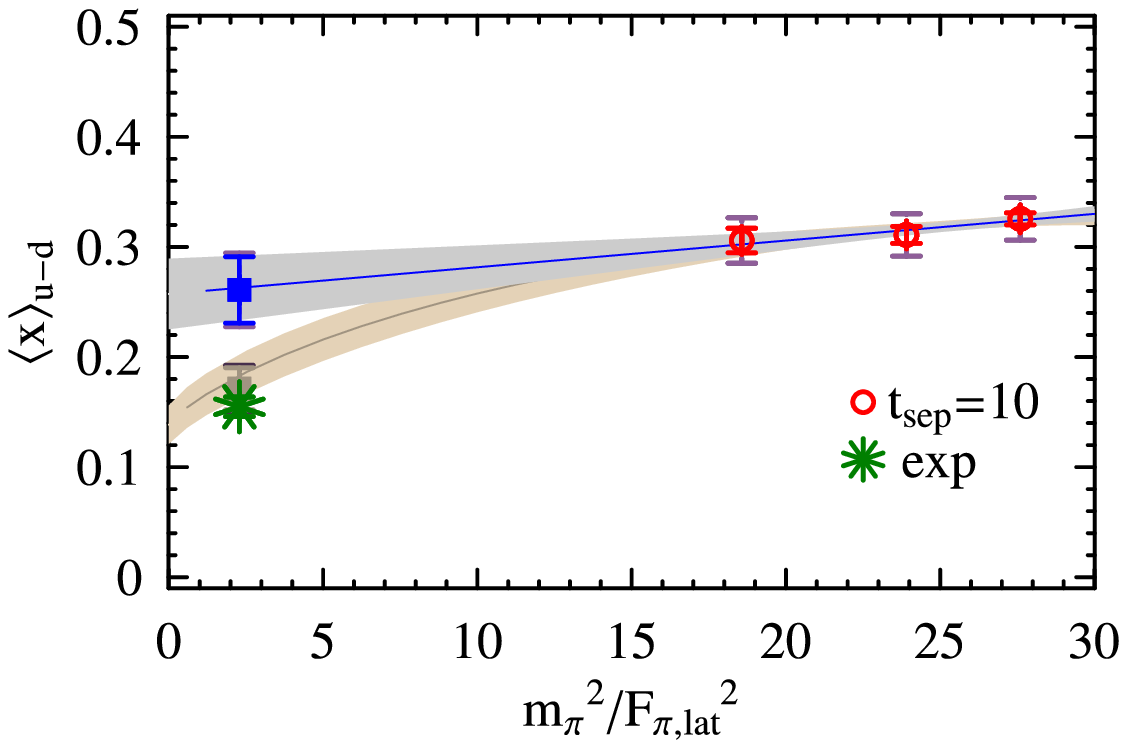}
\includegraphics[width=0.8\columnwidth,clip]{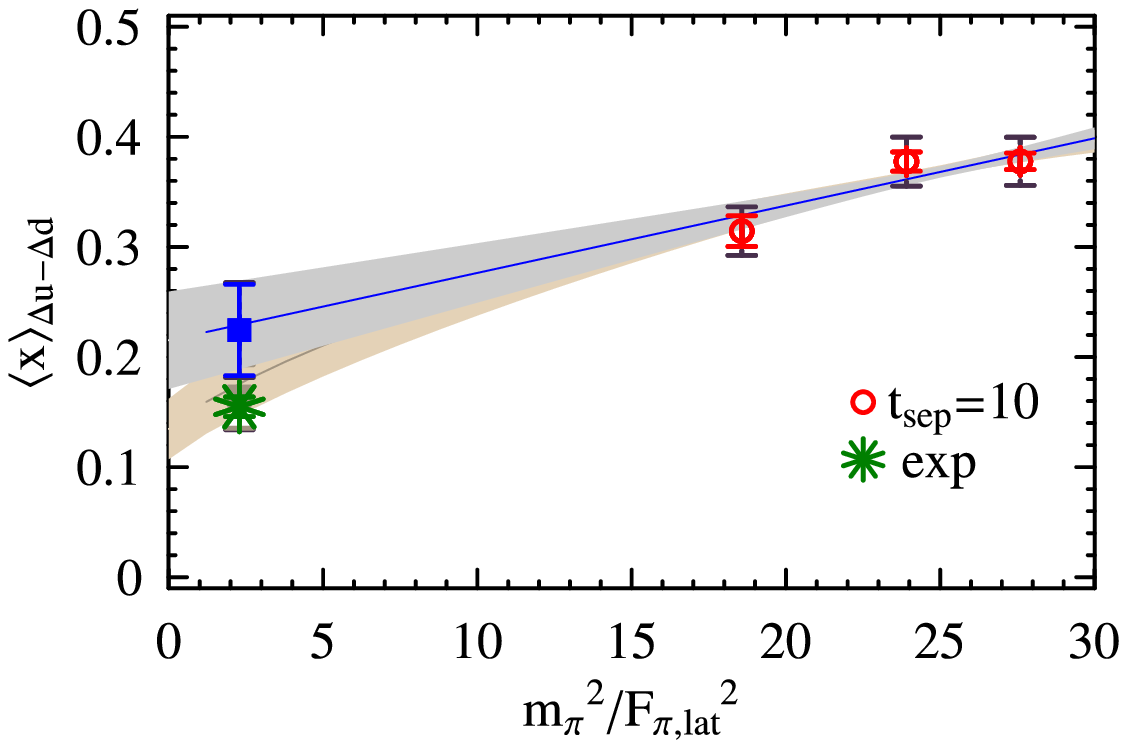}
\caption{Renormalized quark momentum fraction, $\langle x \rangle_{u-d}$, and helicity fraction, $\langle x \rangle_{\Delta u - \Delta d}$, in terms of $m_\pi^2/F_{\pi,{\rm lat}}^2$. The blue band indicates the linear extrapolation and the brown band is the phenomenological fit as described in Refs.~\cite{Chen:2001gr,Chen:2001eg,Detmold:2002nf}. (Graphics conventions as in Fig.~\ref{fig:zxq}.)}
\label{fig:zxqxdqchenmpifpi}
\end{figure}

Another alternative is to consider the dependence in quantities on the dimensionless ratio $m_\pi^2/F_{\pi,{\rm lat}}^2$, as shown in Fig.~\ref{fig:zxqxdqchenmpifpi}. Similar analyses have been done before, for example, in Refs.~\cite{Beane:2005rj,Edwards:2006qx}. (Note that $F_{\pi,{\rm lat}}$ is replaced by  $f_{\pi,{\rm lat}}(m_f)/\sqrt{2}$ at each $m_f$ to be consistent with the chiral extrapolation formulation used in Refs.~\cite{Chen:2001gr,Chen:2001eg,Detmold:2002nf}.) Since this is a dimensionless quantity, there is no systematic error coming from the lattice scale determination; therefore, it is a good way to find out whether we have control over lattice artifacts.

For the momentum fraction, $\langle x \rangle_{u-d}$, the fit yields an extrapolated value of 0.260(30) with an acceptable $\chi^2/{\rm dof} = 0.48$. That this extrapolation is consistent with the earlier ones indicates no further uncontrollable systematics other than the extrapolation itself. The value is about three standard deviations above experiment. Note the larger error on the extrapolated point is caused by the ``re-scaling'' of the extrapolation range. Similar conclusions are drawn for the helicity fraction, $\langle x \rangle_{\Delta u - \Delta d}$: the updated extrapolation brings the value to 0.224(41), within two standard deviations of experiment. However, the $\chi^2$ per degree of freedom for this fit is still large at 5.17.

The fit results from the non-linear chiral extrapolations (see Appendix~\ref{subsec:Chen} for more details) are summarized in Fig.~\ref{fig:zxqxdqchenmpifpi}. The momentum fraction is 0.171(19) at the physical point and agrees with experiment within one standard deviation. The fit has an acceptable $\chi^2$ per degree of freedom of 0.85. The helicity fraction extrapolates to 0.163(28) and is consistent both with the extrapolation in terms of $m_\pi$ and the experimental number. However the fit quality is poor: $\chi^2/{\rm dof} = 4.57$. These values are consistent with the earlier fits.

The trial fits and extrapolations are summarized in Table~\ref{tab:fracextr}. In the cases of both the quark momentum and helicity fraction, the extrapolation with the chiral form seems to give us consistent values with the experiments. However, these extrapolations require a 50\% drop from the magnitude of our measured points and do not always give the smallest $\chi^2/{\rm dof}$. We simply must measure at lower pion masses before such extrapolations can be deemed trustworthy.
Also note that the extrapolations in terms of $m_\pi^2$ or the dimensionless quantity $m_\pi^2/F_{\pi,{\rm lat}}^2$ are consistent with each other.

\begin{table}[b]
\caption{
Summary of the extrapolations of the renormalized first moment of the unpolarized distribution. The chiral perturbation theory formulation can be found in Ref.~\cite{Chen:2001gr,Chen:2001eg,Detmold:2002nf} or appendix~\ref{subsec:Chen}. The fit parameters are in the order of fit slope, intersection for the linear fits and $C$,  $e$ for the chiral formulation.
}
\label{tab:fracextr}
\begin{center}
\begin{tabular}{cllll}
\hline\hline
&
\multicolumn{1}{c}{$\langle x \rangle_{u-d} $}  &
\multicolumn{1}{c}{$\chi^2/{\rm d.o.f.}$} &
\multicolumn{1}{c}{fit parameters} \\
                  \hline
Linear vs. $m_\pi^2$ & 0.282(19)& 0.28 & \{0.09(5), 0.280(19)\}\\
Linear vs. $m_\pi^2/F_{\pi,{\rm lat}}^2$ & 0.260(30)& 0.48&
\{0.0024(12), 0.26(3)\}\\
ChPT vs. $m_\pi^2$ & 0.147(9)& 2.19 & \{0.109(8), 2.64(11)\}\\
ChPT vs. $m_\pi^2/F_{\pi,{\rm lat}}^2$  &0.171(19)&  0.85& \{0.139(17), 3.74(22)\}\\
experiment & 0.154(3)&&\\
\hline
&
\multicolumn{1}{c}{$\langle x \rangle_{\Delta u-\Delta d} $}  &
\multicolumn{1}{c}{$\chi^2/{\rm d.o.f.}$} &
\multicolumn{1}{c}{fit parameters} \\
                  \hline
Linear vs. $m_\pi^2$ & 0.286(25)& 6.78 &\{0.21(6), 0.281(27)\}\\
Linear vs. $m_\pi^2/F_{\pi,{\rm lat}}^2$ & 0.224(41)& 5.17 &\{0.0061(17), 0.22(4)\}\\
ChPT vs. $m_\pi^2$& 0.170(14)& 3.57&\{0.134(12),0.25(12)\}\\
ChPT vs. $m_\pi^2/F_{\pi,{\rm lat}}^2$  & 0.163(28)& 4.57 &\{ 0.135(27),  3.31(20)\}\\
experiment & 0.196(4)&&\\
 \hline\hline
\end{tabular}
\end{center}
\end{table}
We emphasize that the data in general is not consistent with
a simple linear form, but that it does trend toward experiment
as the quark mass is reduced.

\subsection{Transversity}\label{subsec:transversity}

\begin{figure}[t]
\includegraphics[width=0.9\columnwidth,clip]{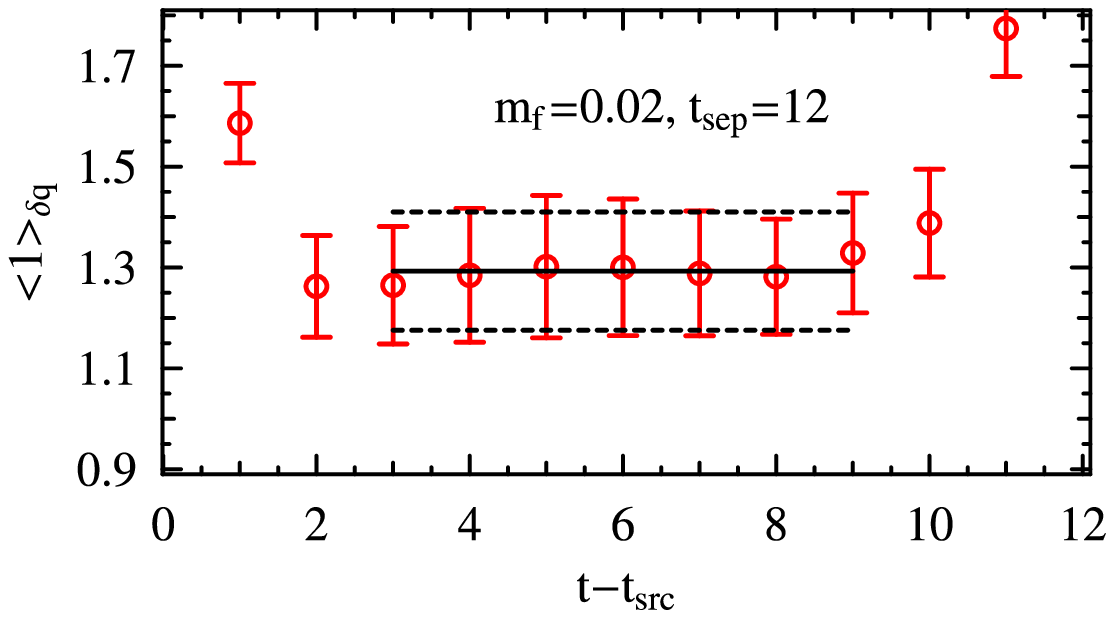}
\includegraphics[width=0.9\columnwidth,clip]{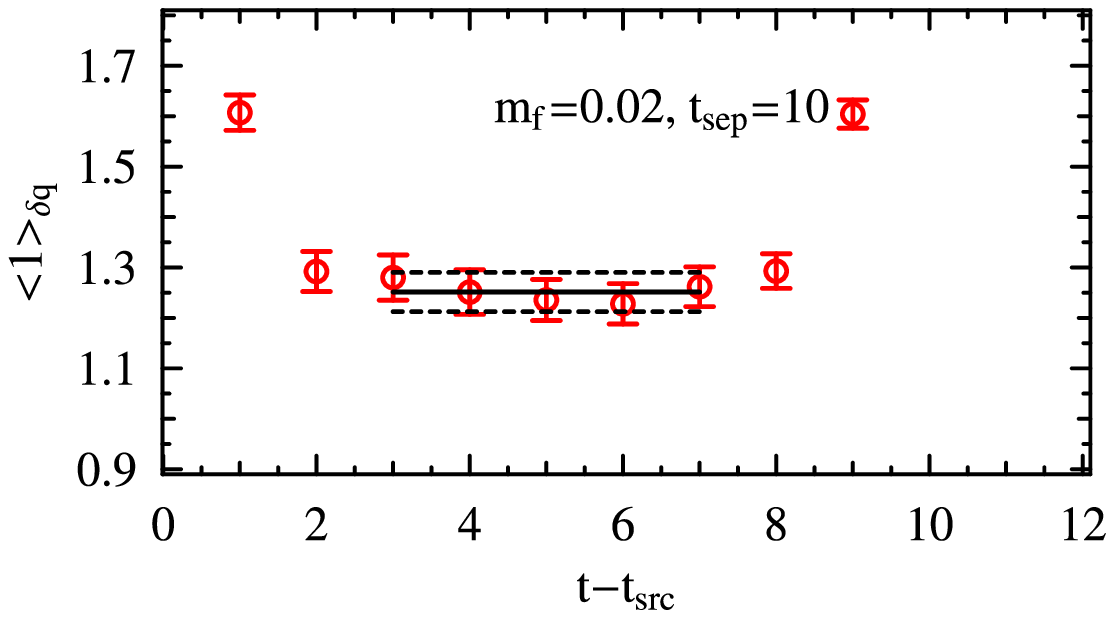}
\includegraphics[width=0.9\columnwidth,clip]{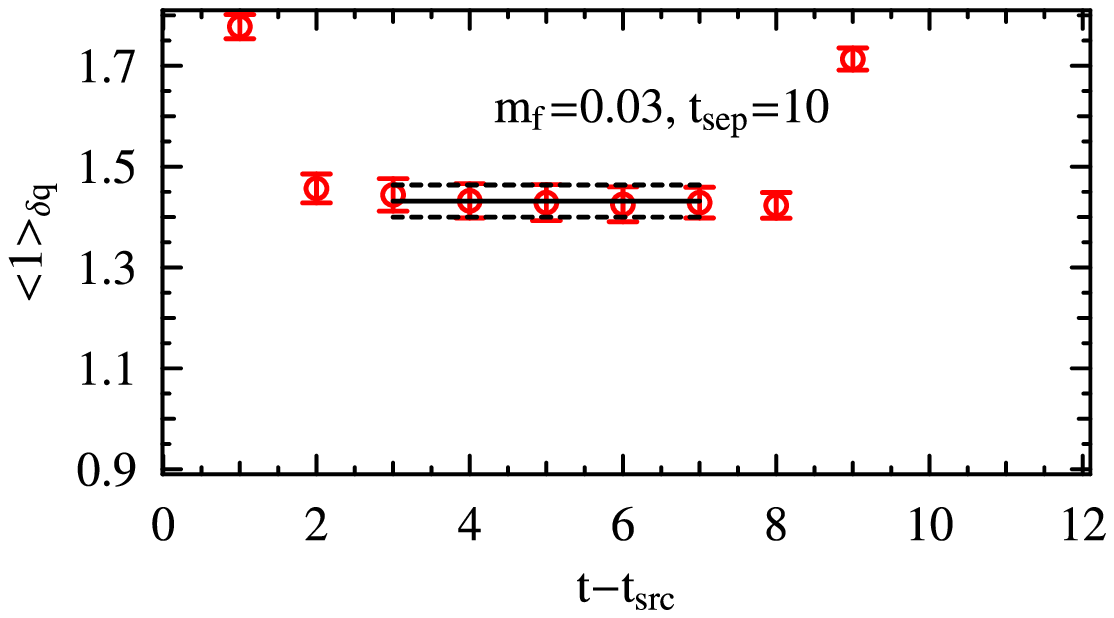}
\includegraphics[width=0.9\columnwidth,clip]{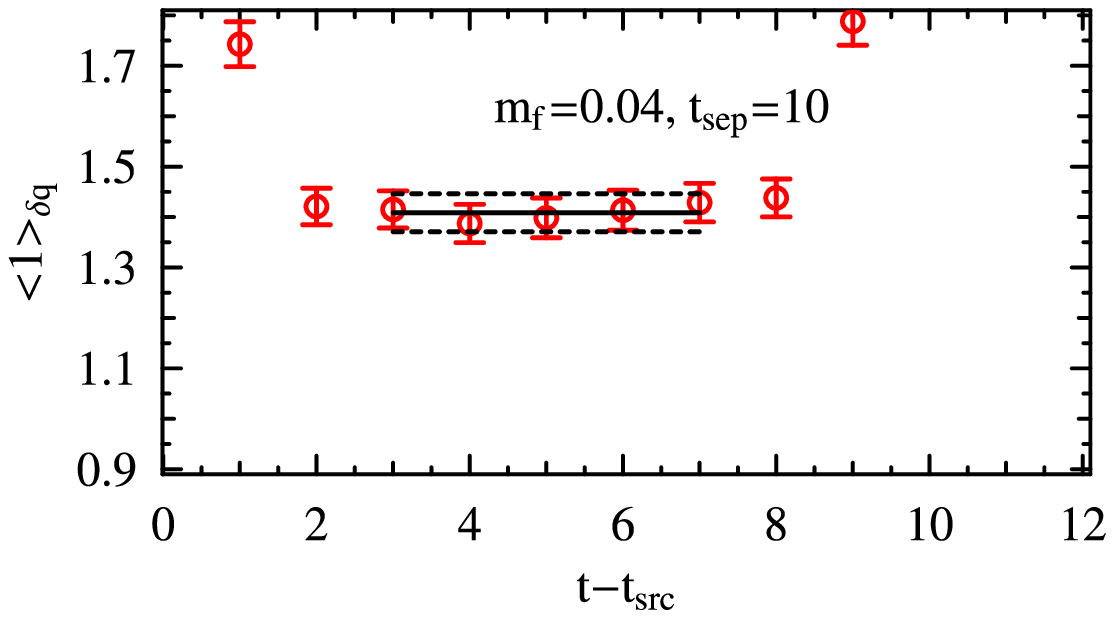}
\caption{The bare values of the transversity; symbols as in Fig.~\ref{fig:rawxq}.} \label{fig:rawtrans}
\end{figure}

\begin{table}[h]
\caption{Bare transversity.}
\label{tab:transversity}
\begin{center}
\begin{tabular}{ccl}
\hline\hline
$m_f$ & $t_{\rm sep}$ &
\multicolumn{1}{c}{$\langle 1 \rangle_{\delta u - \delta d}$}\\
\hline
0.02& 12 & 1.29(11)\\
0.02& 12 & 1.25(4)\\
0.03& 10 & 1.43(3)\\
0.04& 10 & 1.41(4)\\
\hline\hline
\end{tabular}
\end{center}
\end{table}
Another interesting quantity regarding the spin structure of the nucleon is its transversity\cite{Jaffe:1993xb,GrossePerdekamp:2002eb}, $\langle 1\rangle_{\delta q}$.
Here again we have calculated only the isovector quantity. The raw data are shown in Fig.~\ref{fig:rawtrans}, together with the fit range we use; Table~\ref{tab:transversity} summarizes the bare values. Again, at the $m_f=0.02$ point, we use two different source-sink separations to
control possible excited state contributions; in this case the two choices give results that are
comfortably consistent. Therefore, we use the value obtained from $t_{\rm sep}=10$ for all further
analysis and discussion.

\begin{figure}[t]
\includegraphics[width=0.9\columnwidth,clip]{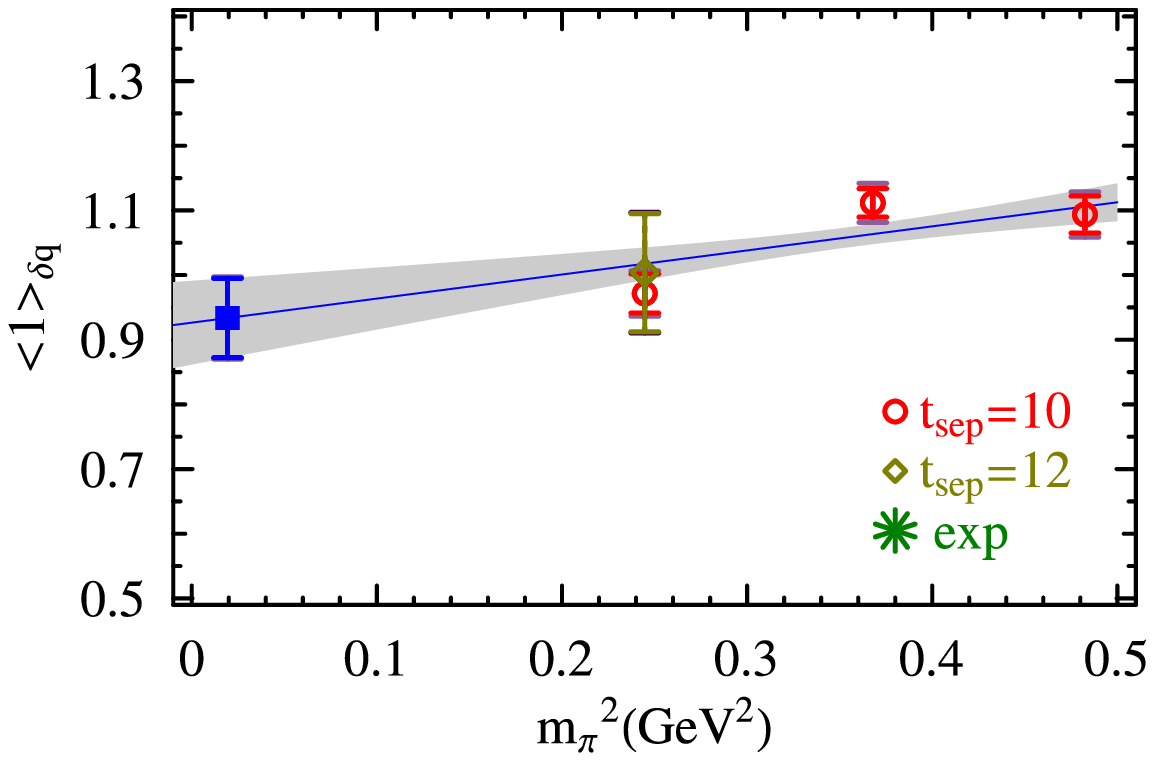}
\caption{Nonperturbatively renormalized transversity with linear extrapolation to the physical pion mass. (Graphic conventions as in Fig.~\ref{fig:zxq}.)}
\label{fig:transversity}
\end{figure}

Using these fit values and the renormalization at the chiral limit listed in Table~\ref{tab:Zfactors} we obtain the quark-mass dependence as shown in Fig.~\ref{fig:transversity}.
Similar to the axial charge radius, we observe a significant deviation at the lightest quark mass. We extrapolate linearly and obtain $\langle 1 \rangle_{\delta u - \delta d} = 0.93(6)+0.37(17) m_\pi^2$ with a $\chi^2$ per degree of freedom of 7.3. This translates into a value at the physical point of 0.93(6). It will be interesting to compare this result with experiment when the latter becomes available.

\subsection{Twist-3 moment}\label{subsec:d1}

\begin{figure}[t]
\includegraphics[width=0.9\columnwidth,clip]{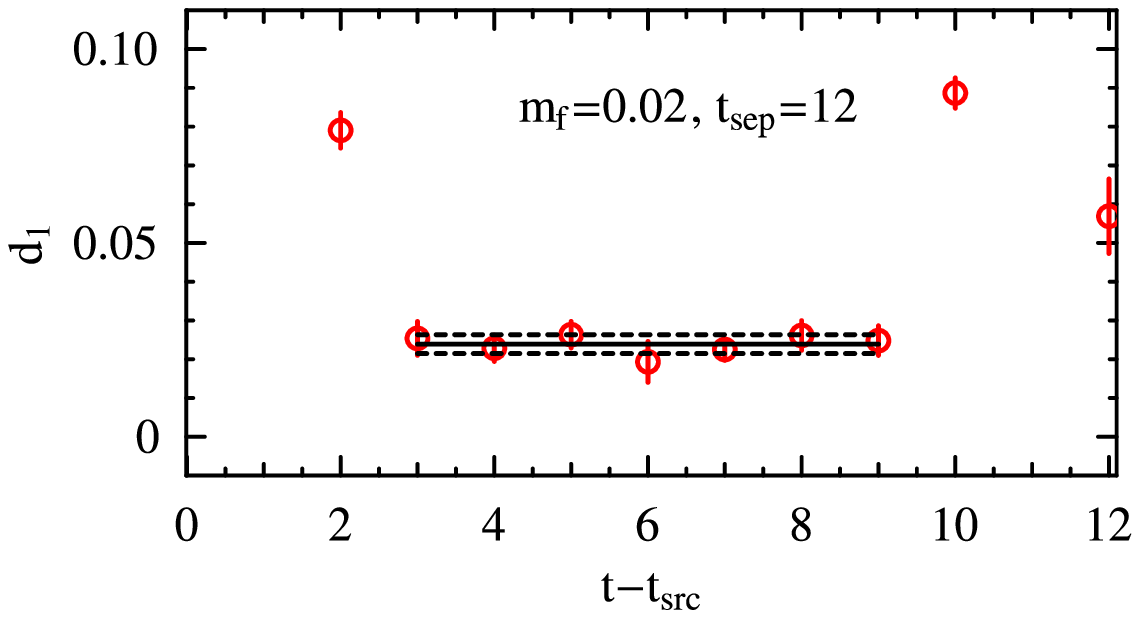}
\includegraphics[width=0.9\columnwidth,clip]{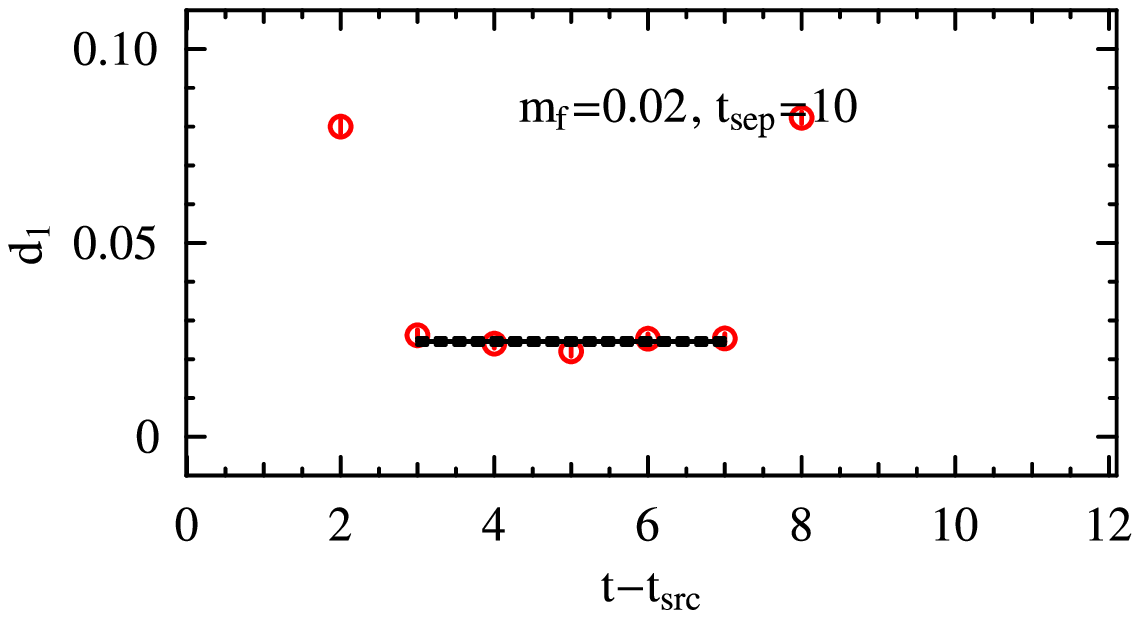}
\includegraphics[width=0.9\columnwidth,clip]{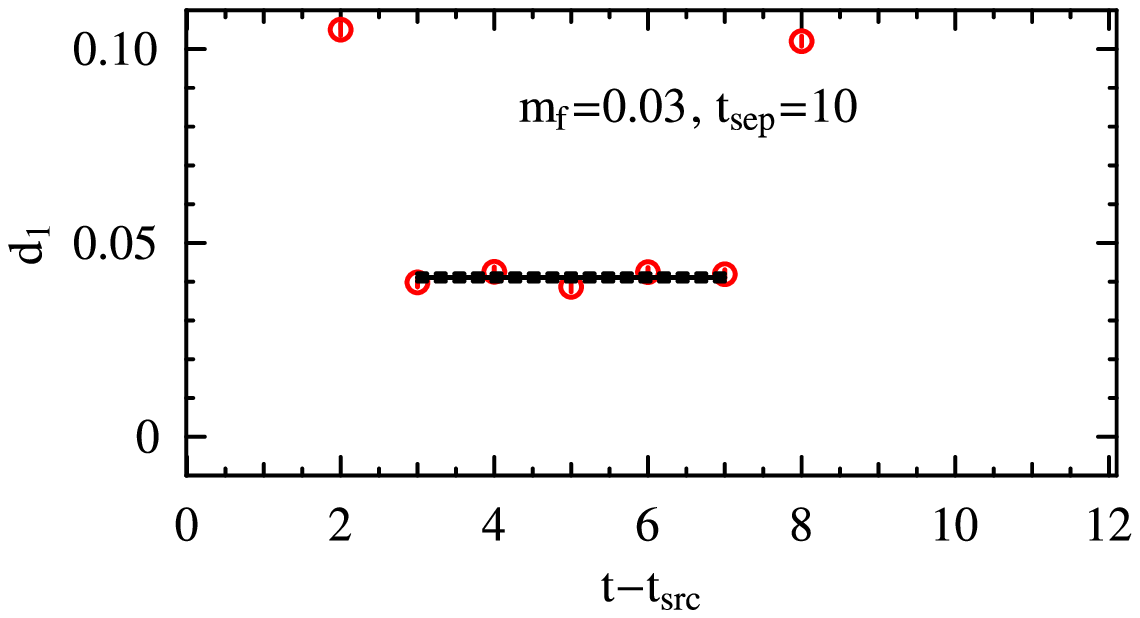}
\includegraphics[width=0.9\columnwidth,clip]{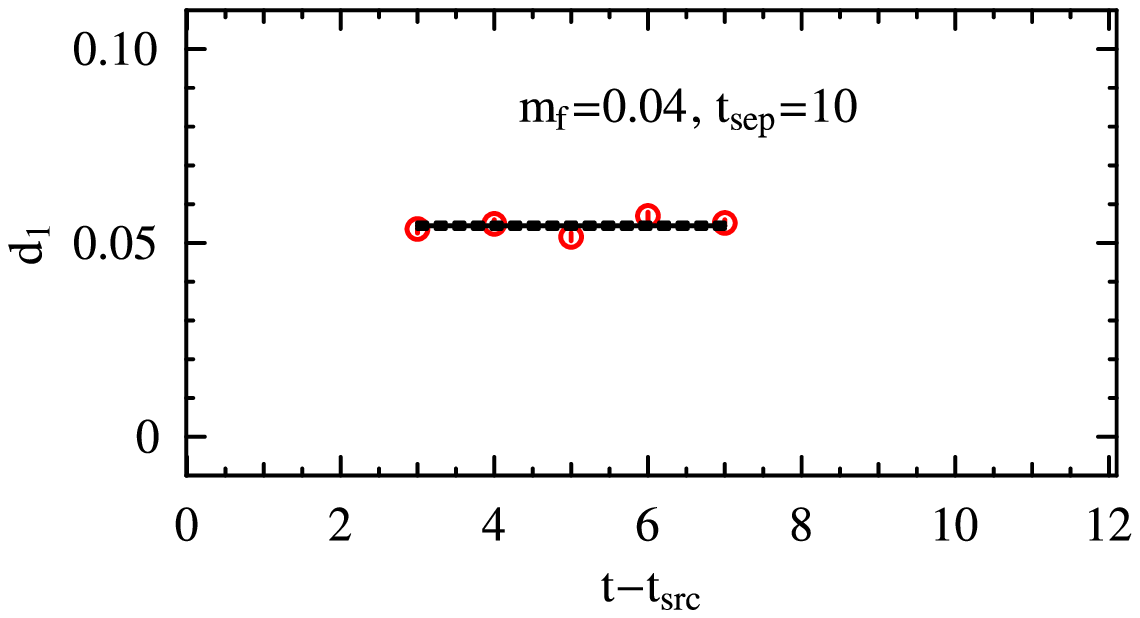}
\caption{The bare twist-3 moment $d_1$; symbols as in Fig.~\ref{fig:rawxq}.}
\label{fig:rawd1}
\end{figure}

\begin{figure}[h]
\includegraphics[width=0.9\columnwidth,clip]{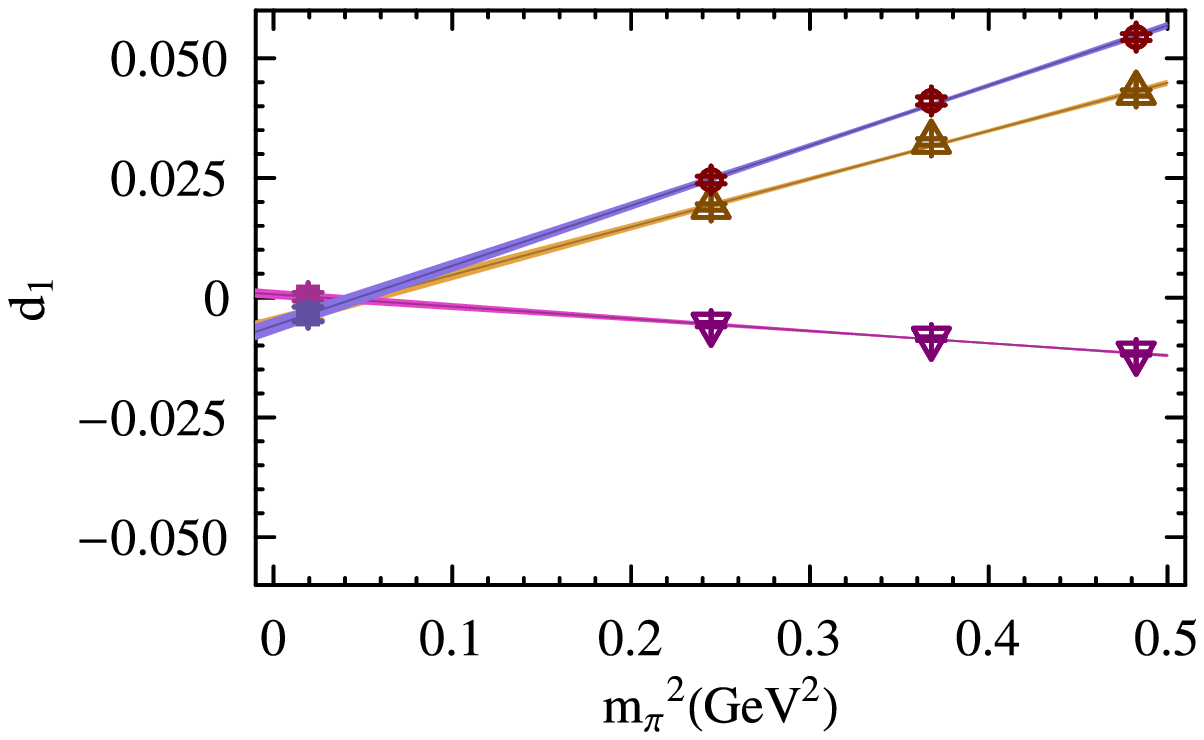}
\caption{Bare values of the twist-3 moment $d_1$,
with linear extrapolation to the physical pion point. Up quark (up triangles), down quark (down triangles), and the isovector combination (circles) are shown.}
\label{fig:d1}
\end{figure}

We also calculated the twist-3 first moment of the polarized structure function, $d_1$.  The good chiral symmetry of DWF prevents lower-dimensional operators from spoiling the calculation, unlike in calculations with fermion discretizations that violate chiral symmetry, such as Wilson fermions. The calculated bare values are summarized in Fig.~\ref{fig:rawd1} and Table~\ref{tab:d1}.
The quark-mass dependence of the bare values is plotted in Fig.~\ref{fig:d1}. (Note that we do not intend to compare these calculations with experiment so do not renormalize them.)
A linear fit yields $-0.0059(16)+0.126(4) m_\pi^2$ and an extrapolated value $d_1^{\rm bare}=-0.0035(13)$ at the physical pion mass, which is about three standard deviations away from zero,
but small compared to the values at non-zero mass.
This suggests the Wandzura-Wilczek relation~\cite{Wandzura:1977qf}, which need not hold in a confining theory, nevertheless holds in QCD.

\begin{table}[h]
\caption{Bare values of the twist-3 moment $d_1$ of the polarized structure function.}
\label{tab:d1}
\begin{center}
\begin{tabular}{cclll}
\hline\hline
$m_f$  & $t_{\rm sep}$ &
\multicolumn{1}{c}{$d_1^{u-d}$} &
\multicolumn{1}{c}{$d_1^u$}&
\multicolumn{1}{c}{$d_1^d$}\\
                  \hline
0.02  &12&  0.0240(25) & 0.0185(19)& $-$0.0055(13)\\
0.02  &10&  0.0246(8)  & 0.0189(7) & $-$0.0056(4)\\
0.03  &10&  0.0411(8)  & 0.0325(6) & $-$0.0085(3)\\
0.04  &10&  0.0544(7)  & 0.0428(6) & $-$0.0117(3)\\
\hline\hline
\end{tabular}
\end{center}
\end{table}

\section{Conclusion and Outlook}
\label{sec:conclusion}
In this work, we have presented numerical lattice QCD calculations using two degenerate flavors of dynamical (DWF) quarks. We calculated isovector vector, axial, tensor and pseudoscalar form factors and some low-order moments of the unpolarized and polarized structure functions.

We found that the ratio of the axial charge to vector charge, $g_A/g_V$, has a significant drop at the lightest pion mass point (about 500~MeV) which may be a sign of finite-volume effect; the heavier pion mass points (around 600 and 700~MeV) are in rough agreement with experiment and show small dependence on the quark mass. A linear extrapolation in pion mass-squared gives 0.89(6), which deviates from experiment by more than five standard deviations; leaving out the lightest point gives 1.23(12). If the downward trend at the lightest point, away from experiment, is  indeed a finite-volume effect, then even $m_\pi L\approx 5$ is not large enough for nucleon calculations with dynamical fermions, in contrast to quenched calculations~\cite{Sasaki:2003jh}\footnote{This effect may in fact be illusory due to the presence of quenched chiral logs as recently argued in~\cite{Yamazaki:2008py}}.

We studied the momentum-transfer dependence of the Dirac, Pauli, axial and induced pesudoscalar form factors. Using a conventional dipole extrapolation, we found masses  $M_V$, $M_T$ and $M_A$ at each pion mass points that are more than 50\% higher than the experimental ones, while the mean-square charge radii, $\langle r_V^2 \rangle$, $\langle r_T^2 \rangle$, $\langle r_A^2 \rangle$ are far below experiment. Similar effects have been observed in the past dynamical calculations~\cite{Edwards:2006qx}. Most phenomenological studies expect to see a dramatic increase at lower pion mass; although this trend is not observed in our current study, it may be resolved in the future with lighter pion calculations.

The magnetic moments of nucleon were calculated in two different ways: first, directly from extrapolation of the Pauli form factor and second, by looking at the ratio of the isovector electric and magnetic moments $G^v_M(0)/G_E^v(0)$ in the forward limit; the latter expression has milder momentum dependence. Using the ratio method, we see decreasing $m_\pi^2$ dependence, and with a simple linear extrapolation we find  $\mu_p-\mu_n=3.4(7)$, roughly consistent with the experimental value.

In studying the axial current form factors, we find that the renormalized axial form factor at finite momenta appears to be in broad agreement with experimental values from neutrino scattering. The Goldberger-Treiman relation appears to hold for low momentum transfer, $q^2\le 0.5\mbox{ GeV}^2$. Assuming the relation, we obtain an estimate for the pion-nucleon coupling of $g_{\pi NN}=15.5(1.4)$ and $g_P=\frac{m_{\mu}}{2m_{N}} G_{P}(0.88 m_\mu^2) = 7.68 \pm 1.03$, which is consistent with experiment.

For the structure functions,
the renormalized values of $\langle x\rangle_{u-d}$ and  $\langle x\rangle_{\Delta u-\Delta d}$ show a trend toward their respective experimental values at the lightest quark mass point in our study.
The quark momentum fraction, $\langle x \rangle_{u-d}(2\mbox{ GeV})$, is 0.282(19) by simple linear chiral extrapolation, which overshoots the experimental value by more than three standard deviations. The quark helicity fraction, $\langle x \rangle_{\Delta u-\Delta d}(2\mbox{ GeV})$, is 0.286(25) by simple linear extrapolation, which overshoots the experimental value by almost four standard deviations.
The former does not deviate significantly from a linear ansatz when the source-sink separation is set to 10 time slices. When increased to 12, the lightest point lies below the straight line fit, suggesting
possible excited state contamination for $t_{\rm sep}=10$, and perhaps the desired physical quark mass dependence for $t_{\rm sep}=12$.
However, the statistics for the larger separation are relatively poor, and definitive conclusions can not be drawn. A similar pattern holds for the helicity fraction, except that at $t_{\rm sep}=10$ a non-linear quark mass dependence is already observed. Non-linear fits motivated by chiral perturbation theory were also used to fit the data. Again, these indicated favorable trends in some cases, but definitive results require simulations with lighter quark masses and multiple volumes, work that is now well underway.

We also calculated  transversity and twist-3 matrix elements. The transversity renormalized at 2~GeV, $\langle 1 \rangle_{\delta u - \delta d}(2\mbox{ GeV})$,  is 0.93(6) by simple linear extrapolation  in $m_\pi^2$. However, there is noticeable non-linearity in the data.
A linear extrapolation to the physical point yields $d_1^{\rm bare}=-0.0035(13)$ for the twist-3 operator, which is about three standard deviations away from zero, indicating only small breaking of the Wandzura-Wilczek relation~\cite{Wandzura:1977qf}.

Unfortunately, even with our dynamical calculations, we cannot resolve the long-standing differences between the predictions of QCD and experimental measurements, in such quantities as the charge radii, and quark momentum and helicity fractions. If both QCD and the relevant experiments are correct, we expect to see a dramatic shift in these values as the constraints of high quark mass, small volume and coarse lattice spacing are lifted. The limitations of the current gauge ensembles do not allow such a study at this stage. However, we are currently studying these quantities with a new series of ensembles produced by the RBC and UKQCD Collaborations~\cite{Antonio:2006px,Allton:2007hx}.  And another
ensemble is being generated by these groups and LHPC.
They are respectively at $a^{-1}=1.7$ and 2.1~GeV lattice cutoff with up to ($2.75\mbox{ fm})^3$ volumes; they feature physical dynamical strange quark mass and two degenerate up and down quarks as light as 1/7 the strange mass. This will give us better control of the systematic errors in a calculation with much lighter pion masses in the valence and the sea sectors; whether we will start to see the curvature suggested by the chiral perturbation theory or other phenomenological models will be very interesting to discover.

\section*{ACKNOWLEDGMENTS}

The authors would like to acknowledge S.~Choi for his private communication providing actual values of the form factor $g_P(q^2)$ in his experiment and Kostas Orginos for his useful discussions on this work.  We thank RBC members for physics discussions and RIKEN,
Brookhaven National Laboratory and the U.S. Department of Energy
for providing the facilities essential for the completion of this
work. In addition, TB and TY were partially funded by US Department of Energy Outstanding Junior Investigator Grant DE-FG02-92ER40716; SS is supported by the JSPS for a Grant-in-Aid for Scientific Research (C) (No. 19540265); HWL is supported by DOE contract DE-AC05-06OR23177 under which the Jefferson Science Associates, LLC operates the Thomas Jefferson National
Accelerator Facility.

\appendix
\section*{Appendices}

\subsection{Nonperturbative renormalization}
\label{subsec:renorm}
In this appendix, we describe the formulation we used in the
nonperturbative renormalization in more detail. In subsection~\ref{subsec:NPR}, we obtain renormalization constants in RI/MOM scheme. In order to compare our results with the experimental ones, we need to further to convert the renormalization
constant to $\overline{\rm MS}$ scheme.
In the case of the first moments of the momentum fraction and helicity operators, one finds the conversion equations in the
continuum via~\cite{Gockeler:1998ye}
\begin{eqnarray}
\frac{Z^{{\rm \overline{MS}}}}{Z^{{\rm RI}}}& = &
    1 + \frac{g^2}{16 \pi^2} C_{F} \Bigg[ G_n + (1-\xi) S_{n-1}
\nonumber \\
&+&  \left( - \frac{4}{n+1} + (1-\xi) \frac{2}{n} \right)
      \frac{\left( \sum_\mu p_\mu h_\mu (p) \right)^2 }
      {p^2 \sum_\mu h_\mu (p)^2}    \Bigg]
      \nonumber \\
\end{eqnarray}
with
\begin{eqnarray}
G_n & = & \frac{2}{n(n+1)} \left( -3-S_{n-1}+ 2 S_{n+1} \right)
 \nonumber \\
& +& \frac{2}{n+1} - 4 \sum_{j=2}^n \frac{1}{j} \left( 2S_j -
S_{j-1} \right) -1 \,,\\
S_n & = & \sum_{j=1}^n \frac{1}{j} \,,\\
h_\mu (p) & =& \sum_{\mu_2,\ldots,\mu_n} c_{\mu \mu_2 \ldots \mu_n}
              p_{\mu_2} \cdots p_{\mu_n} \,,
\end{eqnarray}
where $\xi=0$ in Landau gauge.
 $c_{\mu \mu_2 \ldots
\mu_n}=c_{\mu \nu}$ are $\delta_{\mu 4}\delta_{\nu
4}-\frac{1}{3}\sum_{k=1}^3\delta_{\mu k}\delta_{\nu k}$ for momentum fraction and
$\delta_{\mu 3}\delta_{\nu 4}+\delta_{\mu 4}\delta_{\nu 3}$ for helicity operators. In the tensor-current case, we use the matching factor  calculated from Ref.~\cite{Aoki:2007xm}:
\begin{eqnarray}
\frac{Z^{{\rm \overline{MS}}}}{Z^{{\rm RI}}} &=&
1+\left(\frac{\alpha_s}{4\pi}\right)^2\frac{1}{81}(-4866+1656\zeta (3)+259 n_f).\nonumber \\
\end{eqnarray}

We use the three-loop solution for the running of
$\alpha_s$~\cite{Gimenez:1998ue}:
\begin{eqnarray}
\frac{\alpha_s}{4 \pi} &=& \frac{1}{\beta_0 \ln \left( \mu^2 /
\Lambda^2_{\rm QCD} \right) } - \frac{\beta_1 \ln \ln \left( \mu^2 /
\Lambda^2_{\rm QCD} \right) }{ \beta_0^3 \ln^2 \left( \mu^2 /
\Lambda^2_{\rm QCD} \right)}
\nonumber \\
&+&  \frac{1}{\beta_0^5 \ln ^3\left( \mu^2 / \Lambda^2_{\rm QCD}
\right) }\left\{ \beta_1^2 \ln^2 \ln \left( \mu^2 /
\Lambda^2_{\rm QCD} \right) \right. \nonumber \\
&-& \left.   \beta_1^2 \ln \ln \left( \mu^2 / \Lambda^2_{\rm QCD}
\right) + \beta_2 \beta_0 - \beta_1^2 \right\}
\end{eqnarray}
where $\alpha = \frac{g^2}{4 \pi} \,$ and $C_{F} = \frac{N_C^2 -1}{2
N_C}$, $\Lambda_{\rm QCD}=300$~MeV\cite{Izubuchi:2003rp}. The
various $\beta_i$ are
\begin{eqnarray*}
\beta_0&=& \frac{11 N_c-2 n_f}{3}
\end{eqnarray*}
\begin{eqnarray*}
\beta_1&=& \frac{34 N_c^2}{3}-\frac{10}{3} N_c n_f - \frac{(N_c^2-1) n_f}{N_c}
\end{eqnarray*}
\begin{eqnarray}
\beta_2&=& \frac{2857}{54} N_c^3 + \frac{(N_c^2 - 1)^2}{4
N_c^2}n_f - \frac{205}{36} (N_c^2 - 1) n_f \nonumber \\ &-&
\frac{1415}{54}N_c^2n_f\nonumber \\
&+& \frac{11}{18} \frac{N_c^2 - 1}{N_c} n_f^2 + \frac{79}{54} N_c
n_f^2\quad  \mbox{in $\overline{\rm MS}$ scheme},
\end{eqnarray}
where $N_c$ is the number of colors and $n_f$ is the number of
quark flavors. In the quenched approximation, $n_f=0$; in this
work, $n_f=2$.

The running of the $Z$-factors to two loops~\cite{Buras:1998ra} is
\begin{eqnarray} C(\mu^2) =
 \alpha_s(\mu)^{\overline{\gamma}_0}
\left\{ 1 + \frac{\alpha_s(\mu)}{4 \pi} \left( \overline{\gamma}_1
-\overline{\beta}_1 \overline{\gamma}_0 \right) \right\},
\end{eqnarray}
where $\overline{\gamma}_i$ is defined by the anomalous dimensions
$\gamma_i$ divided by $2\beta_0$. The anomalous dimensions for
twist-two operators of interest were calculated by Floratos {\it
et~al.}~\cite{Floratos:1977au}:
\begin{eqnarray}
\gamma_0 &=& 2  C_{F} \left(1 - \frac{2}{n (n + 1)} + 4
\sum_{j=2}^{n}  \frac{1}{j}\right) \nonumber \\
\gamma_1 &=& (-8.30 C_{F}^2+27.85 C_{F} C_{A} -9.48 C_{F} T_{F}),
\end{eqnarray}
for twist-${n}$ operator. (Note that where $T_{F}(n_f)=\frac{n_f}{2}$, $C_F=\frac{N_c-1}{2N_c}$
and $C_{A}=N_c$.) The anomalous dimensions for the operator, such as ${\cal O}_{34}^{\sigma q} $, can be
found in Ref.~\cite{Broadhurst:1994se}
\begin{eqnarray}
\gamma_0 &=& 2  C_{F}  \nonumber \\
\gamma_1 &=& \frac{C_{F}}{9}(257 C_{A}-171 C_{F} -52 T_{F}).
\end{eqnarray}

\subsection{Chiral extrapolation models}

In this appendix, we collect some relevant formulae used in this paper from various models proposed to conduct chiral extrapolations for baryonic observables.

\subsubsection{Finite-volume correction on $g_A$}
\label{subsec:BeaneSavage}
The small-scale expansion scheme~\cite{Hemmert:1997ye} tries to use explicit degrees of freedom from the pion, nucleon and $\Delta(1232)$ resonance, expanding in terms of $\Delta_0$, the mass
splitting between the $N$ and $\Delta$ in the chiral limit.
This splitting is treated as a small parameter, of the order $O(\epsilon)$. They define $g_A^0$ and $F_{\pi}^0$ as the value of the axial charge $g_A$ and the pion decay constant $F_{\pi}$ in the chiral limit, $c_A$ the $N \Delta$ and $g_1$ the $\Delta \Delta$ axial coupling constants respectively. We simply abbreviate $\Delta_0$, $g_A^0$ and $F_{\pi}^0$ as $\Delta$, $g_A$ and $F_{\pi}^0$ in the following formula.

At the leading order $O(\epsilon^3)$, the finite-volume correction to the nucleon axial charge, $g_A$, was proposed in Ref.~\cite{Beane:2004rf} within the SSE scheme:
\begin{eqnarray}
\delta g_A & = &
{m_\pi^2\over 3\pi^2 F_{\pi}^2}\left[\ g_A^3 {\bf F_1}
 +2c_A^2\left( g_A - {25\over 81}g_{1}\right) {\bf F_2} \right. \nonumber\\
& +& \left. g_A  {\bf F_3} + 2c_A^2 g_A {\bf F_4} \right], \label{eq:gA_FV}
\end{eqnarray}
where
\begin{widetext}
\begin{eqnarray}
{\bf F_1}(m,L)  &=&  \sum_{{\bf n}\ne {\bf 0}} \left[\ K_0(m L
|{\bf n}|) - {K_1(m L |{\bf n}|)\over m L |{\bf n}|} \ \right]; \nonumber\\
{\bf F_2}(m,\Delta,L)&=& -\sum_{{\bf n}\ne {\bf 0}} \left[\
{K_1(m L |{\bf n}|)\over m L |{\bf n}|} +
{\Delta^2-m^2\over m^2} K_0(m L |{\bf n}|) -\ {\Delta\over m^2}
\int_m^\infty d\beta  \; { 2\beta K_0(\beta L |{\bf n}|) +(\Delta^2-m^2) L
|{\bf n}| K_1(\beta L |{\bf n}|) \over
\sqrt{\beta^2+\Delta^2-m^2}} \right]; \nonumber\\
{\bf F_3}(m,L) &=& -{3\over 2} \sum_{{\bf n}\ne {\bf 0}} {K_1(m L
|{\bf n}|)\over m L |{\bf n}|}; \nonumber\\
{\bf F_4}(m,\Delta,L)&=& {8\over 9} \sum_{{\bf n}\ne {\bf 0}}
\left[\ {K_1(m L |{\bf n}|)\over m L |{\bf n}|} - {\pi e^{-m L
|{\bf n}|}\over 2\Delta L |{\bf n}|} -
{\Delta^2-m^2\over m^2\Delta} \int_m^\infty d\beta \ { \beta\
K_0(\beta L |{\bf n}|)\over \sqrt{\beta^2+\Delta^2-m^2}} \right],
\end{eqnarray}
\end{widetext}
with modified Bessel function of the second kind $K_\alpha(z)$.
(Note that the $m$ in the above equations are short for $m_\pi$.)
This formula can be reduced to that of LO HBChPT with $c_A=0$.

\subsubsection{The mean-squared Dirac radius in HBChPT}
\label{subsec:MSR-HBChPT}
Heavy baryon chiral perturbation theory (HBChPT)\cite{Bernard:1992qa}
involves only the pion and nucleon, expanding in terms of the momentum $p$. At leading order (LO), once we use the physical values of $g_A$, $F_{\pi}$ and $\langle r_V^2 \rangle$ instead of their values in the chiral limit, the HBChPT formula provides a closed form
for the $m_{\pi}^2$ dependence of the Dirac mean-squared radius without unknown parameters~\cite{{Beg:1973sc},{Bernard:1992qa}}:
\begin{equation}
\langle r_V^2 \rangle(m_{\pi, {\rm lat}})
=\langle r_V^2 \rangle_{\rm exp}-\frac{1+5g^2_{A, {\rm exp}}}{
(4\pi F_{\pi, {\rm exp}})^2} {\rm ln}\left(
\frac{m^2_{\pi, {\rm lat}}}{\mu^2}
\right)
\end{equation}
where the scale $\mu$ should be set by the physical value of the pion mass as $\mu=0.139$~GeV~\footnote{A missing
factor of 2 in the prefactor of $\ln m_{\pi}^2$ term, which
can be found in Ref.~\cite{Leinweber:1992hj}, causes an accidental
agreement between this leading one-loop form and lattice data obtained
in the heavy quark mass region.
}.

\subsubsection{Chen, {\it et~al.}}
\label{subsec:Chen}

Chen~{\it et~al.}~\cite{Chen:2001eg,Chen:2001gr,Detmold:2002nf} calculate a chiral
perturbation expression in the continuum for the moments of the
unpolarized and helicity distributions:
\begin{widetext}
\begin{eqnarray}
\langle x\rangle_{u-d} &=& C \left[ 1
 - \frac{3g_{A,{\rm exp}}^2 + 1}{(4\pi F_{\pi,{\rm exp}})^2}m_{\pi, {\rm lat}}^2
 \ln\left(\frac{m_{\pi, {\rm lat}}^2}{\mu^2}\right)\right]
+e(\mu^2)\frac{m_{\pi, {\rm lat}}^2}{(4\pi F_{\pi,{\rm exp}})^2} \\
\langle x\rangle_{\Delta u- \Delta d} &=& \tilde{C} \left[ 1 -
\frac{2g_{A,{\rm exp}}^2 + 1}{(4\pi F_{\pi,{\rm exp}})^2} m_{\pi, {\rm lat}}^2
\ln\left(\frac{m_{\pi, {\rm lat}}^2}{\mu^2}\right)\right]
+ \tilde{e}(\mu^2)\frac{m_{\pi, {\rm lat}}^2}{(4\pi F_{\pi,{\rm exp}})^2}.
\label{eq:Chenmoments}
\end{eqnarray}
\end{widetext}
We use $\mu=0.139$~MeV and $F_{\pi,{\rm exp}}=92.4$~MeV  in our extrapolation.

There is a variation of this formulation replacing the $F_{\pi,{\rm exp}}$ with $F_{\pi,{\rm lat}}$ and changing the scale $\mu$ to $F_{\pi,{\rm lat}}$. This provides a dimensionless quantity $\frac{m_{\pi,{\rm lat}}^2}{F_{\pi,{\rm lat}}^2}$ in the extrapolation, which is independent of the scale setting in the lattice calculation.


\end{document}